\documentclass{article}

\usepackage{epubtk}
\usepackage{graphicx}
\usepackage{amssymb}
\usepackage{amsfonts}
\usepackage{amsmath}

\showlistoftablesfalse

\def\l{\left}
\def\r{\right}

\def\interv{\l[-1, 1\r]}
\def\fraction{\displaystyle\frac}
\let\ph\varphi
\let\th\theta

\begin{document}

\title{Spectral Methods for Numerical Relativity}

\author{%
\epubtkAuthorData{Philippe Grandcl\'ement}{%
Laboratoire Univers et Th\'eories \\
UMR 8102 du C.N.R.S., Observatoire de Paris \\
F-92195 Meudon Cedex, France}{%
Philippe.Grandclement@obspm.fr}{%
http://www.luth.obspm.fr/minisite.php?nom=Grandclement}%
\\
\and \\
\epubtkAuthorData{J\'er\^ome Novak}{%
Laboratoire Univers et Th\'eories \\
UMR 8102 du C.N.R.S., Observatoire de Paris \\
F-92195 Meudon Cedex, France}{%
Jerome.Novak@obspm.fr}{%
http://www.luth.obspm.fr/minisite.php?nom=Novak}%
}

\date{}
\maketitle

\begin{abstract}
  Equations arising in General Relativity are usually too complicated to be
  solved analytically and one has to rely on numerical methods to solve sets
  of coupled partial differential equations. Among the possible choices, this
  paper focuses on a class called spectral methods where, typically, the
  various functions are expanded onto sets of orthogonal polynomials or
  functions. A theoretical introduction on spectral expansion is first given
  and a particular emphasis is put on the fast convergence of the spectral
  approximation. We present then different approaches to solve partial
  differential equations, first limiting ourselves to the one-dimensional
  case, with one or several domains. Generalization to more dimensions is then
  discussed. In particular, the case of time evolutions is carefully studied
  and the stability of such evolutions investigated. One then turns to results
  obtained by various groups in the field of General Relativity by means of
  spectral methods. First, works which do not involve explicit time-evolutions
  are discussed, going from rapidly rotating strange stars to the computation of
  binary black holes initial data. Finally, the evolutions of various systems
  of astrophysical interest are presented, from supernovae core collapse to
  binary black hole mergers.
\end{abstract}

\epubtkKeywords{Numerical Relativity, Numerical Methods}

\newpage


\section{Introduction}
\label{s:introduction}

Einstein's equations represent a complicated set of nonlinear partial
differential equations for which some exact~\cite{bicak-06} or
approximate~\cite{blanchet-06} analytical solutions are known. But these
solutions are not always suitable for physically or astrophysically
interesting systems, which require an accurate description of their
relativistic gravitational field without any assumption on the symmetry or
with the presence of matter fields, for instance. Therefore, many efforts have
been undertaken to solve Einstein's equations with the help of computers in
order to model relativistic astrophysical objects. Within this field of {\em
  numerical relativity\/}, several numerical methods have been experimented
and a large variety of them are currently being used. Among them, {\em
  spectral methods\/} are now increasingly popular and the goal of this review
is to give an overview (at the moment it is written or updated) of the methods
themselves, the groups using them and the obtained results. Although some
theoretical framework of spectral methods is given in
Sections~\ref{section:theoretical_foundations} to \ref{s:time}, more details
about spectral methods can be found in the books by Gottlieb and
Orszag~\cite{gottlieb-77}, Canuto~et~al.~\cite{canuto-88, canuto-06,
  canuto-07}, Fornberg~\cite{fornberg-95}, Boyd~\cite{boyd-01} and
Hesthaven~et~al.~\cite{hesthaven-07}. While these references have of course
been used for writing this review, they can also help the interested reader to
get deeper understanding of the subject. This review is organized as follows:
hereafter in the introduction, we briefly introduce the spectral methods,
their usage in computational physics and give a simple example.
Section~\ref{section:theoretical_foundations} gives important notions
concerning polynomial interpolation and the solution of ordinary differential
equations (ODE) with spectral methods. Multi-domain approach is also
introduced there, whereas some of the multi-dimensional techniques are
described in Section~\ref{s:space_time}. The cases of time-dependent partial
differential equations (PDE), are treated in Section~\ref{s:time}. The last
two sections are then reviewing results obtained using spectral methods: on
stationary configurations and initial data (Section~\ref{s:station}), and on
the time-evolution (Section~\ref{s:dynamical_evolutions}) of stars,
gravitational waves and black holes.

\subsection{About spectral methods}
\label{ss:about_sm}

When doing simulations and solving PDEs, one faces the problem of representing
and deriving functions on a computer, which deals only with (finite)
integers. Let us take a simple example of a function $f: \interv \to
\mathbb{R}$. The most straightforward way to approximate its derivative is through {\em
  finite-differences methods\/}: first one must setup a {\em grid\/}
$$
\left\{ x_i \right\}_{i=0...N} \subset \interv
$$ 
of $N+1$ points in the interval, and represent $f$ by its $N+1$ values on these
grid points
$$
\left\{ f_i = f(x_i) \right\}_{i=0...N}.
$$
Then, the (approximate) representation of the derivative $f'$ shall be, for
instance
\begin{equation}
  \label{eq:finite_differences}
  \forall i<N, \, f'_i = f'(x_i) \simeq \frac{f_{i+1} - f_i}{x_{i+1} - x_i}.
\end{equation}
If we suppose an equidistant grid, so that $\forall i<N, \, x_{i+1} - x_i =
\Delta x=1/N$, the error in the approximation~(\ref{eq:finite_differences}) will
decay as $\Delta x$ (first-order scheme). One can imagine higher-order
schemes, with more points involved for the computation of each derivative and,
for a scheme of order $n$, the accuracy can vary as $\left( \Delta x
\right)^n=1/N^n$.

{\em Spectral methods\/} represent an alternate way: the function $f$ is no
longer represented through its values on a finite number of grid points, but
using its coefficients (coordinates) $\left\{ c_i \right\}_{i=0...N}$ in a
finite basis of known functions $\left\{ \Phi_i \right\}_{i=0...N}$
\begin{equation}
  \label{eq:spec_gene}
  f(x) \simeq \sum_{i=0}^N c_i \Phi_i(x).
\end{equation}
A relatively simple case is, for instance, when $f(x)$ is a periodic function
of period 2, and the $\Phi_{2i}(x) = \cos(\pi ix), \Phi_{2i+1}\sin(\pi ix)$ are
trigonometric functions. Equation~(\ref{eq:spec_gene}) is then nothing but the
truncated Fourier decomposition of $f$. In general, derivatives can be
computed from the $c_i$'s, with the knowledge of the expression for each
derivative $\Phi'_i(x)$ as a function of $\left\{ \Phi_i \right\}_{i=0...N}$.
The decomposition~(\ref{eq:spec_gene}) is approximate in the sense that
$\left\{ \Phi_i \right\}_{i=0...N}$ represent a complete basis of some
finite-dimensional functional space, whereas $f$ usually belongs to some other
infinite-dimensional space. Moreover, the coefficients $c_i$ are computed with
finite accuracy. Among the major advantages of using spectral methods is the
rapid decay of the error (faster than any power of $1/N$, and in practice
often exponential $e^{-N}$), for well-behaved functions (see
 Section~\ref{sss:convergence}); one therefore has an {\em infinite\/}-order
scheme.

In a more formal and mathematical way, it is useful to work within the methods
of weighted residuals (MWR, see also  Section~\ref{ss:sm_for_ODEs}). Let us
consider the PDE
\begin{eqnarray}
  Lu(x) &=& s(x)\quad x\in U \subset \mathbb{R}^d, \\ \label{eq:mwr_oper}
  Bu(x) &=& 0 \quad x\in \partial U,\label{eq:mwr_bound}
\end{eqnarray}
where $L$ is a linear operator, $B$ the operator defining the boundary
conditions and $s$ is a source term. A function $\bar{u}$ is said to be a {\em
  numerical solution\/} of this PDE if it satisfies the boundary
conditions~(\ref{eq:mwr_bound}) and makes ``small'' the residual
\begin{equation}
  \label{eq:mwr_residual}
  R = L\bar{u} - s.
\end{equation}
If the solution is searched in a finite-dimensional subspace of some given
Hilbert space (any relevant $L^2_U$ space) in terms of the
expansion~(\ref{eq:spec_gene}), then the functions
$\left\{\Phi_i(x)\right\}_{i=0...N}$ are called {\em trial functions\/} and,
in addition the choice of a set of {\em test functions\/}
$\left\{\xi_i(x)\right\}_{i=0...N}$ defines the notion of smallness for the
residual by means of the Hilbert space scalar product
\begin{equation}
  \label{eq:mwr_small}
  \forall i=0...N, \quad \left( \xi_i, R \right) = 0.
\end{equation}
Within this framework, various numerical methods can be classified according
to the choice of the trial functions:

\begin{itemize}
  \item {\bf Finite differences}: the trial functions are overlapping
  local polynomials of fixed order (lower than $N$),
  \item {\bf Finite elements}: the trial functions are local smooth
  functions which are non-zero only on sub-domains of $U$,
  \item {\bf Spectral methods}: the trial functions are global smooth
  functions on $U$.
\end{itemize}

Various choices of the test functions define different types of
spectral methods, as detailed in  Section~\ref{ss:sm_for_ODEs}. Usual
choices for the trial functions are (truncated) Fourier series,
spherical harmonics or orthogonal families of polynomials.

\subsection{Spectral methods in physics}
\label{ss:sm_in_physics}

We do not give here all the fields of physics where spectral methods are being
employed, but we sketch the variety of equations and physical models that have
been simulated with such techniques. Spectral methods originally appeared in
numerical fluid dynamics, where large spectral hydrodynamic codes have been
regularly used to study turbulence and transition to the turbulence, since the
seventies. For fully resolved, direct numerical calculations of Navier--Stokes
equations, spectral methods were often preferred for their high accuracy.
Historically, they also allowed for two- or three-dimensional simulations of
fluid flows, because of their reasonable computer memory requirements. Many
applications of spectral methods in fluid dynamics have been discussed by
Canuto~et~al.~\cite{canuto-88, canuto-07}, and the techniques developed in
that field can be of some interest for numerical relativity.

From pure fluid-dynamics simulations, spectral methods have rapidly been used
in connected fields of research: geophysics~\cite{shen-99}, meteorology and
climate modeling~\cite{temperton-91}. In this last domain of research, they
provide global circulation models that are then used as boundary conditions to
more specific (lower-scale) models, with improved micro-physics. In this way,
spectral methods are only a part of the global numerical model, combined with
other techniques to bring the highest accuracy, for a given computational
power. Solution to the Maxwell equations can, of course, also be obtained with
spectral methods and therefore, magneto-hydrodynamics (MHD) have been studied
with these techniques (see e.g.\ Hollerbach~\cite{hollerbach-00}). This
has been the case in astrophysics too, where for example spectral
three-dimensional numerical models of solar magnetic dynamo action realized by
turbulent convection have been computed~\cite{brun-04}. Still in astrophysics,
the Kompaneet's equation, describing the evolution of photon distribution
function in a bath of plasma at thermal equilibrium within the Fokker-Planck
approximation, has been solved using spectral methods to model the X-ray
emission of {\it Her\/}~X-1~\cite{bonazzola-85, bonazzola-99}. In the
simulations of cosmological structure formation or galaxy evolution, many
N-body codes rely on a spectral solver for the computation of the
gravitational force by the so-called particle-mesh algorithm. The mass
corresponding to each particle is decomposed onto neighboring grid points,
thus defining a density field. The Poisson equation giving the Newtonian
gravitational potential is then usually solved in Fourier space for both
fields~\cite{hockney-81}.

To our knowledge, the first published results on the numerical solution of
Einstein's equations, using spectral methods is the spherically-symmetric
collapse of a neutron star to a black hole by Gourgoulhon in
1991~\cite{gourgoulhon-91}. He used the spectral methods as they have been
developed in the Meudon group by Bonazzola and Marck~\cite{bonazzola-90}.
Later studies of fast rotating neutron stars~\cite{bonazzola-93} (stationary
axisymmetric models), the collapse of a neutron star in tensor-scalar theory
of gravity~\cite{novak-98} (spherically-symmetric dynamical spacetime) and
quasi-equilibrium configurations of binary neutron stars~\cite{bonazzola-99b}
and of black holes~\cite{grandclement-02} (three-dimensional and stationary
spacetimes) have grown in complexity until the three-dimensional
time-dependent numerical solution of Einstein's equations~\cite{bonazzola-04}.
On the other hand, the first fully three-dimensional evolution of the whole
Einstein system has been achieved in 2001 by Kidder~et~al.~\cite{kidder-01},
where a single black hole was evolved until $t\simeq 600M-1300M$, using
excision techniques. They used spectral methods as developed in the
Cornell/Caltech group by Kidder~et~al.~\cite{kidder-00a} and
Pfeiffer~et~al.~\cite{pfeiffer-03a}. Since then, they have focused on the
evolution of a binary black hole system, which has recently been simulated
until the merger and the ring-down by Scheel~et~al.~\cite{scheel-08}. Other
groups (for instance Ansorg~et~al.~\cite{ansorg-03}, Bartnik and
Norton~\cite{bartnik-00}, Frauendiener~\cite{frauendiener-99} and
Tichy~\cite{tichy-06}) have also used spectral methods to solve Einstein's
equations; Sections~\ref{s:station} and~\ref{s:dynamical_evolutions} are
devoted to a more detailed review of all these works.

\subsection{A simple example}
\label{ss:simple_example}

Before entering the details of spectral methods in
Sections~\ref{section:theoretical_foundations}, \ref{s:space_time}
and~\ref{s:time}, let us give here their spirit with the simple
example of the Poisson equation in a spherical shell:
\begin{equation}
  \label{eq:simple_poisson}
  \Delta \phi = \sigma ,
\end{equation}
where $\Delta$ is the Laplace operator~(\ref{eq:laplace_spher}) expressed in
spherical coordinates $(r, \theta, \varphi)$ (see also
 Section~\ref{ss:spherical_coordinates_harmonics}). We want to solve
Equation~(\ref{eq:simple_poisson}) in the domain where $0<R_{\rm min} \leq r \leq
R_{\rm max}, \, \theta \in [0, \pi], \, \varphi \in [0, 2\pi)$. This Poisson
equation naturally arises in numerical relativity when, for example, solving
for initial conditions or the Hamiltonian constraint in the 3+1
formalism~\cite{gourgoulhon-07}: the linear part of these equations can be
cast into the form~(\ref{eq:simple_poisson}), and the non-linearities put into
the source $\sigma$, with an iterative scheme on $\phi$.

First, the angular parts of both fields are decomposed onto a (finite)
set of spherical harmonics $\left\{ Y_\ell^m \right\}$ (see
 Section~\ref{sss:spherical_harmonics}):
\begin{equation}
  \label{eq:simple_decomp_ylm}
  \sigma(r, \theta, \varphi) \simeq \sum_{\ell=0}^{\ell_{\rm max}} \sum_{m =
    -\ell}^{m =\ell} s_{\ell m}(r) Y_\ell^m(\theta, \varphi),
\end{equation}
with a similar formula relating $\phi$ to the radial functions
$f_{\ell m}(r)$. Because spherical harmonics are eigenfunctions of the
angular part of the Laplace operator, the Poisson equation can be
equivalently solved as a set of ordinary differential equations for
each couple $(\ell, m)$, in terms of the coordinate $r$:
\begin{equation}
  \label{eq:simple_eq_lm}
  \forall (\ell, m), \quad  \frac{d^2 f_{\ell m}}{dr^2} + \frac{2}{r} \frac{d f_{\ell
      m}}{dr} - \frac{\ell(\ell+1)f_{\ell m}}{r^2} = s_{\ell m}(r). 
\end{equation}
We then map
\begin{eqnarray}
  [R_{\rm min}, R_{\rm max}] &\to& [-1, 1] \nonumber\\
  r &\mapsto & \xi = \frac{2r - R_{\rm max} - R_{\rm min}}{R_{\rm max} -
    R_{\rm min}},  \label{eq:simple_map}
\end{eqnarray}
and decompose each field onto a (finite) basis of Chebyshev polynomials
$\left\{ T_i \right\}_{i=0...N}$ (see  Section~\ref{sss:cheby}):
\begin{eqnarray}
  s_{\ell m} (\xi) = \sum_{i=0}^N c_{i\ell m} T_i(\xi), \nonumber\\
  f_{\ell m} (\xi) = \sum_{i=0}^N a_{i\ell m} T_i(\xi).  \label{eq:simple_cheb}
\end{eqnarray}
Each function $f_{\ell m}(r)$ can be regarded as a column-vector $A_{\ell m}$
of its $N+1$ coefficients $a_{i\ell m}$ in this basis; the linear differential
operator on the left-hand side of Equation~(\ref{eq:simple_eq_lm}) being thus a
matrix $L_{\ell m}$ acting on this vector:
\begin{equation}
  \label{eq:simple_system}
  L_{\ell m} A_{\ell m} = S_{\ell m},
\end{equation}
with $S_{\ell m}$ being the vector of the $N+1$ coefficients $c_{i\ell m}$ of
$s_{\ell m}(r)$. This matrix can be computed from the recurrence relations
fulfilled by the Chebyshev polynomials and their derivatives (see
 Section~\ref{sss:cheby} for details).

The matrix $L$ is singular, because the
problem~(\ref{eq:simple_poisson}) is ill-posed. One must indeed
specify boundary conditions at $r=R_{\rm min}$ and $r=R_{\rm
  max}$. For simplicity, let us suppose
\begin{equation}
  \label{eq:simple_bc}
  \forall (\theta, \varphi), \quad \phi(r=R_{\rm min}, \theta, \varphi) =
  \phi(r=R_{\rm max}, \theta, \varphi) = 0. 
\end{equation}
To impose these boundary conditions, we shall adopt the tau methods (see
 Section~\ref{s:tau}): we build the matrix $\bar{L}$, taking $L$
and replacing the last two lines by the boundary conditions, expressed in
terms of the coefficients from the properties of Chebyshev polynomials:
\begin{equation}
  \label{eq:simple_coef_bc}
  \forall(\ell, m), \quad \sum_{i=0}^N (-1)^i a_{i\ell m} = \sum_{i=0}^N
  a_{i\ell m} = 0.
\end{equation}
Equations~(\ref{eq:simple_coef_bc}) are equivalent to the boundary
conditions~(\ref{eq:simple_bc}), within the considered spectral
approximation, and they represent the last two lines of $\bar{L}$,
which can now be inverted and give the coefficients of the solution $\phi$.

If one summarizes the steps:

\begin{enumerate}
\item Setup an adapted grid for the computation of spectral
  coefficients (e.g.\ equidistant in the angular directions and
  Chebyshev--Gauss--Lobatto collocation points, see
  Section~\ref{sss:cheby});
\item Get the values of the source $\sigma$ on these grid points;
\item Perform a spherical-harmonics transform (for example using some
  available library~\cite{spharmonicskit-04}), followed by the Chebyshev
  transform (using a Fast Fourier Transform-FFT, or a Gauss--Lobatto
  quadrature) of the source $\sigma$;
\item For each couple of values $(\ell,m)$, build the corresponding matrix
  $\bar{L}$, with the boundary conditions and invert the system
  (using any available linear-algebra package) with
  the coefficients of $\sigma$ as the right-hand side;
\item Perform the inverse spectral transform to get the values of $\phi$ on
  the grid points, from its coefficients.
\end{enumerate}

A numerical implementation of this algorithm has been reported by
Grandcl\'ement~et~al.~\cite{grandclement-01}, who have observed that
the error decayed as $e^{-\ell_{\rm max}}\cdot e^{-N}$, provided that the
source $\sigma$ is smooth. Machine round-off accuracy can be reached with
$\ell_{\rm max} \sim N \sim 30$, which makes the matrix inversions of step~4
very cheap in terms of CPU, and the whole method affordable in terms of memory
usage. These are the main advantages of using spectral methods, as it shall be
shown in the following sections.

\newpage


\section{Concepts in One Dimension}
\label{section:theoretical_foundations}

In this section the basic concepts of spectral methods in one spatial
dimension are presented. Some generalities about the approximation of
functions by polynomials are first exposed. The basic formulas of the spectral
expansion are then given and two sets of polynomials are discussed (Legendre
and Chebyshev polynomials). A particular emphasis is put on convergence
properties (i.e., the way the spectral approximation converges to the real
function).

In Section~\ref{ss:sm_for_ODEs}, three different methods of solving an ordinary
differential equation (ODE) are exhibited and applied to a simple
problem. Section~\ref{ss:multidomain_techniques} is concerned with
multi-domain techniques. After giving some motivations for the use of
multi-domain decomposition, four different implementations are
discussed, as well as their respective merits. One simple example is
given, which uses only two domains.

For problems in more than one dimension see Section~\ref{s:space_time}.

\subsection{Best polynomial approximation}
\label{ss:best_approximation}

Polynomials are the only functions that a computer can exactly evaluate and so
it is natural to try to approximate any function by a polynomial. When
considering spectral methods, one will use global polynomials on a few
domains. This is to be contrasted with finite difference schemes, for
instance, where only local polynomials are considered.

In this particular section, real functions of $\interv$ are
considered. A theorem due to Weierstrass (see for
instance~\cite{courant-53}) states that the set $\mathbb {P}$ of all
polynomials is a dense subspace of all the continuous functions on
$\interv$, with the norm $\l\|\cdot\r\|_\infty$. This maximum norm is
defined as 
\begin{equation}
  \l\|f\r\|_\infty = \max_{x\in\interv} \l|f\l(x\r)\r|.
\end{equation}

This means that, for any continuous function $f$ of $\interv$, there
exists a sequence of polynomials $\l(p_i\r), i\in \mathbb{N}$ that
converges {\em uniformly} towards $f$:
\begin{equation}
  \lim_{i\rightarrow\infty} \l\|f-p_i\r\|_\infty = 0.
\end{equation}
This theorem shows that it is probably a good idea to approximate
continuous functions by polynomials.

Given a continuous function $f$, the best polynomial approximation of
degree $N$, is the polynomial $p^\star_N$ that minimizes the norm of
the difference between $f$ and itself:
\begin{equation}
  \l\|f - p_N^\star\r\|_\infty = {\rm min} \l\{ \l\|f - p\r\|_\infty,
  p \in \mathbb {P}_N \r\}.
\end{equation}

{\em Chebyshev alternate theorem} states that for any continuous
function $f$, $p^\star_N$ is unique (Theorem~9.1
of~\cite{quarteroni-07} and theorem 23 of~\cite{meinardus-67}). There
exist $N+2$ points $x_i \in \interv$ such, that the error is exactly
attained at those points, in an alternate manner: 
\begin{equation}
  f\l(x_i\r) - p^\star_N\l(x_i\r) = \l(-1\r)^{i+\delta} \l\|f -
  p^\star_N\r\|_\infty,
\end{equation}
where $\delta = 0$ or $\delta = 1$. An example of a function and its
best polynomial approximation is shown on Figure~\ref{figure:best_N2}.

\epubtkImage{}{%
\begin{figure}
  \centerline{\includegraphics[height=8cm]{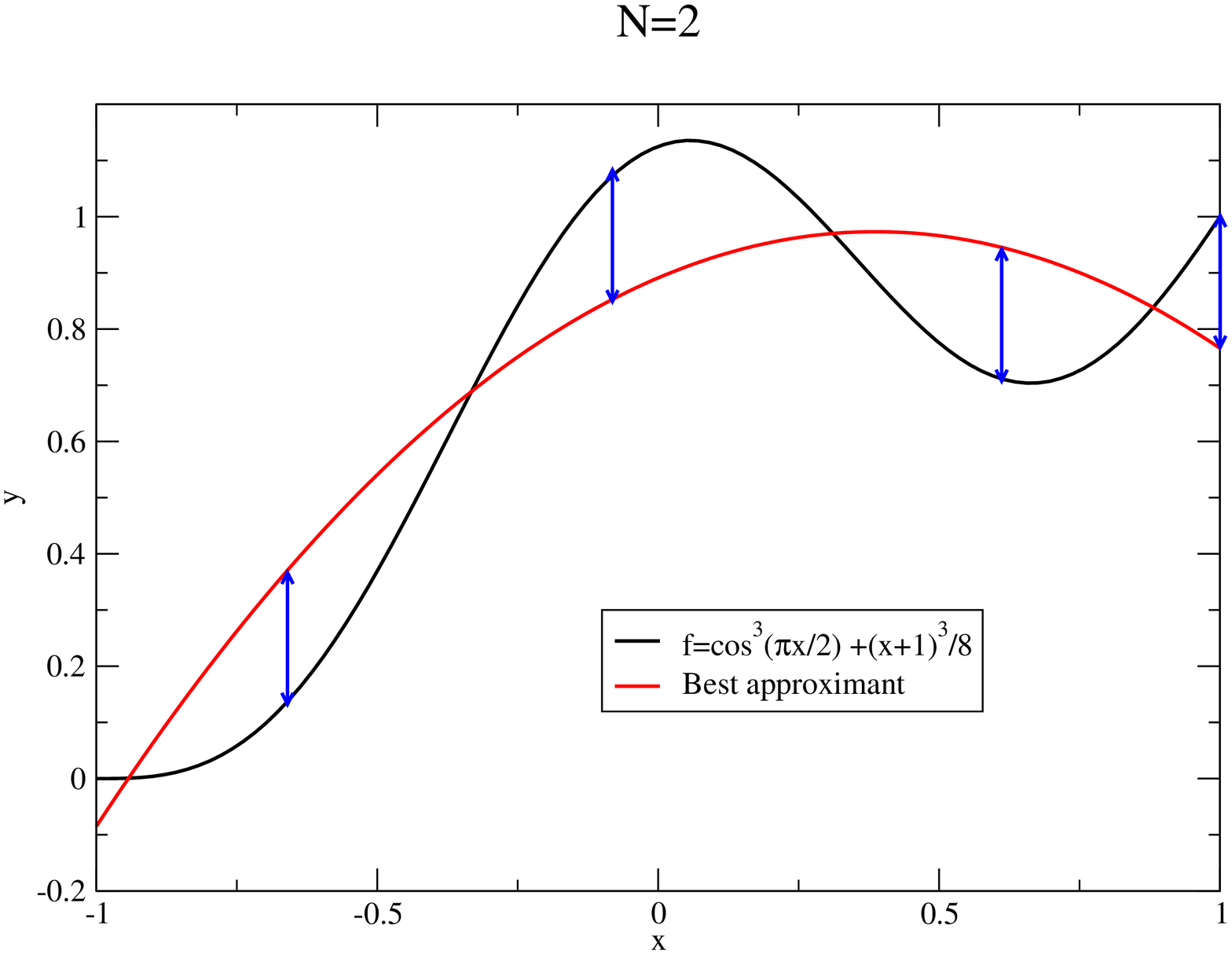}}
  \caption{Function $f=\cos^3\l(\pi x/2\r) + \l(x+1\r)^3/8$ (black
  curve) and its best approximation of degree $2$ (red curve). The
  blue arrows denote the $4$ points where the maximum error is
  attained.}
  \label{figure:best_N2}
\end{figure}}

\subsection{Interpolation on a grid}
\label{ss::interpolation_grid}

A {\em grid} $X$ on the interval $\interv$ is a set of $N+1$ points
$x_i \in \interv$, $ 0 \leq i \leq N$. These points are called the
{\it nodes} of the grid $X$.

Let us consider a continuous function $f$ and a grid family of grids $X$ with
$N+1$ nodes $x_i$. Then, there exists a unique polynomial of degree $N$,
$I_N^X f$, that coincides with $f$ at each node:
\begin{equation}
  I_N^Xf \l(x_i\r) = f\l(x_i\r) \quad 0 \leq i \leq N.
\end{equation}

$I_N^Xf$ is called the interpolant of $f$ through the grid
$X$. $I_N^Xf$ can be expressed in terms of the Lagrange cardinal
polynomials:
\begin{equation}
  I_N^Xf = \sum_{i=0}^N f\l(x_i\r) \ell_i^X\l(x\r),
\end{equation}
where the $\ell_i^X$ are the Lagrange cardinal polynomials. By
definition, $\ell_i^X$ is the unique polynomial of degree $N$, that
vanishes at all nodes of the grid $X$ {\it but} at $x_i$, where it is
equal to $1$. It is easy to show that the Lagrange cardinal
polynomials can be written as
\begin{equation}
  \ell_i^X \l(x\r) = \prod_{j=0, j\not= i}^N \frac{x-x_j}{x_i-x_j}.
\end{equation}
Figure \ref{figure:lagrange} shows some examples of Lagrange cardinal
polynomials. An example of a function and its interpolant on a uniform
grid can be seen on Figure~\ref{figure:interpole}.

\epubtkImage{}{%
\begin{figure}
  \centerline{\includegraphics[height=8cm]{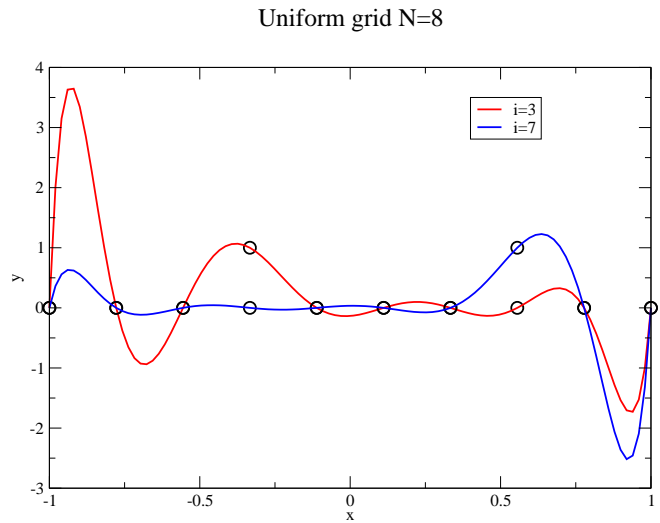}}
  \caption{Lagrange cardinal polynomials $\ell_3^X$ (red curve) and
  $\ell_7^X$ on an uniform grid with $N=8$. The black circles denote
  the nodes of the grid.}
  \label{figure:lagrange}
\end{figure}}

\epubtkImage{}{%
\begin{figure}
  \centerline{\includegraphics[height=8cm]{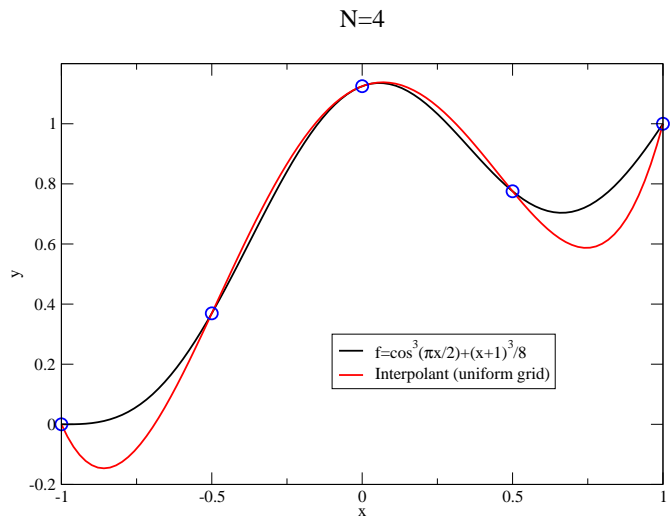}}
  \caption{Function $f=\cos^3\l(\pi x/2\r) + \l(x+1\r)^3/8$ (black
  curve) and its interpolant (red curve)on a uniform grid of $5$
  nodes. The blue circles show the position of the nodes.}
  \label{figure:interpole}
\end{figure}}

Thanks to Chebyshev alternate theorem, one can see that the best
approximation of degree $N$ is an interpolant of the function at $N+1$
nodes. However, in general, the associated grid is not known. The
difference between the error made by interpolating on a given grid $X$
can be compared to the smallest possible error for the best
approximation. One can show that (see Prop.~7.1
of~\cite{quarteroni-07}):
\begin{equation}
  \l\|f - I_N^X f \r\|_\infty \leq \l(1+\Lambda_N\l(X\r)\r) \l\|f - p^\star_N\r\|_\infty,
\end{equation}
where $\Lambda$ is the {\it Lebesgue constant} of the grid $X$ and is defined as:
\begin{equation}
  \Lambda_N \l(X\r) = {\rm max}_{x\in\interv} \sum_{i=0}^N \l|\ell_i^X\l(x\r)\r|.
\end{equation}

A theorem by Erd\"os~\cite{erdos-61} states that, for any choice of
grid $X$, there exists a constant $C>0$ such that:
\begin{equation}
  \Lambda_N\l(X\r) > \frac{2}{\pi} \ln\l(N+1\r) - C.
\end{equation}
It immediately follows that $\Lambda_N \rightarrow \infty$ when
$N\rightarrow \infty$. This is related to a result from 1914 by Faber~\cite{faber-14}
that states that for any grid, there always exists at
least one continuous function $f$  which interpolant does not converge
uniformly to $f$. An example of such failure of the convergence is
show on Figure~\ref{figure:runge}, where the convergence of the interpolant
to the function $f=\displaystyle\frac{1}{1+16x^2}$ is
clearly not uniform (see the behavior near the boundaries of the
interval). This is known as the Runge phenomenon.

\epubtkImage{}{%
  \begin{figure}
    \centerline{
      \includegraphics[width=7.5cm]{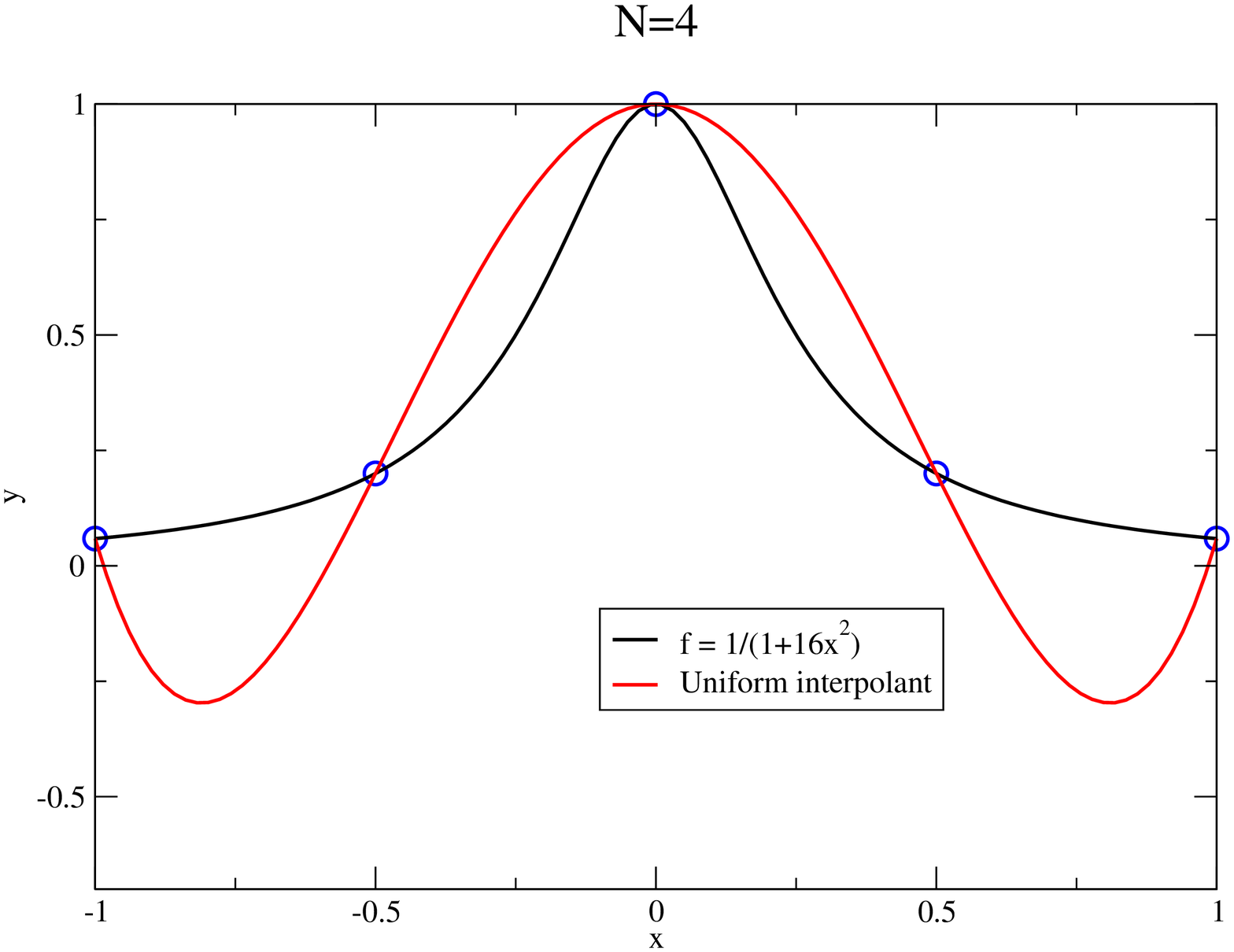}
      \includegraphics[width=7.5cm]{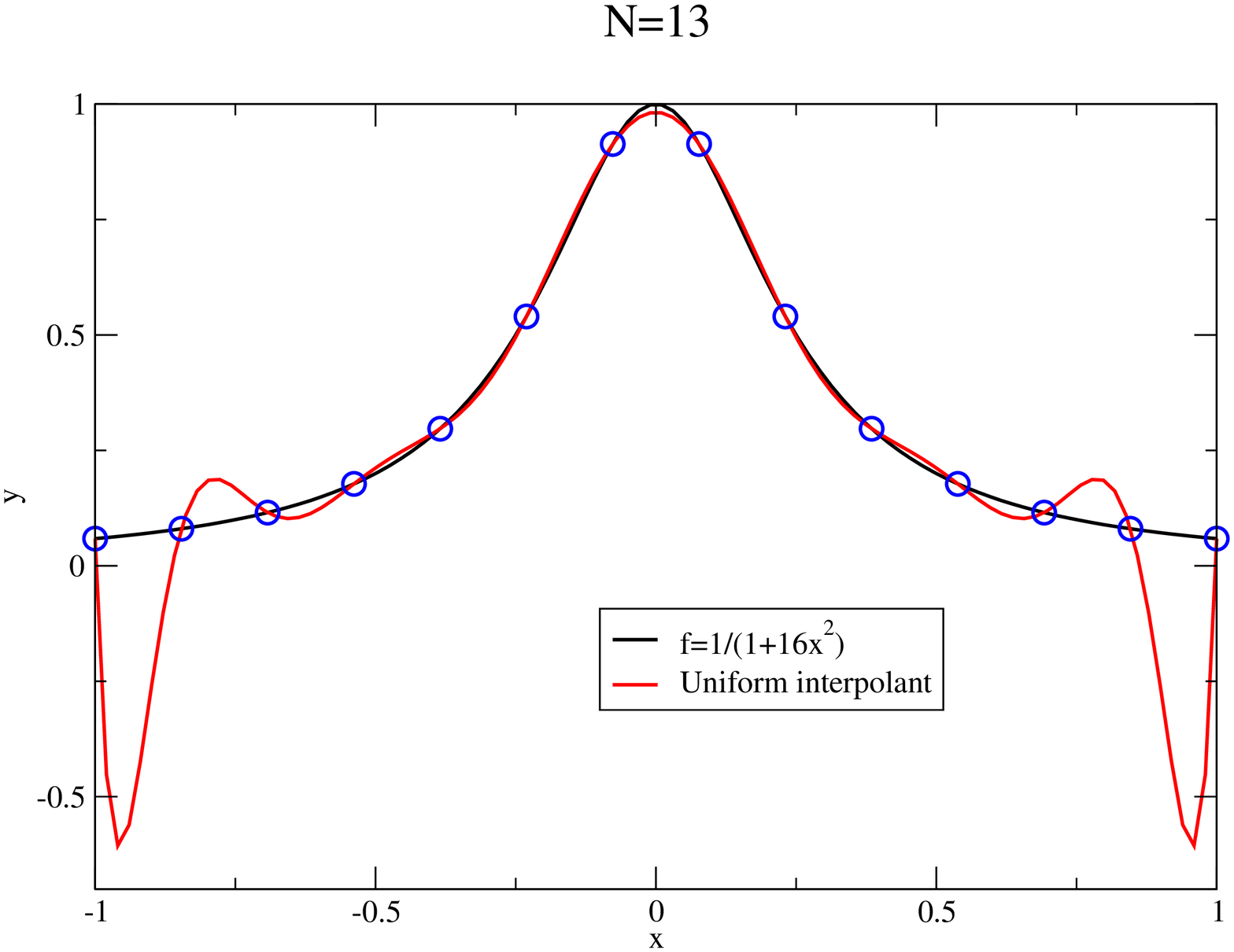}
    }
    \caption{Function $f=\displaystyle\frac{1}{1+16x^2}$ (black curve)
    and its interpolant (red curve) on a uniform grid of 5 nodes (left
    panel) and 14 nodes (right panel). The blue circles show the
    position of the nodes.}
  \label{figure:runge}
\end{figure}}

Moreover, a theorem by Cauchy (Theorem~7.2 of~\cite{quarteroni-07})
states that, for all functions $f \in \mathcal{C}^{\l(N+1\r)}$, the
interpolation error, on a grid $X$ of $N+1$ nodes is given by
\begin{equation}
  \label{equation:interpol_error}
  f\l(x\r) - I_N^X f \l(x\r) = \frac{f^{N+1} \l(\epsilon\r)}{\l(N+1\r) !} w^X_{N+1}\l(x\r),
\end{equation}
where $\epsilon \in\interv$. $w^X_{N+1}$ is the nodal polynomial of
$X$, being the only polynomial of degree $N+1$, with a leading
coefficient $1$ and that vanishes on the nodes of $X$. It is then easy
to show that 
\begin{equation}
w^X_{N+1} \l(x\r) = \prod_{i=0}^N \l(x - x_i\r).
\end{equation}

On Equation~(\ref{equation:interpol_error}), one has a priori no control on
the term involving $f^{N+1}$. For a given function, it can be rather large and
this is indeed the case for the function $f$ shown on
Figure~\ref{figure:runge} (one can check, for instance that,
$\l|f^{N+1}\l(1\r)\r|$ becomes larger and larger). However, one can hope to
minimize the interpolation error by choosing a grid such that the nodal
polynomial is as small as possible. A theorem by Chebyshev states that this
choice is unique and is given by a grid which nodes are the zeros of the
Chebyshev polynomial $T_{N+1}$ (see Section~\ref{ss:polynomial_interpolation}
for more details on Chebyshev polynomials). With such a grid, one can achieve
\begin{equation}
\l\|w^X_{N+1}\r\|_\infty = \frac{1}{2^N},
\end{equation}
which is the smallest possible value (see Equation~(18), Section~4.2,
Chapter~5 of~\cite{isaacson-66}). So, a grid based on nodes of Chebyshev
polynomials can be expected to perform better that a standard uniform one.
This is what can be seen on Figure~\ref{figure:no_runge}, which shows the same
function, and its interpolants, as on Figure~\ref{figure:runge} but with a
Chebyshev grid. Clearly, the Runge phenomenon is no longer present. It can be
checked, that, for this choice of function $f$, the uniform convergence of the
interpolant to the function is recovered. The reason is that
$\l\|w^X_{N+1}\r\|_\infty$ decreases faster than $f^{N+1}/\l(N+1\r)!$
increases. Of course Faber's result implies that this can not be true for all
the functions. There still must exist some functions for which the interpolant
does not converge uniformly to the function itself (it is actually the class
of functions that are not absolutely continuous, like the Cantor function).

\epubtkImage{}{%
  \begin{figure}
    \centerline{
      \includegraphics[width=7.5cm]{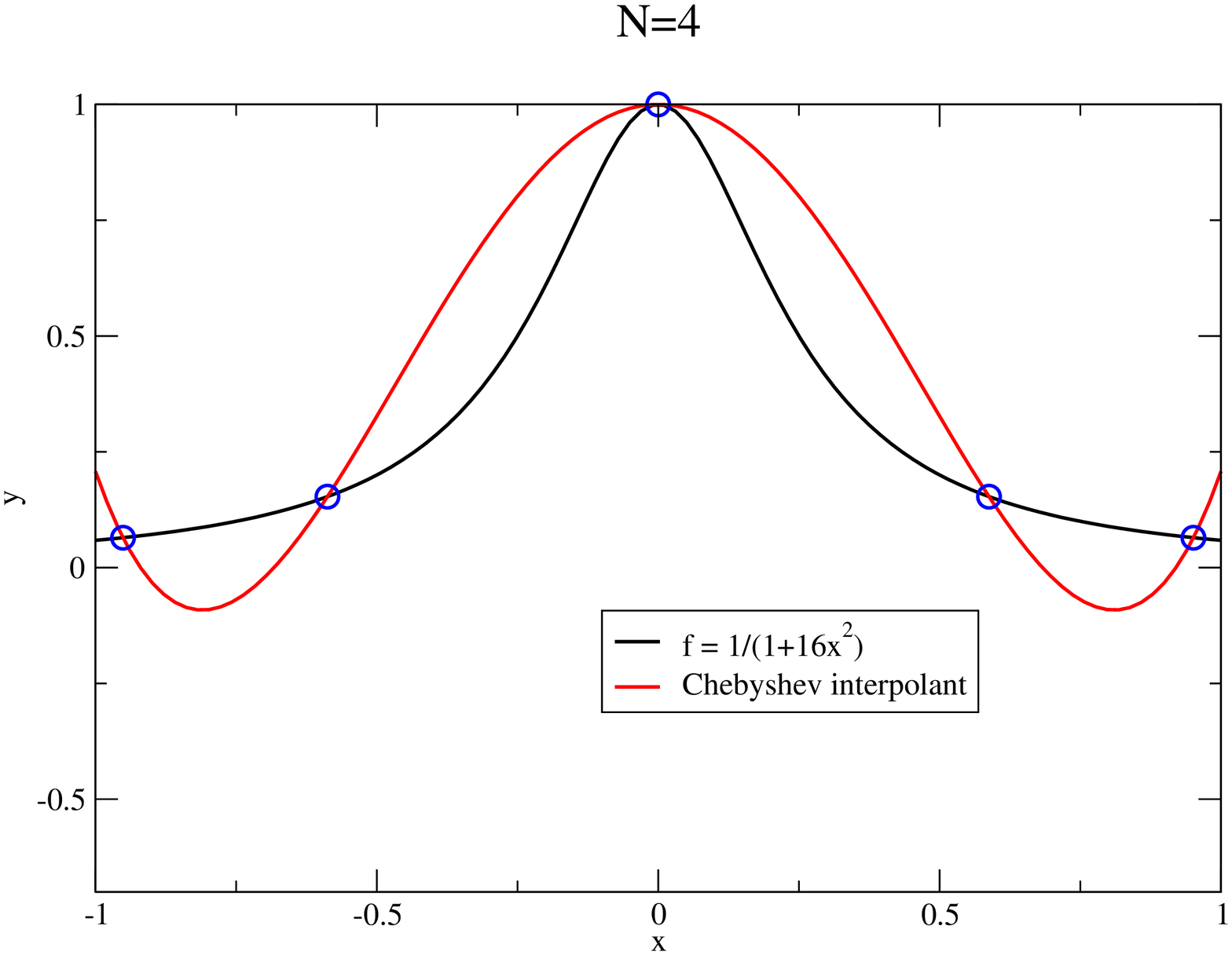}
      \includegraphics[width=7.5cm]{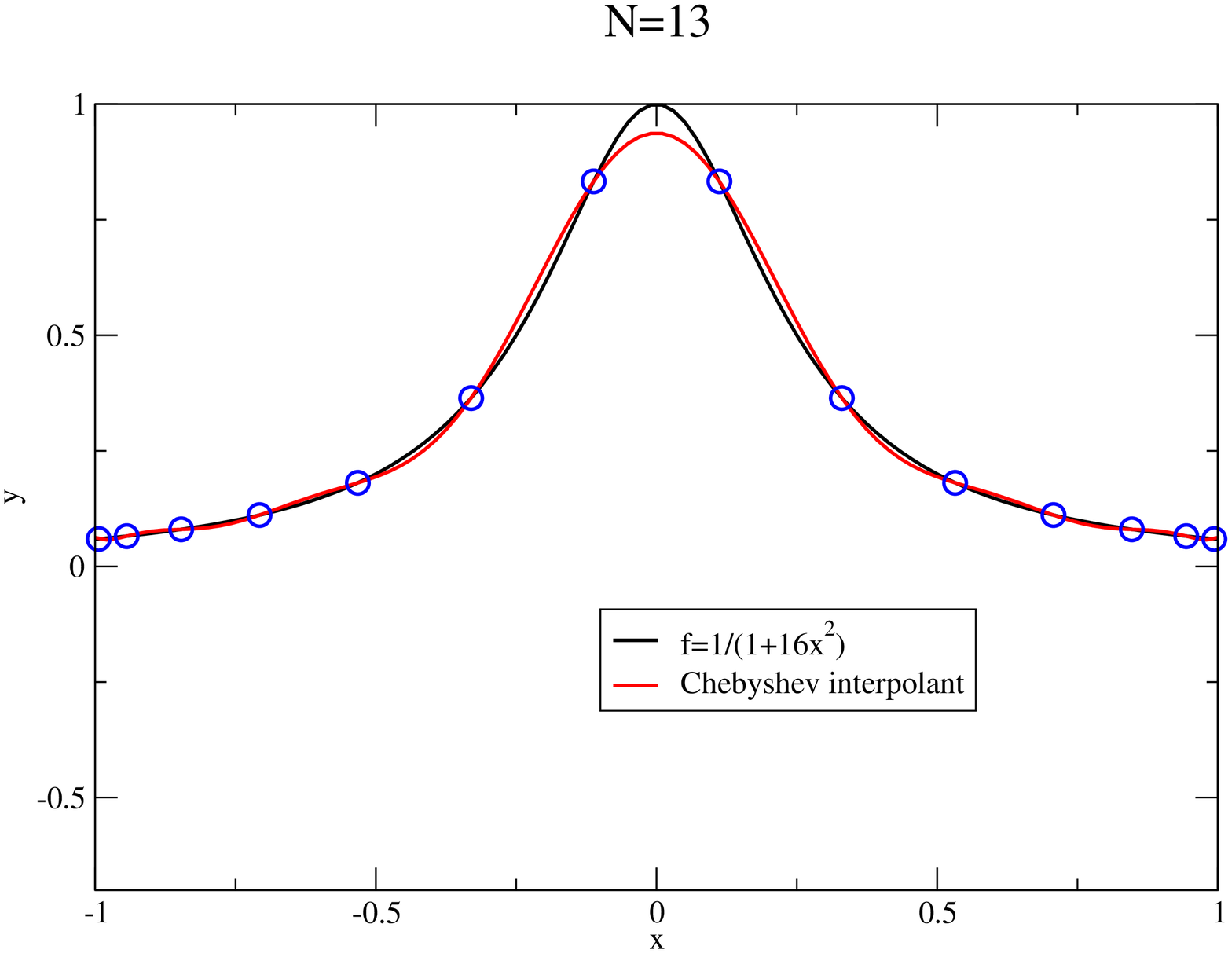}
      }
    \caption{Same as Figure~\ref{figure:runge} but using a grid
    based on the zeros of Chebyshev polynomials. The Runge phenomenon
    is no longer present.}
    \label{figure:no_runge}
\end{figure}}

\subsection{Polynomial interpolation}
\label{ss:polynomial_interpolation}

\subsubsection{Orthogonal polynomials}
\label{sss:ortho}

Spectral methods are often based on the notion of {\it orthogonal
  polynomials}. In order to define orthogonality, one has to define
the scalar product of two functions on an interval $\interv$. Let us
consider a positive function $w$ of $\interv$ called the {\it
  measure}. The scalar product of $f$ and $g$ with respect to this
measure is defined as:
\begin{equation}
\label{equation:scalar_prod}
\l(f, g\r)_w = \int_{x\in\interv} f\l(x\r)g\l(x\r)w\l(x\r) {\rm d}x.
\end{equation}
A basis of $P_{\mathbb N}$ is then a set of $N+1$ polynomials $\left\{
p_n \right\}_{n=0\dots N}$. $p_n$ is of degree $n$ and the polynomials
are orthogonal: $\l(p_i, p_j\r)_w = 0$ for $i\not=j$.

The projection $P_N f$ of a function $f$ on this basis is then 
\begin{equation}
P_N f= \sum_{n=0}^N \hat{f}_n p_n,
\end{equation}
where the coefficients of the projection are given by 
\begin{equation}
\label{equation:coef_proj}
\hat{f}_n = \displaystyle\frac{\l(f, p_n\r)}{\l(p_n,p_n\r)}.
\end{equation}
The difference between $f$ and its projection goes to zero when $N$ increases:
\begin{equation}
\l\|f-P_Nf\r\|_\infty \rightarrow 0 \quad {\rm when} \quad N\rightarrow \infty.
\end{equation}
Figure \ref{figure:projection} shows the function $f=\cos^3\l(\pi
x/2\r) + \l(x+1\r)^3/8$ and its projection on Chebyshev polynomials
(see Section~\ref{sss:cheby}), for $N=4$ and $N=8$, illustrating the
rapid convergence of $P_Nf$ to~$f$.

\epubtkImage{}{%
  \begin{figure}
    \centerline{
      \includegraphics[width=7.5cm]{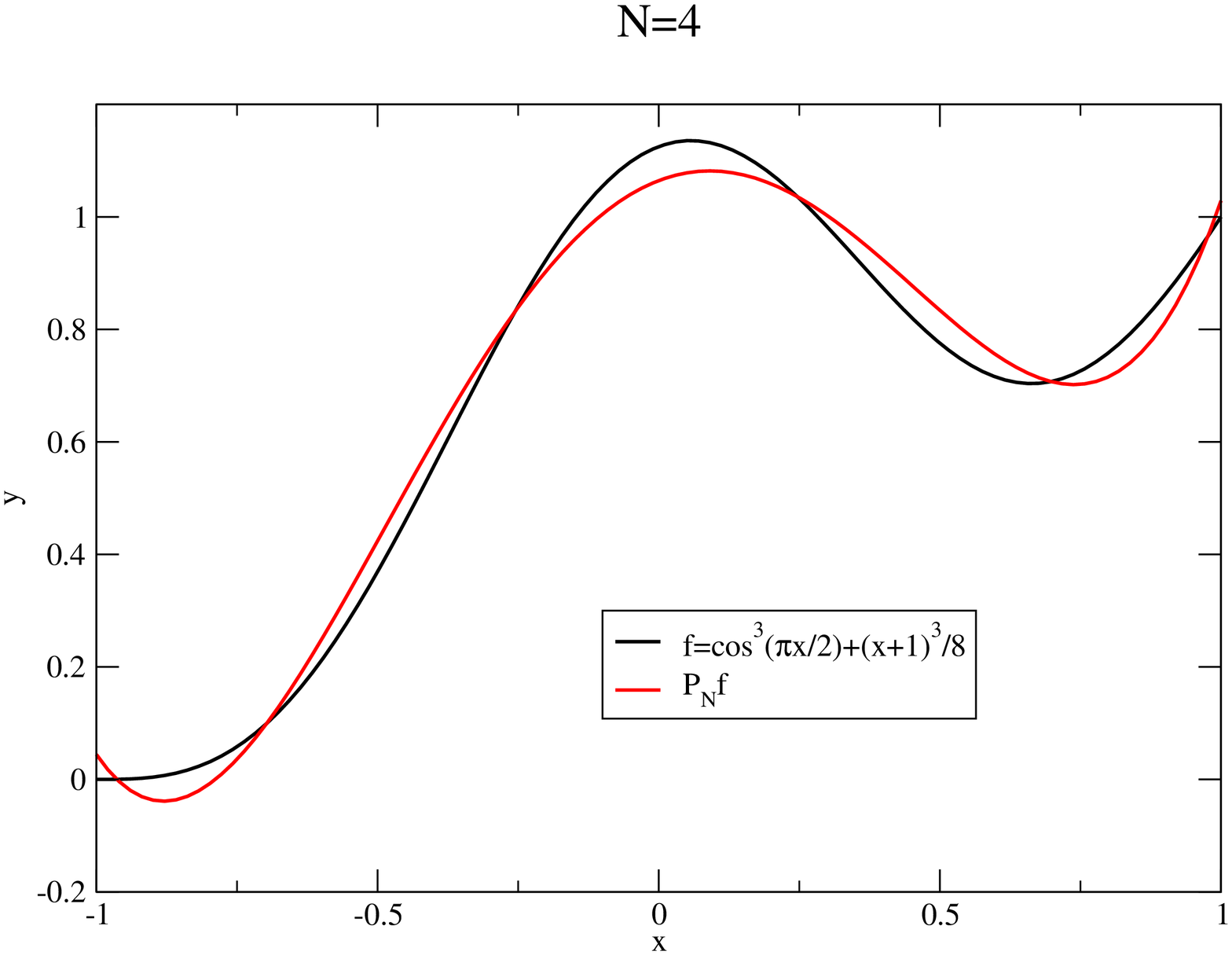}
      \includegraphics[width=7.5cm]{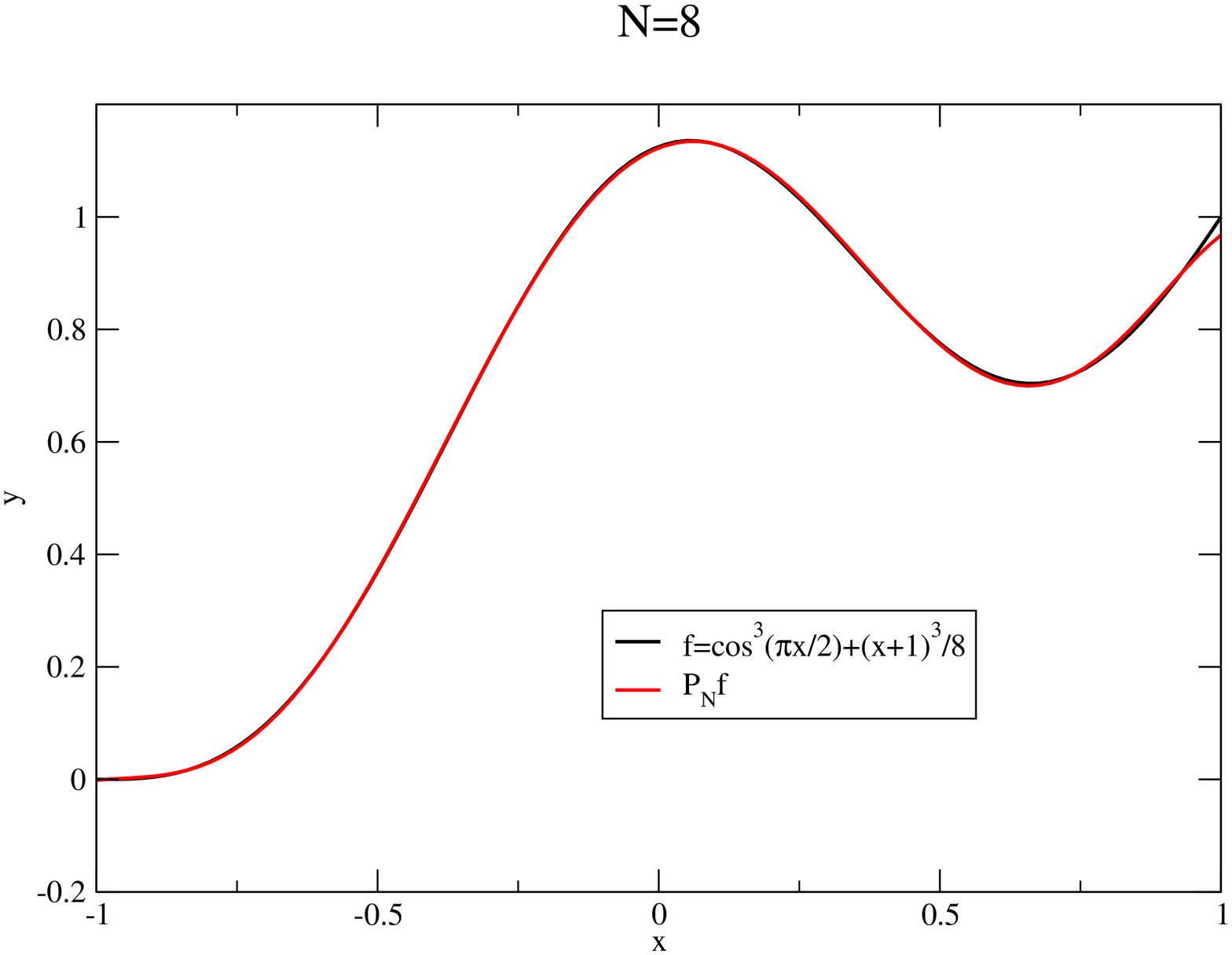}
    }
    \caption{Function $f=\cos^3\l(\pi x/2\r) + \l(x+1\r)^3/8$ (black
    curve) and its projection on Chebyshev polynomials (red curve),
    for $N=4$ (left panel) and $N=8$ (right panel).}
    \label{figure:projection}
\end{figure}}

At first sight, the projection seems to be an interesting means of
numerically representing a function. However, in practice this is not
the case. Indeed, to determine the projection of a function, one needs
to compute the integrals~(\ref{equation:coef_proj}), which requires
the evaluation of $f$ at a great number of points, making the
whole numerical scheme impractical.

\subsubsection{Gaussian quadratures}
\label{sss:gauss_quad}

The main theorem of Gaussian quadratures (see for
instance~\cite{canuto-06}) states that, given a measure $w$, there
exist $N+1$ positive real numbers $w_n$ and $N+1$ real numbers $x_n
\in \interv$ such that:
\begin{equation}
  \forall f \in {\mathbb P}_{2N+\delta}, \quad \int_{\interv} f\l(x\r)
  w\l(x\r) {\rm d}x = \sum_{n=0}^N f\l(x_n\r) w_n.
\end{equation}
The $w_n$ are called the {\it weights} and the $x_n$ are the
collocation points. The integer $\delta$ can take several values
depending on the exact quadrature considered:

\begin{itemize}
  \item Gauss quadrature: $\delta=1$.
  \item Gauss--Radau: $\delta=0$ and $x_0 = -1$.
  \item Gauss--Lobatto: $\delta=-1$ and $x_0=-1,\ x_N=1$.
\end{itemize}

Gauss quadrature is the best choice because it applies to polynomials
of higher degree but Gauss--Lobatto quadrature is often more useful for
numerical purposes because the outermost collocation points coincide
with the boundaries of the interval, making it easier to impose
matching or boundary conditions. More detailed results and
demonstrations about those quadratures can be found for instance
in~\cite{canuto-06}.

\subsubsection{Spectral interpolation}

As already stated in~\ref{sss:ortho}, the main drawback of
projecting a function onto orthogonal polynomials comes from the
difficulty to compute the integrals~(\ref{equation:coef_proj}). The
idea of spectral methods is to approximate the coefficients of the
projection by making use of the Gaussian quadratures. By doing so, one
can define the {\it interpolant} of a function $f$ by
\begin{equation}
  \label{equation:def_interpole}
  I_N f = \sum_{n=0}^N \tilde{f}_n p_n \l(x\r),
\end{equation}
where
\begin{equation}
  \label{equation:def_coef}
  \tilde{f}_n = \frac{1}{\gamma_n}\sum_{i=0}^N f\l(x_i\r) p_n\l(x_i\r)
  w_i \quad {\rm and} \quad \gamma_n = \sum_{i=0}^N p_n^2\l(x_i\r)
  w_i.
\end{equation}
The $\tilde{f}_n$ exactly coincides with the coefficients $\hat{f}_n$,
if the Gaussian quadrature is applicable for
computing~\ref{equation:coef_proj}), that is for all $f \in{\mathbb
  P}_{N+\delta}$. So, in general, $I_Nf \not= P_Nf$ and the difference
between the two is called the {\em aliasing error}. The advantage of
using the $\tilde{f}_n$ is that they are computed by estimating $f$ at
the $N+1$ collocation points only.

One can show that $I_Nf$ and $f$ coincide at the collocation points:
$I_Nf\l(x_i\r) = f\l(x_i\r)$ so that $I_N$ interpolates $f$ on the
grid which nodes are the collocation
points. Figure~\ref{figure:spectral_interpol} shows the function
$f=\cos^3\l(\pi /2\r) + \l(x+1\r)^3/8$ and its spectral interpolation
using Chebyshev polynomials, for $N=4$ and $N=6$.

\epubtkImage{}{%
  \begin{figure}
    \centerline{
      \includegraphics[width=7.5cm]{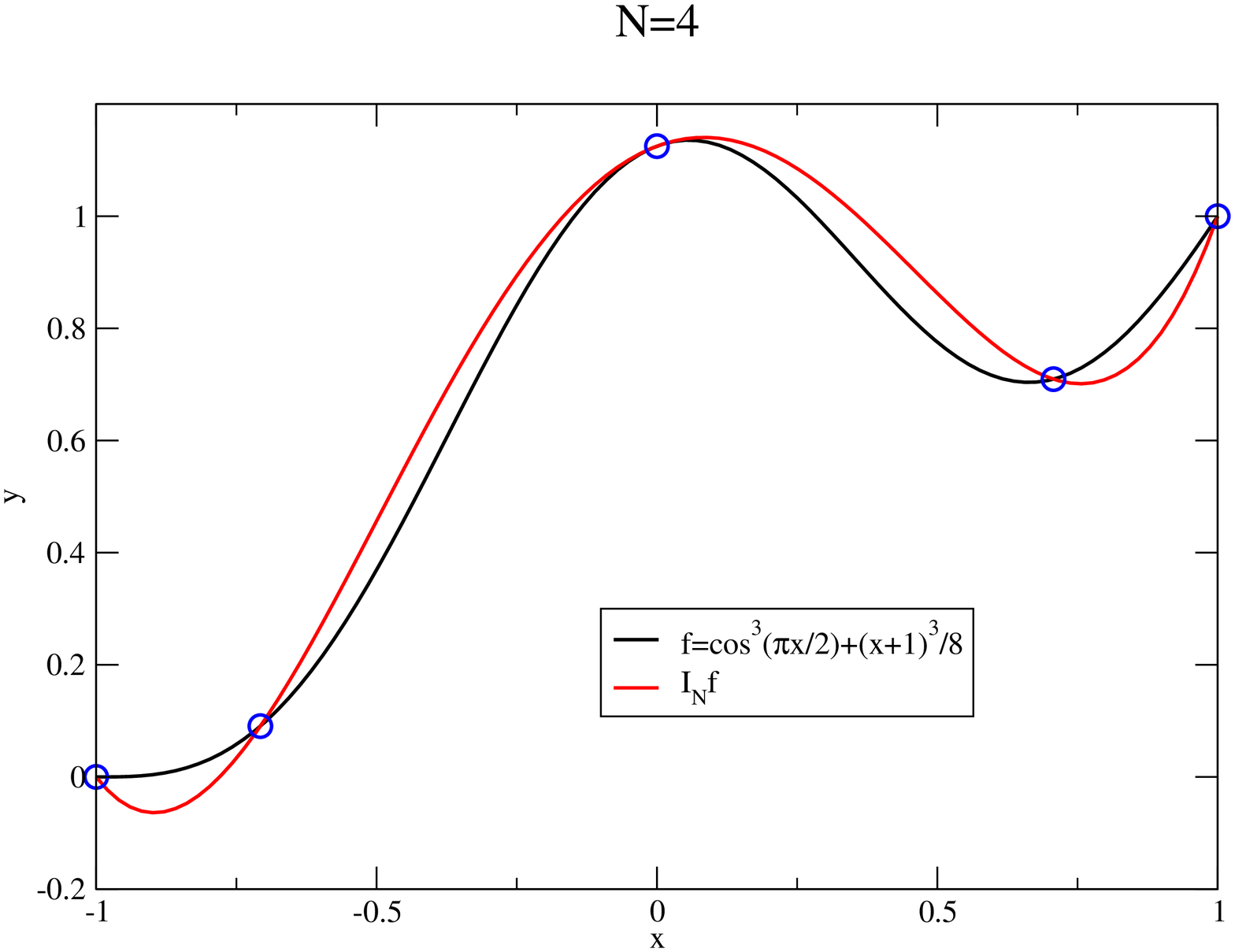}
      \includegraphics[width=7.5cm]{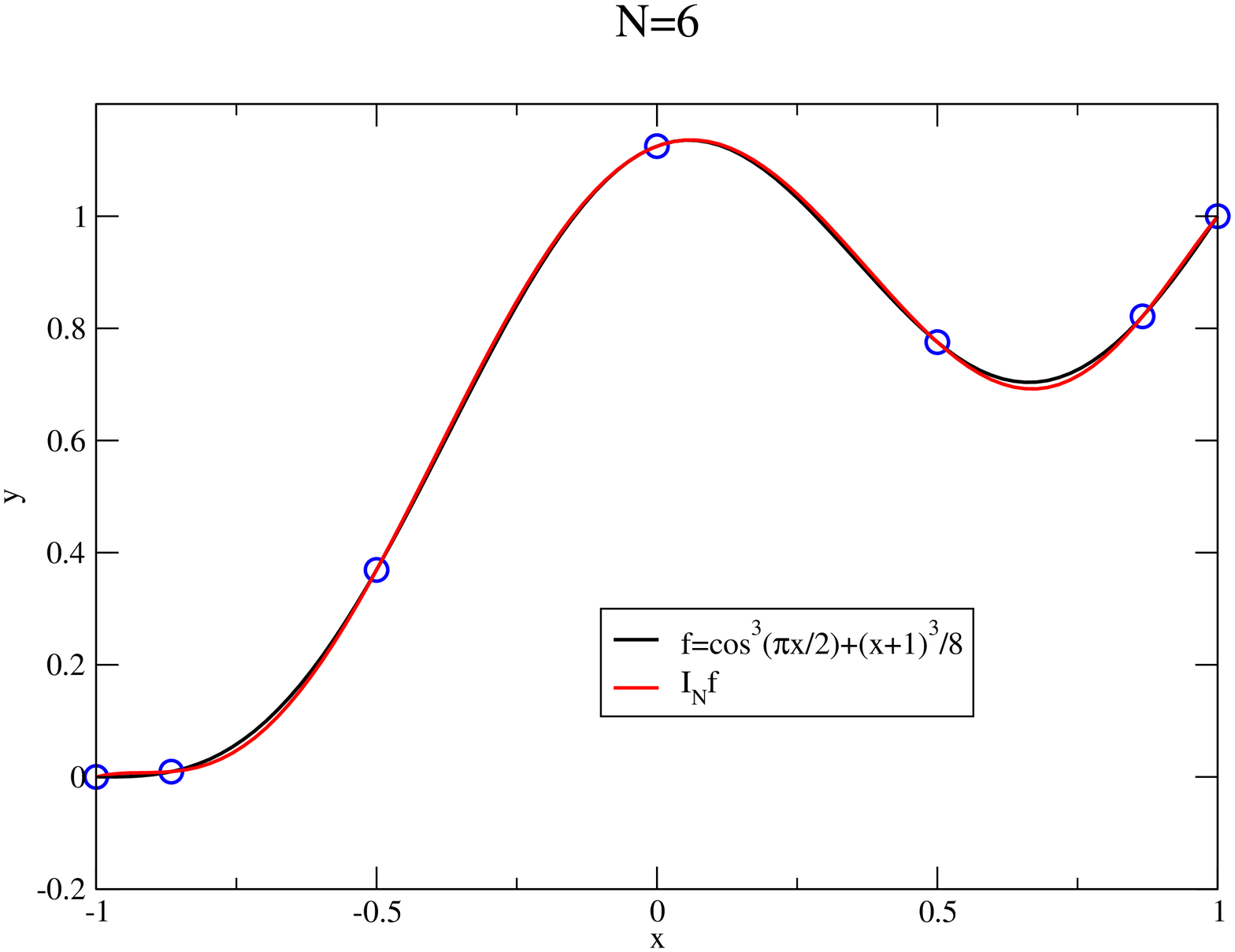}
    }
    \caption{Function $f=\cos^3\l(\pi x/2\r) + \l(x+1\r)^3/8$ (black
    curve) and its interpolant $I_Nf$ on Chebyshev polynomials (red
    curve), for $N=4$ (left panel) and $N=6$ (right panel). The
    collocation points are denoted by the blue circles and correspond
    to Gauss--Lobatto quadrature.}
    \label{figure:spectral_interpol}
\end{figure}}

\subsubsection{Two equivalent descriptions}

The description of a function $f$ in terms of its spectral
interpolation can be given in two different, but equivalent spaces:

\begin{itemize}
  \item in the configuration space if the function is described by its
  value at the $N+1$ collocation points $f\l(x_i\r)$.
  \item in the coefficient space if one works with the $N+1$
  coefficients $\tilde{f}_i$.
\end{itemize}

There is a bijection between both spaces and the following relations
enable us to go from one to the other:

\begin{itemize}
  \item the coefficients can be computed from the values of
  $f\l(x_i\r)$ using Equation~(\ref{equation:def_coef}).
  \item the values at the collocation points are expressed in terms of
  the coefficients by making use of the definition of the
  interpolant~(\ref{equation:def_interpole}):
  \begin{equation}
    f\l(x_i\r) = \sum_{n=0}^N \tilde{f}_n p_n\l(x_i\r).
  \end{equation}
\end{itemize}

Depending on the operation one has to perform on a given function, it
may be more clever to work in one space or the other. For instance,
the square root of a function is very easily given in the collocation
space by $\sqrt{f\l(x_i\r)}$, whereas the derivative can be computed
in the coefficient space if, and this is generally the case, the
derivatives of the basis polynomials are known, by $f'\l(x\r) =
\displaystyle\sum_{n=0}^N \tilde{f}_n p'_n\l(x\r)$.

\subsection{Usual polynomials}

\subsubsection{Sturm--Liouville problems and convergence}
\label{sss:strum}

The Sturm--Liouville problems are eigenvalue problems of the form:
\begin{equation}
\label{e:sturm}
-\l(pu'\r)' + qu = \lambda w u,
\end{equation}
on the interval $\l[-1, 1\r]$. $p$ $q$ and $w$ are real valued
functions such that:

\begin{itemize}
  \item $p\l(x\r)$ is continuously differentiable, strictly positive and continuous at $x=\pm 1$.
  \item $q\l(x\r)$ is continuous, non negative and bounded.
  \item $w\l(x\r)$ is continuous, non negative and integrable.
\end{itemize}

The solutions are then the eigenvalues $\lambda_i$ and the eigenfunctions $u_i\l(x\r)$.
The eigenfunctions are orthogonal with respect to the measure $w$:
\begin{equation}
  \int_{-1}^1 u_m\l(x\r) u_n\l(x\r) w\l(x\r) {\rm d}x = 0 \quad {\rm for} \quad m\not=n.
\end{equation}

Singular Sturm--Liouville problems are particularly important for
spectral methods. A Sturm--Liouville problem is singular if and only if
the function $p$ vanishes at the boundaries $x=\pm 1$. One can
show that, if the functions of the spectral basis are chosen to be the
solutions of a singular Sturm--Liouville problem, then the convergence
of the function to its interpolant is spectral, that is faster than
any power-law of $N$, $N$ being the order of the expansion (see
Section~5.2 of~\cite{canuto-06}). Let us precise that this does not
necessarily imply that the convergence is exponential. Convergence
properties are discussed in more details for Legendre and Chebyshev
polynomials in Section~\ref{sss:convergence}.

Conversely, it can be shown that spectral convergence is not ensured
when considering solutions of regular Sturm--Liouville
problems~\cite{canuto-06}.

In what follows two usual types of solutions of singular
Sturm--Liouville problems are considered: Legendre and Chebyshev
polynomials.

\subsubsection{Legendre polynomials}
\label{sss:leg}

Legendre polynomials $P_n$ are eigenfunctions of the following
singular Sturm--Liouville problem:
\begin{equation}
  \l(\l(1-x^2\r)P'_n\r)' + n\l(n+1\r) P_n = 0.
\end{equation}
In the notations of Equation~(\ref{e:sturm}), $p=1-x^2$, $q=0$, $w=1$
and $\lambda_n = -n\l(n+1\r)$.

It follows that Legendre polynomials are orthogonal on $\interv$ with
respect to the measure $w\l(x\r) = 1$. Moreover, the scalar product of
two polynomials is given by:
\begin{equation}
  \l(P_n, P_m\r) = \int_{-1}^1 P_n P_m {\rm d}x = \frac{2}{2n+1} \delta_{mn}.
\end{equation}

Starting from $P_0 = 1$ and $P_1=x$, the successive polynomials can be
computed by the following recurrence expression: 
\begin{equation}
\label{e:recur_leg}
\l(n+1\r) P_{n+1}\l(x\r) = \l(2n+1\r) x P_n\l(x\r)  - n P_{n-1}\l(x\r).
\end{equation}

Among the various properties of Legendre polynomials, one can note
that i) $P_n$ has the same parity as $n$. ii) $P_n$ is of degree
$n$. iii) $P_n\l(\pm 1\r) = \l(\pm 1\r)^n$. iv) $P_n$ has exactly $n$
zeros on $\interv$. The first polynomials are shown on
Figure~\ref{figure:legendre}.

\epubtkImage{}{%
\begin{figure}
  \centerline{\includegraphics[height=8cm]{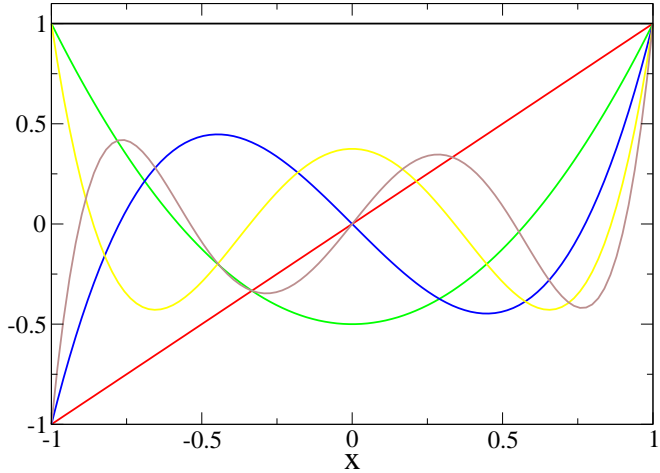}}
  \caption{First Legendre polynomials, from $P_0$ to $P_5$.}
  \label{figure:legendre}
\end{figure}}

The weights and locations of the collocation points associated with
Legendre polynomials depend on the choice of quadrature.

\begin{itemize}
  \item Legendre--Gauss: $x_i$ are the nodes of $P_{N+1}$ and $w_i =
  \fraction{2}{\l(1-x_i^2\r)\l[P'_{N+1} \l(x_i\r)\r]^2}$.
  \item Legendre--Gauss--Radau: $x_0=-1$ and the $x_i$ are the nodes of
  $P_N + P_{N+1}$. The weights are given by $w_0 =
  \fraction{2}{\l(N+1\r)^2}$ and $w_i = \fraction{1}{\l(N+1\r)^2}$.
  \item Legendre--Gauss--Lobatto: $x_0=-1$, $x_N=1$ and $x_i$ are the
  nodes of $P'_N$. The weights are $w_i =
  \fraction{2}{N\l(N+1\r)}\fraction{1}{\l[P_N\l(x_i\r)\r]^2}$.
\end{itemize}

These values have no analytic expression, but they can be computed
numerically in an efficient way.

Some elementary operations can easily be performed on the coefficient
space. Let us assume that a function $f$ is given by its coefficients
$a_n$ so that $f = \displaystyle\sum_{n=0}^N a_n P_n$. Then, the
coefficients $b_n$ of $Hf = \displaystyle\sum_{n=0}^N b_n P_n$ can be
found as a function of the $a_n$, for various operators $H$. For
instance,

\begin{itemize}
\item if $H$ is the multiplication by $x$ then: 
\begin{equation}
  b_n = \frac{n}{2n-1}a_{n-1} + \frac{n+1}{2n+3} a_{n+1}\quad \l(n \geq 1\r).
\end{equation}
\item if $H$ is the derivative: 
\begin{equation}
  b_n = \l(2n+1\r) \sum_{p=n+1, p+n \, {\rm odd}}^N a_p.
\end{equation}
\item if $H$ is the second derivative: 
\begin{equation}
  b_n = \l(n+1/2\r) \sum_{p=n+2, p+n \, {\rm even}}^N \l[p\l(p+1\r) -
  n \l(n+1\r)\r] a_p.
\end{equation}
\end{itemize}

Those kind of relations enable to represent the action of $H$ as a
matrix acting on the vector of the $a_n$, the product being the
coefficients of $Hf$, i.e.\ the $b_n$.

\subsubsection{Chebyshev polynomials}
\label{sss:cheby}

Chebyshev polynomials $T_n$ are eigenfunctions of the following
singular Sturm-Liouville problem:
\begin{equation}
  \l(\sqrt{1-x^2}T'_n\r)' + \frac{n}{\sqrt{1-x^2}} T_n = 0.
\end{equation}
In the notations of Equation~(\ref{e:sturm}), $p=\sqrt{1-x^2}$, $q=0$,
$w=1/\sqrt{1-x^2}$ and $\lambda_n = -n$.

It follows that Chebyshev polynomials are orthogonal on $\interv$ with
respect to the measure $w=1/\sqrt{1-x^2}$ and the scalar product of
two polynomials is 
\begin{equation}
  \l(T_n, T_m\r) = \int_{-1}^{1} \frac{T_n T_m}{\sqrt{1-x^2}} {\rm d}
  x = \frac{\pi}{2} \l(1+\delta_{0n}\r) \delta_{mn}.
\end{equation}
Given that $T_0 = 1$ and $T_1=x$, the higher order polynomials can be
obtained by making use of the recurrence
\begin{equation}
  \label{eq:recurrence_cheb}
  T_{n+1}\l(x\r) = 2xT_n\l(x\r) - T_{n-1}\l(x\r).
\end{equation}
This implies the following simple properties. i) $T_n$ has the same
parity as $n$. ii) $T_n$ is of degree $n$. iii) $T_n\l(\pm 1\r) =
\l(\pm 1\r)^n$. iv) $T_n$ has exactly $n$ zeros on $\interv$. The first
polynomials are shown on Figure~\ref{figure:cheby}.

\epubtkImage{}{%
\begin{figure}
  \centerline{\includegraphics[height=8cm]{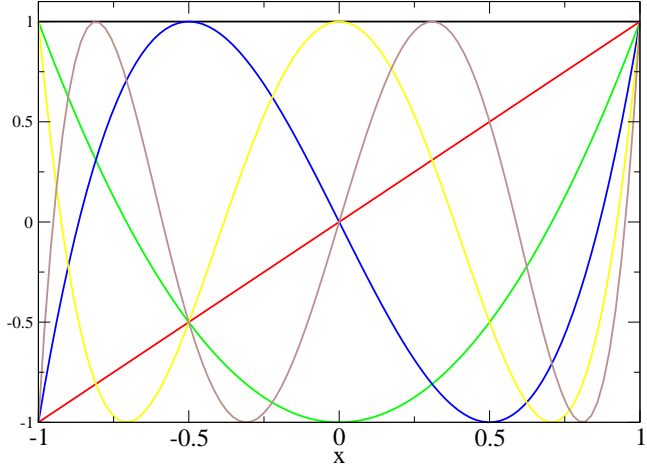}}
  \caption{First Chebyshev polynomials, from $T_0$ to $T_5$.}
  \label{figure:cheby}
\end{figure}}

Contrary to the Legendre case, both the weights and positions of the
collocation points are given by analytic formulas:

\begin{itemize}
  \item Chebyshev--Gauss: $x_i=\cos\fraction{\l(2i+1\r)\pi}{2N+2}$ and
  $w_i = \fraction{\pi}{N+1}$.
  \item Chebyshev--Gauss--Radau: $x_i = \cos\fraction{2\pi i
  }{2N+1}$. The weights are $w_0= \fraction{\pi}{2N+1}$ and $w_i =
  \fraction{2\pi}{2N+1}$
  \item Chebyshev--Gauss--Lobatto: $x_i = \cos\fraction{\pi i
  }{N}$. The weights are $w_0= w_N= \fraction{\pi}{2N}$ and $w_i =
  \fraction{\pi}{N}$.
\end{itemize}

As for the Legendre case, the action of various linear operators $H$
can be expressed in the coefficient space. This means that the
coefficients $b_n$ of $Hf$ are given as functions of the coefficients
$a_n$ of $f$. For instance, 

\begin{itemize}
\item if $H$ is the multiplication by $x$ then: 
\begin{equation}
  b_n = \frac{1}{2}\l[\l(1+\delta_{0\, n-1}\r) a_{n-1} + a_{n+1}\r]
  \quad \l(n \geq 1\r).
\end{equation}
\item if $H$ is the derivative: 
\begin{equation}
b_n = \frac{2}{\l(1+\delta_{0\, n}\r)} \sum_{p=n+1, p+n \, {\rm odd}}^N pa_p.
\end{equation}
\item if $H$ is the second derivative: 
\begin{equation}
  b_n = \frac{1}{\l(1+\delta_{0\, n}\r)} \sum_{p=n+2, p+n \, {\rm
  even}}^N p\l(p^2-n^2\r) a_p.
\end{equation}
\end{itemize}

\subsubsection{Convergence properties}
\label{sss:convergence}

One of the main advantage of spectral method is the very fast
convergence of the interpolant $I_Nf$ to the function $f$, at least
for smooth enough functions. Let us consider a ${\mathcal C}^m$
function $f$, one can place the following upper bounds on the
difference between $f$ and its interpolant $I_N f$:

\begin{itemize}
\item For Legendre: 
\begin{equation}
\label{e:error_leg}
\l\| I_Nf-f \r\|_{L^2} \leq \frac{C_1}{N^{m-1/2}}\sum_{k=0}^m \l\|f^{\l(k\r)}\r\|_{L^2}.
\end{equation}
\item For Chebyshev: 
\begin{equation}
\label{e:error_cheb2}
\l\| I_Nf-f \r\|_{L_w^2} \leq \frac{C_2}{N^m}\sum_{k=0}^m \l\|f^{\l(k\r)}\r\|_{L_w^2}.
\end{equation}
\begin{equation}
\label{e:error_cheb}
\l\| I_Nf-f \r\|_\infty \leq \frac{C_3}{N^{m-1/2}}\sum_{k=0}^m \l\|f^{\l(k\r)}\r\|_{L_w^2}.
\end{equation}
\end{itemize}

The $C_i$ are positive constants. An interesting limit of the
above estimates concerns a ${\mathcal C}^\infty$ function. One can
then see that the difference between $f$ and $I_N f$ decays faster
than any power of $N$. This is the so-called spectral convergence. Let
us precise that this does not necessarily imply that the error decays
exponentially (think about $\exp\l(-\sqrt{N}\r)$ for
instance). Exponential convergence is achieved only for analytic
functions, i.e.\ functions that are locally given by a convergent power
series.

An example of this very fast convergence is shown on
Figure~\ref{figure:evanescent}. The error clearly decays as an
exponential, the function being analytic, until the level of
$10^{-14}$ of the precision of the computation is reached (one is
working in double precision in this particular
case). Figure~\ref{figure:evanescent} illustrates the fact that, with
spectral methods, very good accuracy can be reached with only a
moderate number of coefficients.

\epubtkImage{}{%
\begin{figure}
  \centerline{\includegraphics[height=8cm]{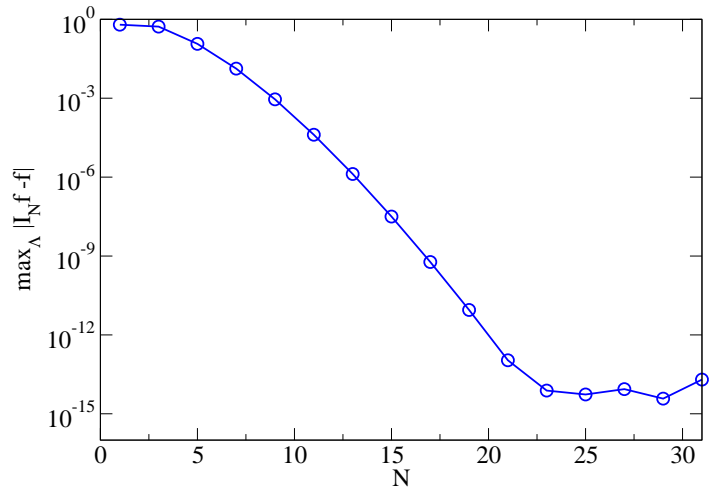}}
  \caption{Maximum difference between $f=\cos^3\l(\pi x/2\r) +
  \l(x+1\r)^3/8$ and its interpolant $I_N f$, as a function of $N$.}
  \label{figure:evanescent}
\end{figure}}

If the function is less regular (i.e. not ${\mathcal C}^{\infty}$),
the error only decays as a power-law, thus making the use of spectral
method less appealing. It can be easily seen on the worst possible
case: the one of a discontinuous function. In this case, the estimates
(\ref{e:error_leg}-\ref{e:error_cheb}) do not even ensure convergence
at all. On Figure~\ref{figure:gibbs} one shows a step function and its
interpolant, for various values of $N$. One can see that the maximum
difference between the function and its interpolant is not even going
to zero when $N$ is increasing. This is known as the Gibbs phenomenon.

Finally, equation~(\ref{e:error_cheb}) shows that if $m >0$, the
interpolant converges uniformly to the function. The continuous
functions that do not converge uniformly to their interpolant, which
existence are proved by Faber~\cite{faber-14} (see
Section~\ref{ss::interpolation_grid}), must belong to the ${\mathcal
  C}^0$ functions. Indeed, for the case $m=0$,
Equation~(\ref{e:error_cheb}) does not prove convergence (neither do
Equations~(\ref{e:error_leg}) and (\ref{e:error_cheb2})).

\epubtkImage{}{%
\begin{figure}
  \centerline{\includegraphics[height=8cm]{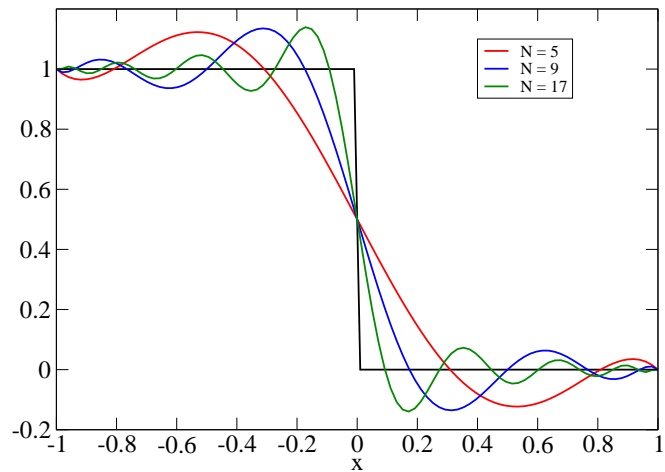}}
  \caption{Step function (black curve) and its interpolant, for
  various values of $N$.}
  \label{figure:gibbs}
\end{figure}}

\subsubsection{Trigonometric functions}
\label{sss:trigo}

A detailed presentation of the theory of Fourier transform is beyond
the scope of this work. However, there is a close link between the
so-called {\em discrete Fourier transform} and the spectral
interpolation and this is briefly outlined here. More details can be
found, for instance, in~\cite{canuto-06}.

The Fourier transform $Pf$ of a function $f$ of $\l[0, 2\pi\r]$ is given by:
\begin{equation}
  Pf\l(x\r) = a_0 + \sum_{n=1}^\infty a_n \cos\l(nx\r) +
  \sum_{n=1}^\infty b_n \sin\l(nx\r).
\end{equation}
The Fourier transform is known to converge rather rapidly to the
function itself, if $f$ is periodic. However, the coefficients $a_n$
and $b_n$ are given by integrals of the form $\displaystyle\int_0^{2\pi}
f\l(x\r) \cos\l(nx\r) {\rm d}x$, that cannot easily be computed (as it
was the case for the projection of a function on orthogonal
polynomials in Section~\ref{sss:ortho}).

The solution to this problem is also very similar to the use of the
Gaussian quadratures. Let us introduce $N+1$ collocation points $x_i =
2\pi i/(N+1)$. Then, the {\em discrete Fourier coefficients} with
respect to those points are:
\begin{eqnarray}
\tilde{a}_0 &=& \frac{1}{N} \sum_{k=1}^N f\l(x_k\r) \\
\tilde{a}_n &=& \frac{2}{N} \sum_{k=1}^N f\l(x_k\r) \cos\l(nx_k\r) \\
\tilde{b}_n &=& \frac{2}{N} \sum_{k=1}^N f\l(x_k\r) \sin\l(nx_k\r)
\end{eqnarray}
and the interpolant $I_N f$ is then given by:
\begin{equation}
  I_N f\l(x\r) = \tilde{a}_0 + \sum_{n=1}^N \tilde{a}_n \cos\l(nx\r) +
  \sum_{n=1}^N \tilde{b}_n \sin\l(nx\r).
\end{equation}

The approximation made by using the discrete coefficients in place of the real
ones is of the same nature as the one made when computing the coefficients
of the projection (\ref{equation:coef_proj}) by means of the Gaussian
quadratures. Let us mention that, in the case of a discrete Fourier transform,
the first and last collocation points lie on the boundaries of the interval, as
for a Gauss-Lobatto quadrature. As for the polynomial interpolation, the
convergence of $I_N f$ to $f$ is spectral for all periodic and ${\mathcal
  C}^\infty$ functions.

\subsubsection{Choice of basis}

For periodic functions of $\l[0, 2\pi\r[$, the discrete Fourier transform is
the natural choice of basis. If the considered function has also some
symmetries, one can use a subset of the trigonometric polynomials. For
instance, if the function is i) periodic on $\l[0, 2\pi\r[$ and is also odd
with respect to $x=\pi$, then it can be expanded on sines only. If the
function is not periodic, then it is natural to expand it either on Chebyshev
or Legendre polynomials. Using Legendre polynomials can be motivated by the
fact that the associated measure is very simple $w\l(x\r) = 1$. The
multi-domain technique presented in Section~\ref{ss:weak} is one particular
example where such property is required. In practice, Legendre and Chebyshev
polynomials usually give very similar results.

\subsection{Spectral methods for ODEs}
\label{ss:sm_for_ODEs}

\subsubsection{Weighted residual method}
\label{sss:wrm}

Let us consider a differential equation of the following form 
\begin{equation}
  Lu\l(x\r) = S\l(x\r), \quad x\in\interv,
\end{equation}
where $L$ is a linear second-order differential operator. The
problem admits a unique solution once appropriate boundary conditions are
prescribed at $x=1$ and $x=-1$. Typically, one can specify i) the
value of $u$ (Dirichlet-type) ii) the value of its derivative
$\partial_x u$ (Neumann-type) iii) a linear combination of both
(Robin-type).

As for the elementary operations presented in Section~\ref{sss:leg} and
\ref{sss:cheby}, the action of $L$ on $u$ can be expressed by a matrix
$L_{ij}$. If the coefficients of $u$ with respect to a given basis are
the $\tilde{u}_i$, then the coefficients of $Lu$ are 
\begin{equation}
\label{e:action_L}
\sum_{j=0}^N L_{ij} \tilde{u}_j.
\end{equation}
Usually, the $L_{ij}$ can easily be computed by combining the action of
elementary operations like the second derivative, the first
derivative, the multiplication or division by $x$ (see
Sec. \ref{sss:leg} and \ref{sss:cheby} for some examples).

A function $u$ is an admissible solution of the problem if and only if
i) it fulfills the boundary conditions exactly (up to machine
accuracy) ii) it makes the {\em residual} $R = Lu - S$ small. In the
weighted residual method, one considers a set of $N+1$ test functions
$\left\{ \xi_n\right\}_{n=0\dots N}$ on $\interv$. The smallness of $R$ is enforced by demanding
that 
\begin{equation}
\label{e:residual}
\l(R, \xi_k\r) = 0, \forall k\leq N.
\end{equation}
As $N$ increases, the obtained solution is closer and closer to the real one.
Depending on the choice of the test functions and the way the boundary
conditions are enforced, one gets various solvers. Three classical examples
are presented below.

\subsubsection{The Tau-method}
\label{s:tau}

In this particular method, the test functions coincide with the basis
used for the spectral expansion, for instance the Chebyshev
polynomials. Let us denote $\tilde{u}_i$ and $\tilde{s}_i$ the
coefficients of the solution $u$ and of the source $S$.

Given the expression of $Lu$ in the coefficient space
(\ref{e:action_L}) and the fact that the basis polynomials are
orthogonal, the residual equations (\ref{e:residual}) are expressed as 
\begin{equation}
\label{e:tau}
\sum_{i=0}^N L_{ni} \tilde{u_i} = \tilde{s}_n, \quad \forall n\leq N,
\end{equation}
the unknowns being the $\tilde{u}_i$. However, as such, this system does not
admit a unique solution, due to the homogeneous solutions of $L$ (i.e. the
matrix associated with $L$ is not invertible) and one has to impose the
boundary conditions. In the Tau-method, this is done by relaxing the {\em last
  two} equations (\ref{e:tau}) (i.e. for $n=N-1$ and $n=N$) and by replacing
them by the boundary conditions at $x=-1$ and $x=1$.

The Tau-method thus ensures that $Lu$ and $S$ have the same coefficients, excepted
 the last ones. If the functions are smooth, then their coefficients should
decrease in a spectral manner and so the ``forgotten'' conditions are less and
less stringent as $N$ increases, ensuring that the computed solution converges
rapidly to the real one.

As an illustration, let us consider the following equation:
\begin{equation}
  \label{e:test_pb}
  \frac{{\rm d}^2u}{{\rm d}x^2} - 4 \frac{{\rm d}u}{{\rm d}x} + 4 u = \exp\l(x\r) - \frac{4e}{\l(1+x^2\r)}
\end{equation}
with the following boundary conditions 
\begin{equation}
  \label{e:bc}
  u\l(x=-1\r)=0 \quad {\rm and}\quad u\l(x=1\r)=0. 
\end{equation}
The exact solution is analytic and is given by 
\begin{equation}
  \label{e:test_sol}
  u\l(x\r) = \exp\l(x\r) - \frac{\sinh\l(1\r)}{\sinh\l(2\r)} \exp\l(2x\r) - \frac{e}{\l(1+x^2\r)}.
\end{equation}

Figure~\ref{f:tau} shows the exact solution and the numerical one, for two
different values of $N$. One can note that the numerical solution converges
rapidly to the exact one, the two being almost indistinguishable for $N$ as
small as $N=8$. The numerical solution exactly fulfills the boundary
conditions, no matter the value of $N$.

\epubtkImage{}{%
  \begin{figure}
    \centerline{
      \includegraphics[width=7.5cm]{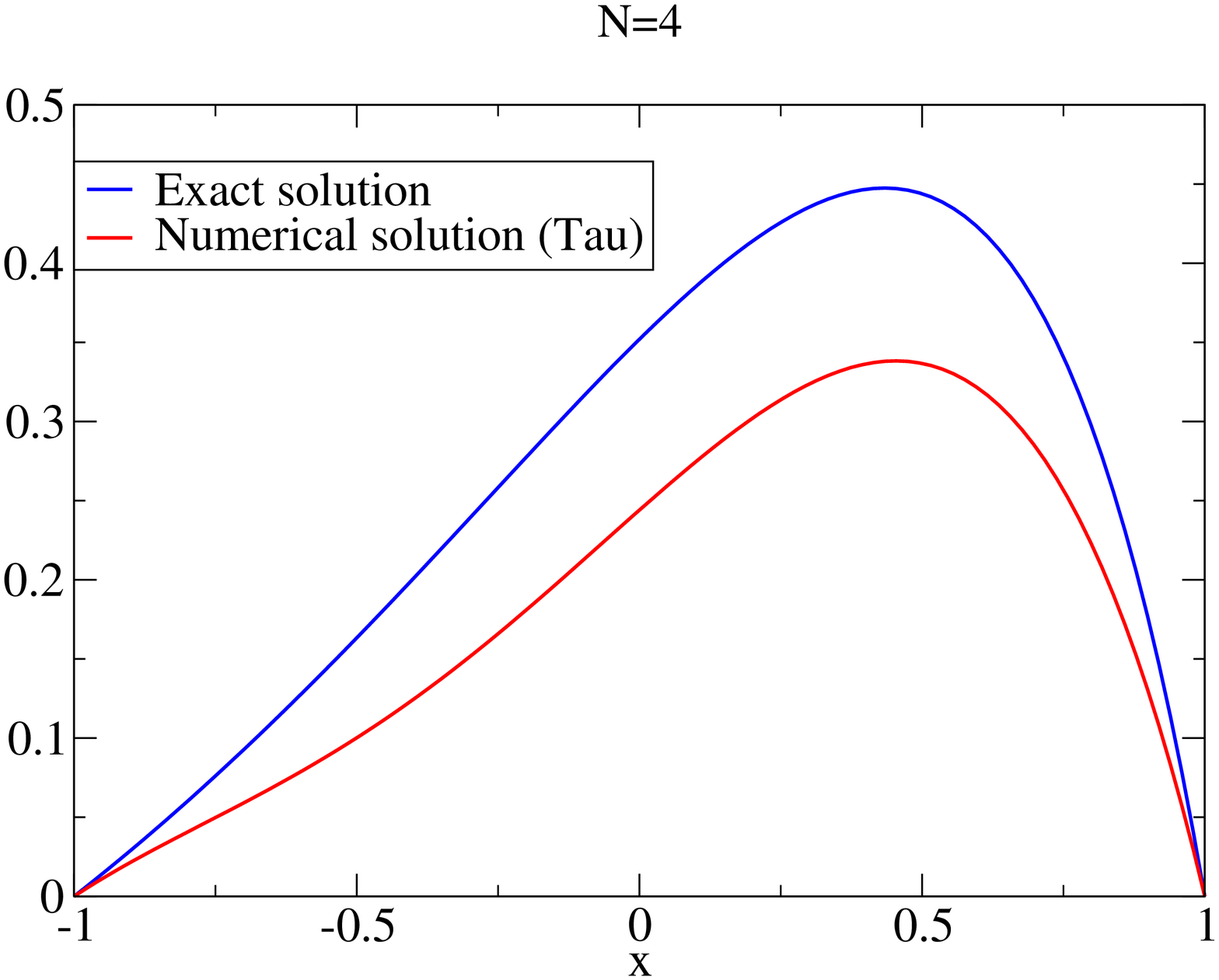}
      \includegraphics[width=7.5cm]{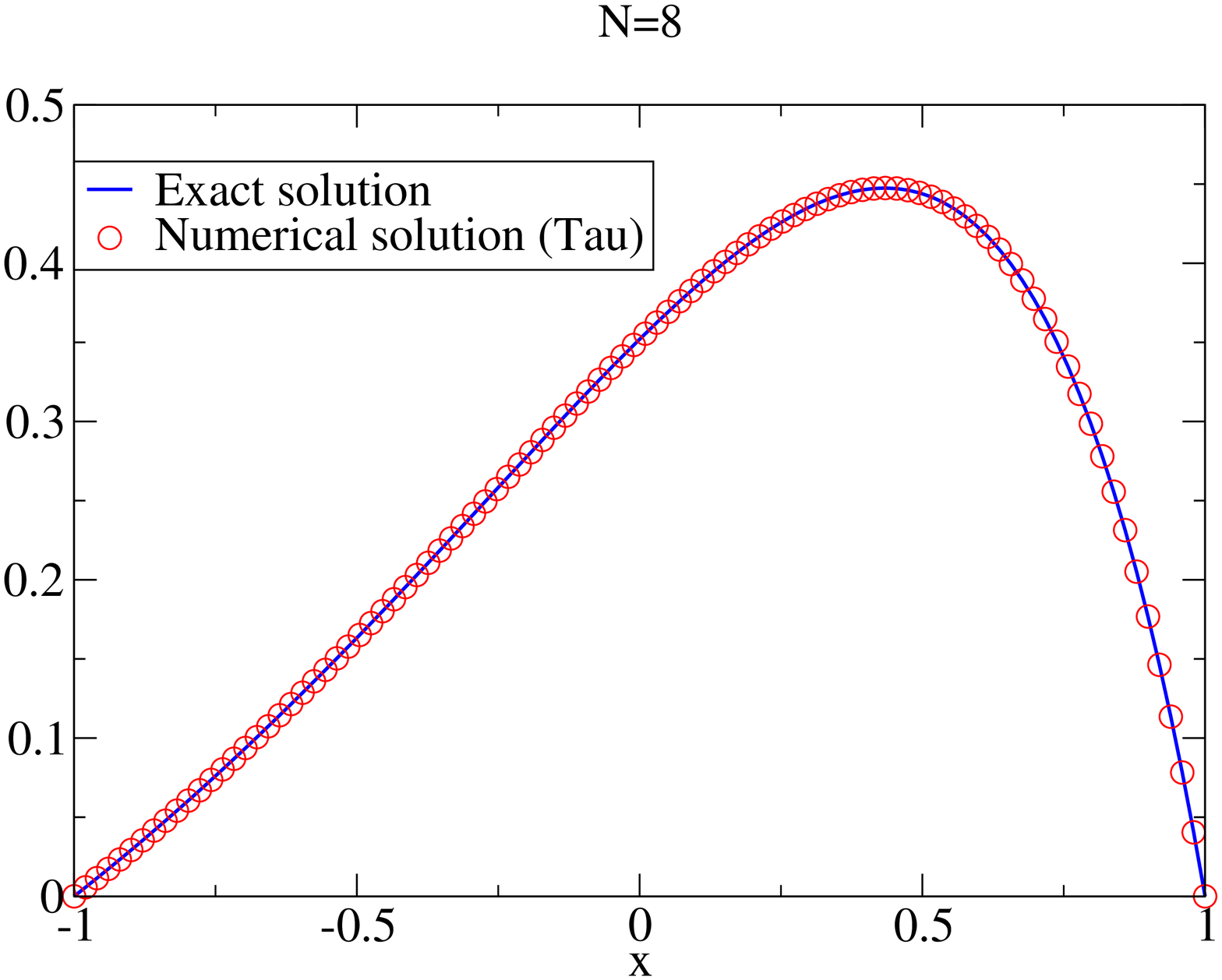}
    }
    \caption{Exact solution (\ref{e:test_sol}) of
    Equation~(\ref{e:test_pb}) (blue curve) and the numerical solution
    (red curves) computed by means of a Tau-method, for $N=4$ (left
    panel) and $N=8$ (right panel).}
    \label{f:tau}
\end{figure}}

\subsubsection{The collocation method}
\label{s:colloc}

The collocation method is very similar to the Tau-method. They only differ in
the choice of test functions. Indeed, in the collocation method one uses
continuous function that are zero at each but one collocation point. They are
indeed the Lagrange cardinal polynomials already seen in
Section~\ref{ss::interpolation_grid} and can be written as $\xi_i\l(x_j\r) =
\delta_{ij}$. With such test functions, the residual equations
(\ref{e:residual}) are
\begin{equation}
  Lu\l(x_n\r) = S\l(x_n\r), \quad \forall n\leq N.
\end{equation}

The value of $Lu$ at each collocation point is easily expressed in
terms of $\tilde{u}$ by making use of (\ref{e:action_L}) and one
gets:
\begin{equation}
  \label{e:colloc}
  \sum_{i=0}^N\sum_{j=0}^N L_{ij}\tilde{u}_j T_i\l(x_n\r) =
  S\l(x_n\r), \quad \forall n\leq N.
\end{equation}

Let us note that even if the collocation method imposes that $Lu$ and $S$
coincide at each collocation point, the unknowns of the system written in the
form (\ref{e:colloc}) are the coefficients $\tilde{u}_n$ and not the
$u\l(x_n\r)$. As for the Tau-method, the system (\ref{e:colloc}) is not
invertible and boundary conditions must be enforced by additional equations.
In this case, the relaxed conditions are the two associated with the outermost
points, i.e. $n=0$ and $n=N$, which are replaced by appropriate boundary
conditions to get an invertible system.

Figure~\ref{f:colloc} shows both the exact and numerical solutions for
Equation~(\ref{e:test_pb}).

\epubtkImage{}{%
  \begin{figure}
    \centerline{
      \includegraphics[width=7.5cm]{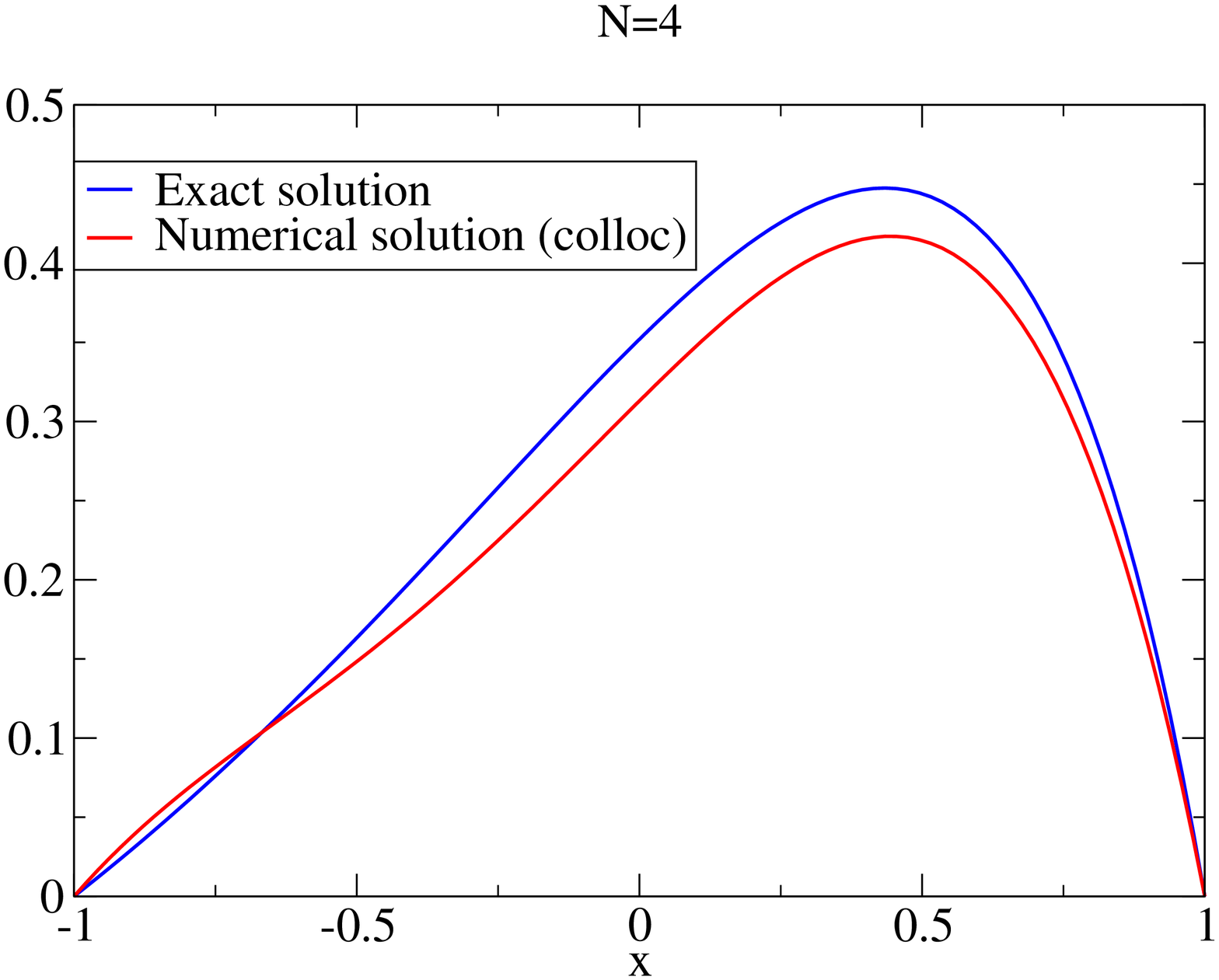}
      \includegraphics[width=7.5cm]{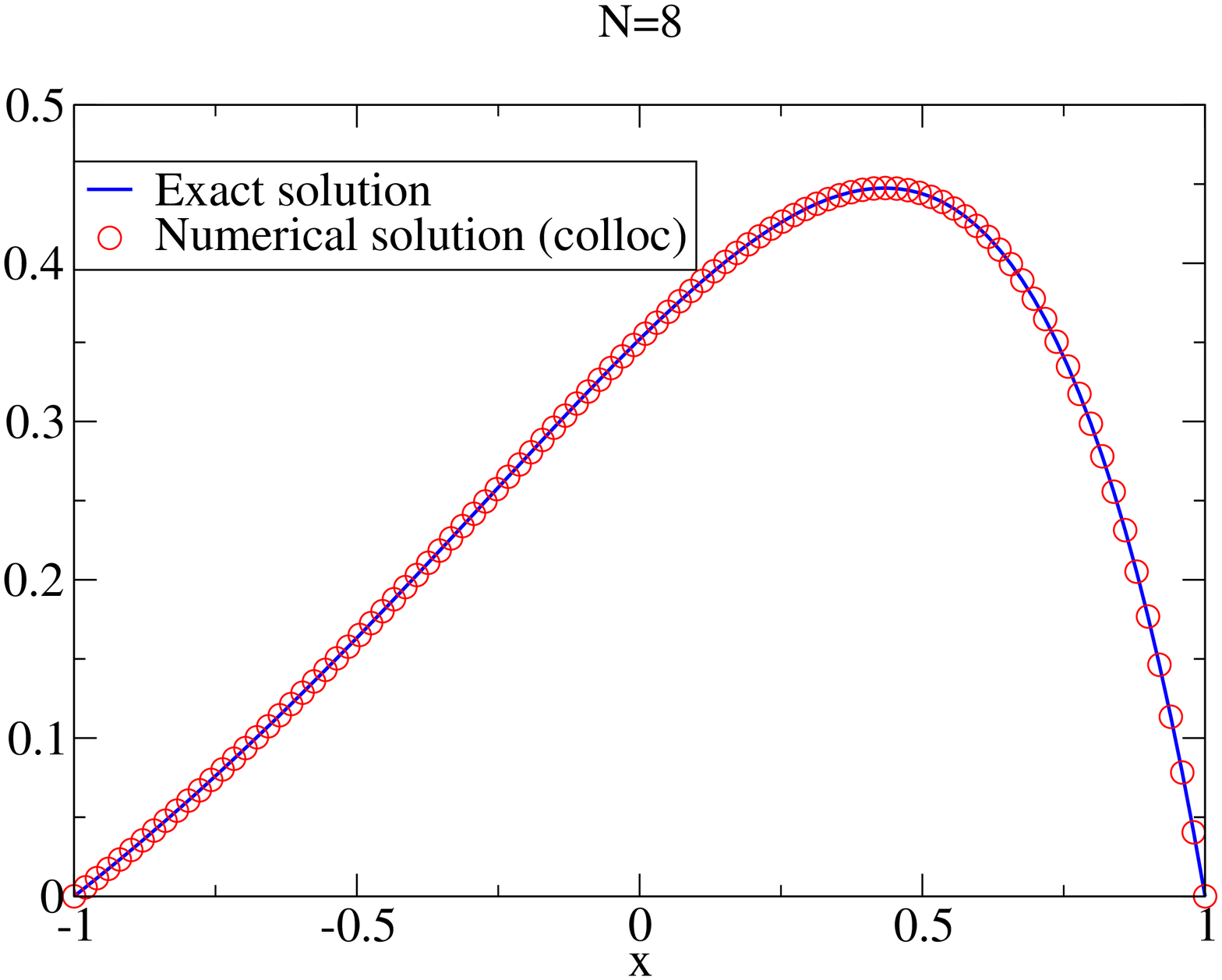}
    }
    \caption{Same as Figure~\ref{f:tau} but for the collocation
    method.}
    \label{f:colloc}
\end{figure}}

\subsubsection{Galerkin method}

The basic idea of Galerkin method is to seek the solution $u$ as a sum of
polynomials $G_i$ that {\em individually} verify the boundary conditions.
Doing so, $u$ automatically fulfills those conditions and they do not have to
be imposed by additional equations. Such polynomials constitute a Galerkin
basis of the problem. For practical reasons, it is better to chose a Galerkin
basis that can easily be expressed in terms of the original orthogonal
polynomials.

For instance, with the boundary conditions (\ref{e:bc}), one can
choose:
\begin{eqnarray}
\label{e:galerkin_example1}
G_{2k}\l(x\r) &=& T_{2k+2}\l(x\r) - T_0 \l(x\r) \\
\label{e:galerkin_example2}
G_{2k+1}\l(x\r) &=& T_{2k+3}\l(x\r) - T_1\l(x\r)
\end{eqnarray}

More generally, the Galerkin basis relates to the usual ones by means
of a transformation matrix 
\begin{equation}
  \label{e:transfo}
  G_i = \sum_{j=0}^N M_{ji} T_j, \quad \forall i\leq N-2.
\end{equation}
Let us mention that the matrix $M$ is not square. Indeed, to maintain the same
degree of approximation, one can consider only $N-1$ Galerkin polynomials, due
to the two additional conditions they have to fulfill (see for instance
Equations~(\ref{e:galerkin_example1}-\ref{e:galerkin_example2})). One can also
note that, in general, the $G_i$ are {\em not} orthogonal polynomials.

The solution $u$ is sought in terms of the coefficients
$\tilde{u}^G_i$ on the Galerkin basis:
\begin{equation}
  u\l(x\r) = \sum_{k=0}^{N-2} \tilde{u}^G_k G_k\l(x\r).
\end{equation}
By making use of Equations~(\ref{e:action_L}) and (\ref{e:transfo})
one can express $Lu$ in terms of the $\tilde{u}^G_i$:
\begin{equation}
  \label{e:lu_galerkin}
  Lu\l(x\r) = \sum_{k=0}^{N-2} \tilde{u}^G_k \sum_{i=0}^N\sum_{j=0}^N
  M_{jk} L_{ij} T_i \l(x\r).
\end{equation}

The test functions used in the Galerkin method are the $G_i$ themselves, so
that the residual system reads:
\begin{equation}
  \l(Lu, G_n\r) = \l(S, G_n\r), \quad \forall n\leq N-2
\end{equation}
where the left-hand-side is computed by means of (\ref{e:lu_galerkin})
and by expressing the $G_i$ in terms of the $T_i$ by
(\ref{e:transfo}). Concerning the right-hand-side, the source itself
{\em is not} expanded on the Galerkin basis, given that it does not
fulfill the boundary conditions. Putting all the pieces together, the
Galerkin system reads:
\begin{equation}
  \sum_{k=0}^{N-2} \tilde{u}^G_k \sum_{i=0}^N\sum_{j=0}^N
  M_{in}M_{jk}L_{ij} \l(T_i|T_i\r) = \sum_{i=0}^N M_{in} \tilde{s}_i
  \l(T_i|T_i\r), \quad \forall n \leq N-2.
\end{equation}
This is a system of $N-1$ equations for the $N-1$ unknowns
$\tilde{u}^G_i$ and it can be directly solved, because it is well-posed. Once
the $\tilde{u}^G_i$ are known, one can obtain the solution in terms of
the usual basis by making, once again, use of the transformation
matrix:
\begin{equation}
  u\l(x\r) = \sum_{i=0}^N \l(\sum_{n=0}^{N-2} M_{in} \tilde{u}^G_n\r) T_i.
\end{equation}

The solution obtained by the application of this method to the
Equation~(\ref{e:test_pb}) is shown on Figure~\ref{f:galerkin}.

\epubtkImage{}{%
  \begin{figure}
    \centerline{
      \includegraphics[width=7.5cm]{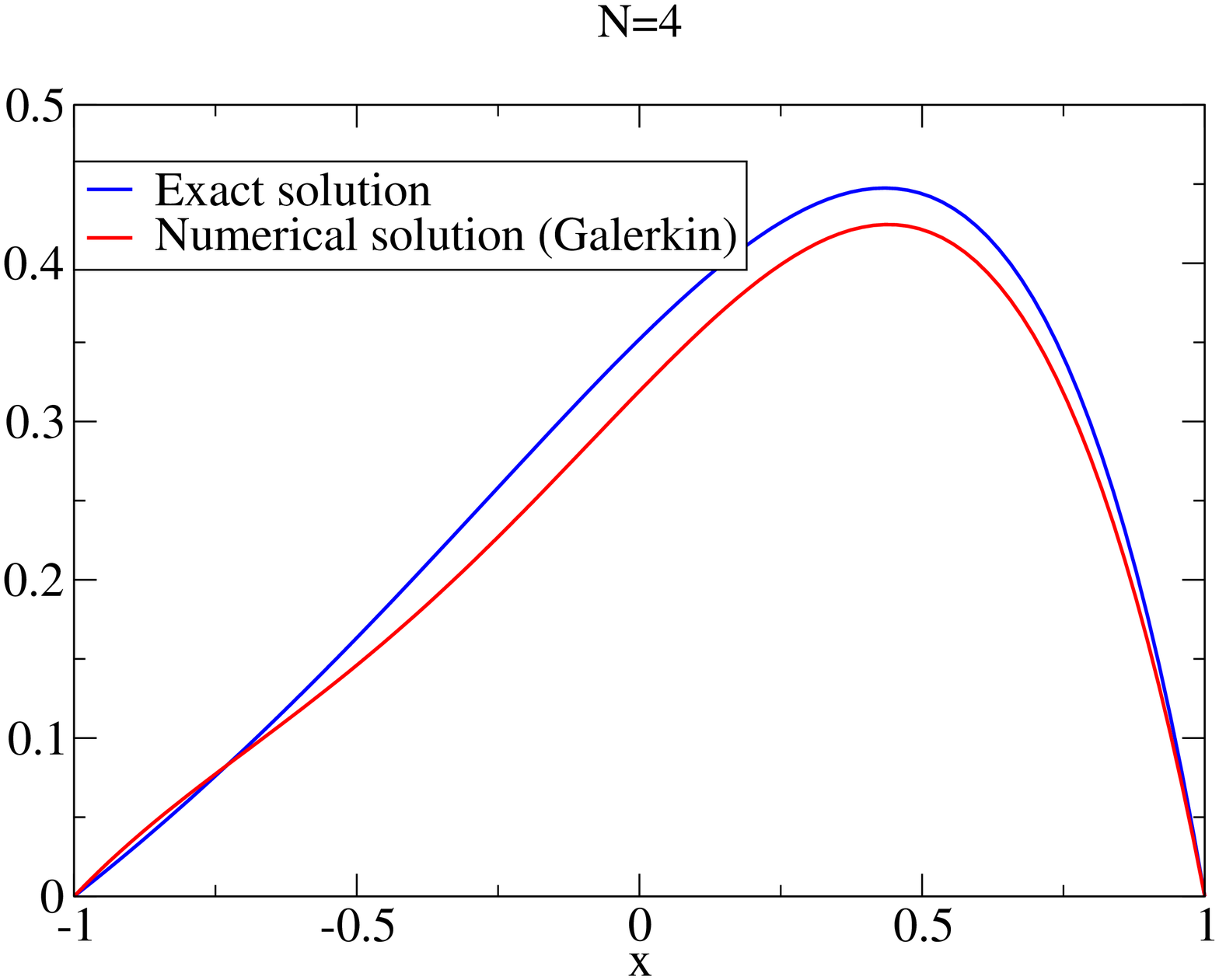}
      \includegraphics[width=7.5cm]{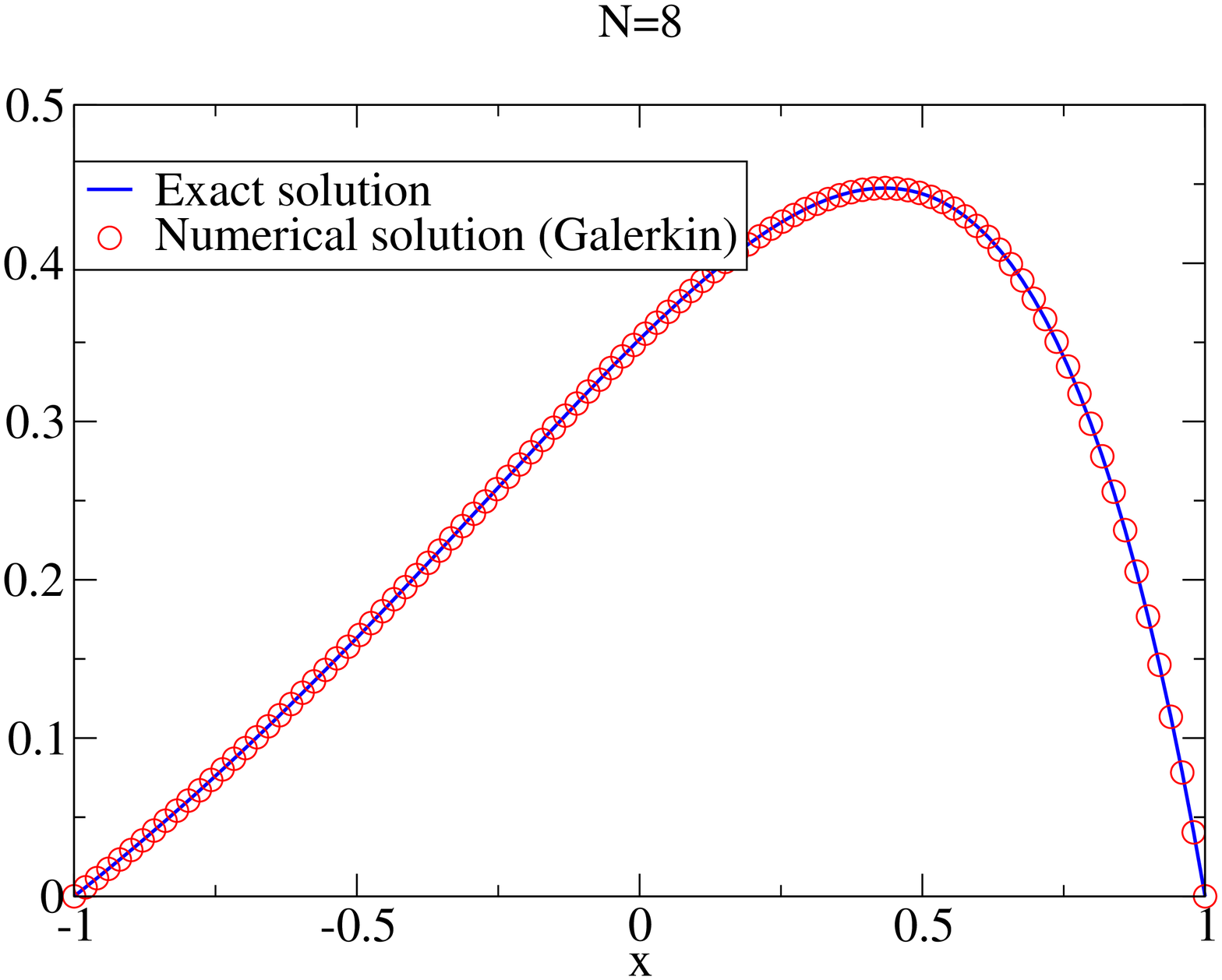}
    }
    \caption{Same as Figure~\ref{f:tau} but for the Galerkin method.}
    \label{f:galerkin}
\end{figure}}

\subsubsection{The methods are optimal}

A numerical method is said to be optimal if it does not introduce an
additional error to the one that would be done by interpolating the exact
solution of a given equation.

Let us call $u_{\rm exact}$ such exact solution, unknown in general. Its
interpolant is $I_N u_{\rm exact}$ and the numerical solution of the equation
is $u_{\rm num}$. The numerical method is then optimal if and only if $\l\|
I_N u_{\rm exact} - u_{\rm exact} \r\|_\infty$ and $\l\|u_{\rm num} - u_{\rm
  exact} \r\|_\infty$ behave in the same manner when $N \rightarrow \infty$.

In general, optimality is difficult to check because both $u_{\rm exact}$ and
its interpolant are unknown. However, for the test problem proposed in
Section~\ref{s:tau} this can be done. Figure~\ref{f:optimal} shows the maximum
relative difference between the exact solution (\ref{e:test_sol}) and its
interpolant and the various numerical solutions. All the curves behave in the
same manner as $N$ increases, indicating that the three methods previously
presented are optimal (at least for this particular case).

\epubtkImage{}{%
\begin{figure}
  \centerline{\includegraphics[height=8cm]{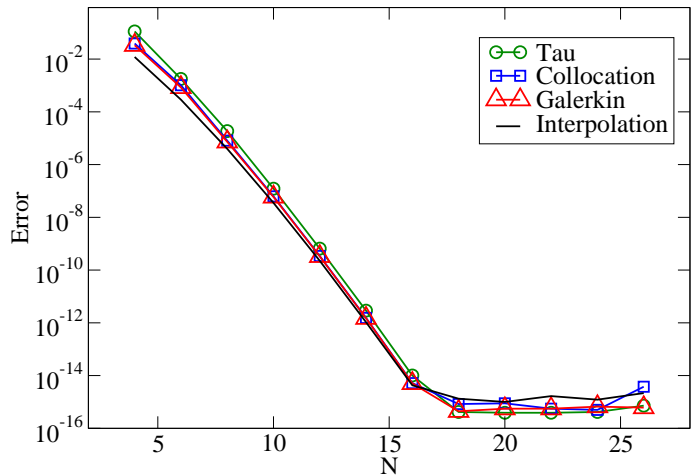}}
  \caption{Difference between the exact solution (\ref{e:test_sol}) of
    Equation~(\ref{e:test_pb}) and its interpolant (black curve) and between
    the exact and numerical solutions for i) the tau method (green curve and
    circle symbols) ii) the collocation method (blue curve and square symbols)
    iii) the Galerkin method (red curve and triangle symbols).}
  \label{f:optimal}
\end{figure}}

\subsection{Multi-domain techniques for ODEs}
\label{ss:multidomain_techniques}

\subsubsection{Motivations and setting}

A seen in Section~\ref{sss:convergence}, spectral methods are very efficient when
dealing with ${\mathcal C}^\infty$ functions. However, they lose some of their
appeal when dealing with less regular functions, the convergence to the exact
functions being substantially slower. Nevertheless, the physicist has
sometimes to deal with such fields. This is the case for the density jump at
the surface of strange stars or the formation of shocks to mention only two
examples. In order to maintain spectral convergence, one then needs to
introduce several computational domains such that the various discontinuities
of the functions lie at the interface between the domains. Doing so, {\em in
  each domain}, one only deals with ${\mathcal C}^\infty$ functions.

Multi-domain techniques can also be valuable when dealing with a physical
space either too complicated or too large to be described by single domain.
Related to that, one can also use several domains to increase the resolution
in some parts of the space where more precision is required. This can be
easily done by using a different number of basis functions in different
domains. One then talks about fixed-mesh refinement.

Parallelism can also be a reason why several domains may be used. Indeed, one
could set a solver, dealing with each domain on a given processor,
and interprocessor communication is then only used for matching the solution
across the various domains. The algorithm of Section~\ref{ss:hom} is well adapted
to such purpose.

In the following, four different multi-domain methods are presented to solve
an equation of the type $Lu=S$ on $\interv$. $L$ is a second order linear
operator and $S$ a given source function. Appropriate boundary conditions are
given at the boundaries $x=-1$ and $x=1$.

For simplicity the physical space is split into two domains: 

\begin{itemize}
  \item first domain: $x \leq 0$ described by $x_1 = 2x+1, \quad x_1 \in\interv$,
  \item second domain: $x\geq 0$ described by $x_2 = 2x-1, \quad x_2 \in\interv$.
\end{itemize}

If $x\leq 0$, a function $u$ is described by its interpolant in terms
of $x_1$: $I_N u \l(x\r) = \displaystyle\sum_{i=0}^N \tilde{u}^1_i
T_i\l(x_1\l(x\r)\r)$. The same is true for $x\geq 0$ with respect to
the variable $x_2$. Such setting is obviously appropriate to deal with
problems where discontinuities occur at $x=0$, that is $x_1=1$ and
$x_2=-1$.

\subsubsection{Multi-domain tau method}

As for the standard tau-method (see Section~\ref{s:tau}) and in each
domain, the test functions are the basis polynomials and one writes
the associated residual equations. For instance in the domain $x \leq
0$ one gets:
\begin{equation}
  \l(T_n, R\r) = 0 \Longrightarrow \sum_{i=0}^N L_{ni} \tilde{u}^1_i =
  \tilde{s}^1_n \quad \forall n\leq N,
\end{equation}
the $\tilde{s}^1$ being the coefficients of the source and $L_{ij}$
the matrix representation of the operator. As for the one-domain case,
one relaxes the last two equations, keeping only $N-1$ equations. The
same is done in the second domain.

Two supplementary equations are enforced to ensure that the boundary
conditions are fulfilled. Finally, the operator $L$ being of second
order, one needs to ensure that the solution {\em and} its first
derivative are continuous at the interface $x=0$. This translates to a
set of two additional equations involving both domains.

So, one considers

\begin{itemize}
  \item $N-1$ residual equations in the first domain,
  \item $N-1$ residual equations in the second domain,
  \item 2 boundary conditions,
  \item 2 matching conditions,
\end{itemize}

for a total of $2N+2$ equations. The unknowns are the coefficients of
$u$ in both domains (i.e. the $\tilde{u}^1_i$ and the
$\tilde{u}^2_i$), that is $2N+2$ unknowns. The system is well posed
and admits a unique solution.

\subsubsection{Multi-domain collocation method}

As for the standard collocation method (see Section~\ref{s:colloc}) and
in each domain, the test functions are the Lagrange cardinal
polynomials. For instance in the domain $x \leq 0$ one gets:
\begin{equation}
  \sum_{i=0}^N \sum_{j=0}^N L_{ij} \tilde{u}^1_j T_i\l(x_{1n}\r)=
  S\l(x_{1n}\r) \quad \forall n\leq N,
\end{equation}
$L_{ij}$ being the matrix representation of the operator and $x_{1n}$
the $n^{\rm th}$ collocation point in the first domain. As for the
one-domain case, one relaxes the two equations corresponding to the
boundaries of the domain, keeping only $N-1$ equations. The same is
done in the second domain.

Two supplementary equations are enforced to ensure that the boundary
conditions are fulfilled. Finally, the operator $L$ being second-order, one
needs to ensure that the solution {\em and} its first derivative are
continuous at the interface $x=0$. This translates as a set of two additional
equations involving the coefficients in both domains.

So, one considers

\begin{itemize}
  \item $N-1$ residual equations in the first domain,
  \item $N-1$ residual equations in the second domain,
  \item $2$ boundary conditions,
  \item $2$ matching conditions,
\end{itemize}

for a total of $2N+2$ equations. The unknowns are the coefficients of
$u$ in both domains (i.e. the $\tilde{u}^1_i$ and the
$\tilde{u}^2_i$), that is $2N+2$ unknowns. The system is well posed
and admits a unique solution.

\subsubsection{Method based on the homogeneous solutions}
\label{ss:hom}

The method exposed here proceeds in two steps. First, particular
solutions are computed in each domain. Then, appropriate linear
combination with the homogeneous solutions of the operator $L$ are
performed to ensure continuity and impose boundary conditions.

In order to compute particular solutions, one can rely on any of the
methods exposed in Section~\ref{ss:sm_for_ODEs}. The boundary conditions
at the boundary of each domain can be chosen (almost) arbitrarily. For
instance one can use, in each domain, a collocation method to solve
$Lu=S$, demanding that the particular solution $u_{\rm part}$ is zero
at both end of each intervals.

Then, in order to have a solution in the whole space, one needs to add
homogeneous solutions to the particular ones. In general, the operator
$L$ is second-order and it admits two independent homogeneous
solutions $g$ and $h$, in each domain. Let us note that, in some
cases, additional regularity conditions can reduce the number of
available homogeneous solutions. The homogeneous solutions can either
be computed analytically if the operator $L$ is simple enough or
numerically but one then needs to have a method for solving $Lu=0$.

In each domain, the physical solution is a combination of the
particular solution and the homogeneous ones of the type:
\begin{equation}
  u = u_{\rm part} + \alpha g + \beta h,
\end{equation}
where $\alpha$ and $\beta$ are constants that must be determined. In
the two domains case, we are left with 4 unknowns. The system they
must verify is composed of i) 2 equations for the boundary conditions
ii) 2 equations for the matching of $u$ and its first derivative
across the boundary between the two domains. The obtained system is
called the matching system and generally admits a unique solution.

\subsubsection{Variational method}
\label{ss:weak}

Contrary to previously presented methods, the variational one is
only applicable with Legendre polynomials. Indeed, the method requires
that the measure be $w\l(x\r) = 1$. It is also useful to extract the
second-order term of the operator $L$ and to rewrite it like $Lu = u''
+ H$, $H$ being of first-order only.

In each domain, one writes the residual equation explicitly:
\begin{equation}
  \l(\xi, R\r) = 0 \Longrightarrow \int \xi u'' {\rm d}x + \int \xi
  \l(Hu\r) {\rm d}x = \int \xi S {\rm d}x.
\end{equation}

The term involving the second derivative of $u$ is then integrated by parts:
\begin{equation}
  \label{e:byparts}
  \l[\xi u'\r] - \int \xi' u' {\rm d} x + \int \xi \l(Hu\r) {\rm d}x =
  \int \xi S {\rm d}x.
\end{equation}

The test functions are the same as the ones used for the collocation method,
i.e. functions being zero at all but one collocation point, in both domains
($d=1, 2$): $\xi_i\l(x_{dj}\r) = \delta_{ij}$. By making use of the Gauss
quadratures, the various parts of Equation~(\ref{e:byparts}) can be expressed
as ($d=1,2$ indicates the domain):
\begin{eqnarray}
  \int \xi_n' u' {\rm d}x &=& \sum_{i=0}^N \xi_n' \l(x_{di}\r) u'\l(x_{di}\r) w_i = 
  \sum_{i=0}^N \sum_{j=0}^N D_{ij} D_{in} w_i u\l(x_{dj}\r) \\
  \int \xi_n \l(Hu\r) {\rm d}x &=& \sum_{i=0}^N \xi_n\l(x_{di}\r) \l(Hu\r)\l(x_{di}\r) w_i = w_n \sum_{i=0}^N H_{ni} u\l(x_{di}\r)\\
  \int \xi_n S {\rm d}x &=& \sum_{i=0}^N \xi_n\l(x_{di}\r) S\l(x_{di}\r) w_i = S\l(x_{dn}\r) w_n,
\end{eqnarray}
where $D_{ij}$ (resp. $H_{ij}$) represent the action of the derivative
(resp. of $H$) {\em in the configuration space}
\begin{eqnarray}
  g'\l(x_{dk}\r) &=& \sum_{j=0}^N D_{kj} g\l(x_{dj}\r) \\
  \l(Hg\r) \l(x_{dk}\r) &=& \sum_{j=0}^N H_{kj} g\l(x_{dj}\r).
\end{eqnarray}

For points {\em strictly} inside each domain, the integrated term $\l[\xi
u'\r]$ of Equation~(\ref{e:byparts}) vanishes and one gets equations of the form:
\begin{equation}
  \label{e:inside}
  -\sum_{i=0}^N \sum_{j=0}^N D_{ij} D_{in} w_i u\l(x_{dj}\r) + w_n
   \sum_{i=0}^N H_{ni} u\l(x_{di}\r) = S\l(x_{dn}\r) w_n.
\end{equation}
This is a set of $N-1$ equations for each domains ($d=1,2$). In the
above form, the unknowns are the $u\l(x_{di}\r)$, i.e. the solution is
sought in the configuration space.

As usual, two additional equations are provided by appropriate boundary
conditions at both ends of the global domain. One also gets an
additional condition by matching the solution across the boundary
between the two domains.

The last equation of the system is the matching of the first
derivative of the solution. However, instead of writing it
``explicitly'', this is done by making use of the integrated term in
Equation~(\ref{e:byparts}) and this is actually the crucial step of the
whole method. Applying Equation~(\ref{e:byparts}) to the last point
$x_{1N}$ of the first domain, one gets:
\begin{equation}
  u'\l(x_1=1\r) =  \sum_{i=0}^N \sum_{j=0}^N D_{ij} D_{iN} w_i
  u\l(x_{1j}\r) - w_N \sum_{i=0}^N H_{Ni} u\l(x_{1i}\r) +
  S\l(x_{1N}\r) w_N.
\end{equation}
The same can be done with the first point of the second domain, to get
$u'\l(x_2=-1\r)$ and the last equation of the system is obtained by
demanding that $u'\l(x_1=1\r) = u'\l(x_2=-1\r)$ and relates the values
of $u$ in both domains.

Before finishing with the variational method, it may be worthwhile to
explain why Legendre polynomials are used. Suppose one wants to work
with Chebyshev polynomials instead. The measure is then $w\l(x\r) =
\fraction{1}{\sqrt{1-x^2}}$. When one integrates the term containing
$u''$ by part one then gets 
\begin{equation}
  \int -u'' f w {\rm d}x = \l[-u' f w\r] + \int u' f' w' {\rm d} x
\end{equation}
Because the measure is divergent at the boundaries, it is difficult,
if not impossible, to isolate the term in $u'$. On the other hand, this
is precisely the term that is needed to impose the appropriate
matching of the solution.

\subsubsection{Merits of the various methods}

From a numerical point of view, the method based on an explicit matching using
the homogeneous solutions is somewhat different from the two others. Indeed,
one has to solve several systems in a row but each one is of the same size
than the number of points in one domain. This splitting of the different
domains can also be useful for designing parallel codes. On the contrary, for
both the variational and the tau method one has to solve only one system, but
its size is the same as the number of points in whole space, which can be
quite large for many domains settings. However, those two methods do not
require to compute the homogeneous solutions, computation that could be tricky
depending on the operators involved and on the number of dimensions.

The variational method may seem more difficult to implement and is only
applicable with Legendre polynomials. However, on the mathematical grounds, it
is the only method which is demonstrated to be optimal. Moreover, some
examples have been found where the others methods are not optimal. It remains
true that the variational method is very dependent on both the shape of the
domains and the type of equation that needs to be solved.

The choice of one method or another thus depends on the particularity of the
situation. As for the mono-domain space, for simple tests problems, the
results are very similar. Figure~\ref{f:errors_multi} shows the maximum error
between the analytical solution and the numerical one for the four different
methods. All errors decay exponentially and reach machine accuracy with the
roughly same number of points.

\epubtkImage{}{%
\begin{figure}
  \centerline{\includegraphics[height=8cm]{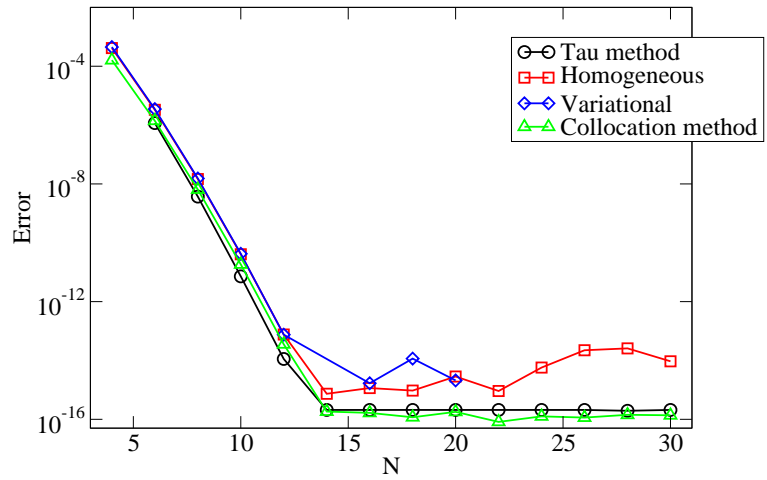}}
  \caption{Difference between the exact and numerical solutions of the
    following test problem.  $\displaystyle\frac{{\rm d}^2u}{{\rm d}x^2} + 4 u
    = S$, with $S\l(x<0\r)=1$ and $S\l(x>0\r)=0$. The boundary conditions are
    $u\l(x=-1\r)=0$ and $u\l(x=1\r)=0$. The black curve and circles denote
    results from the multi-domain Tau method, the red curve and squares from
    the method based on the homogeneous solutions, the blue curve and diamonds
    from the variational one, the green curve and triangles from the
    collocation method.}
  \label{f:errors_multi}
\end{figure}}

\newpage

\section{Multi-Dimensional Cases}
\label{s:space_time}

In principle, the generalization to more than one dimension is rather
straightforward if one uses the tensor product. Let us first take an example,
with the spectral representation of a scalar function $f(x,y)$ defined on the
square $(x,y) \in \interv \times \interv$ in terms of Chebyshev polynomials.
One simply writes
\begin{equation}
  \label{eq:f2d_simple}
  f(x,y) = \sum_{i=0}^M \sum_{j=0}^N a_{ij} T_i(x) T_j(y),
\end{equation}
with $T_i$ being the Chebyshev polynomial of degree $i$. The partial
differential operators can also be generalized, as being linear operators
acting on the space $\mathbb{P}_M \otimes \mathbb{P}_N$. Simple, linear
Partial Differential Equations (PDE) can be solved by one of the methods
presented in Section~\ref{ss:sm_for_ODEs} (Galerkin, tau or collocation), on
this $M N$-dimensional space. The development (\ref{eq:f2d_simple}) can of
course be generalized to any dimension. Some special PDE and spectral basis
examples, where the differential equation decouples for some of the
coordinates, shall be given in Section~\ref{ss:spherical_coordinates_harmonics}.

\subsection{Spatial coordinate systems}
\label{ss:spatial_coordinate_systems}

Most of interesting problems in numerical relativity involve non-symmetric
systems and require the use of a full set of three-dimensional coordinates. We
briefly review hereafter several coordinate sets (all orthogonal) that have
been used in numerical relativity with spectral methods. They are described
through the line element $ds^2$ of the flat metric in the coordinates we
discuss.

\begin{itemize}
\item {\bf Cartesian (rectangular) coordinates} are of course the simplest and
  most straightforward to implement; the line element reads $ds^2 = dx^2 +
  dy^2 + dz^2$. These coordinates are regular in all space, with vanishing
  connection which makes them easy to use, since all differential operators
  have simple expressions and the associated triad is also perfectly regular.
  They are particularly well-adapted to cubic-like domains, see for
  instance~\cite{pfeiffer-03b, pfeiffer-03a}, and~\cite{frauendiener-99} in
  the case of toroidal topology.
\item {\bf Circular cylindrical coordinates} have a line element $ds^2 =
  d\rho^2 + \rho^2\,d\phi^2 + dz^2$ and exhibit a coordinate singularity on
  the $z$-axis ($\rho = 0$). The associated triad being also singular for
  $\rho = 0$, regular vector or tensor fields have components that are
  multi-valued (depending on $\phi$) on any point of the $z$-axis.  As for the
  spherical coordinates, this can be handled quite easily with spectral
  methods. This coordinate system can be useful for axisymmetric or rotating systems,
  see~\cite{ansorg-03}.
\item {\bf Spherical (polar) coordinates} will be discussed more in details in
   Section~\ref{ss:spherical_coordinates_harmonics}. Their line element reads
  $ds^2 = dr^2 + r^2\, d\th^2 + r^2\sin^2\th\, d\ph^2$, showing a
  coordinate singularity at the origin ($r=0$) and on the axis for which
  $\th=0,\pi$. They are very interesting in numerical relativity for the
  numerous spherical-like objects under study (stars or black hole horizons)
  and have been mostly implemented for shell-like domains~\cite{bonazzola-99,
    grandclement-01, pfeiffer-03b, tichy-06} and for spheres including the
  origin~\cite{bonazzola-90, grandclement-01}.
\item {\bf Prolate spheroidal coordinates} consist of a system of confocal
  ellipses and hyperbola, describing an $(x,z)$-plane, and an angle $\ph$
  giving the position, as a rotation with respect to the focal
  axis~\cite{korn-61}. The line element is $ds^2 = a^2\left( \sinh^2 \mu +
    \sin^2 \nu \right)\left( d\mu^2 + d\nu^2 \right) + a^2 \sinh^2 \mu \sin^2
  \nu\, d\ph^2$. The foci are situated at $z=\pm a$ and represent
  coordinate singularities for $\mu =0$ and $\nu = 0, \pi$. These coordinates
  have been used in~\cite{ansorg-04} with black hole punctures data at the foci.
\item {\bf Bispherical coordinates} are obtained by rotation of bipolar
  coordinates around the focal axis, with a line element $ds^2 = a^2 \left(
    \cosh \eta - \cos \chi \right)^{-2} \left( d\eta^2 + d\chi^2 + \sin^2\chi
    d\ph^2 \right)$. As for prolate spheroidal coordinates, the foci
  situated at $z = \pm a$ ($\eta \to \pm \infty, \chi = 0, \pi$) and more
  generally, the focal axis exhibits coordinate singularities. Still, the
  surfaces of constant $\eta$ are spheres situated in the $z>0(<0)$ region for
  $\eta>0(<0)$, respectively. Thus these coordinate are very well adapted for
  the study of binary systems and in particular for excision treatment of binary
  black holes~\cite{ansorg-05}.
\end{itemize}

\subsubsection{Mappings}
\label{sss:mappings}

Choosing a smart set of coordinates is not the end of the story. As for
finite-elements, one would like to be able to cover some complicated
geometries, like distorted stars, tori, etc\dots or even to be able to
cover the whole space. The reason for this last point is that, in numerical
relativity, one often deals with isolated systems for which boundary
conditions are only known at spatial infinity. A quite simple choice is to
perform a mapping from {\em numerical coordinates\/} to {\em physical
  coordinates\/}, generalizing the change of coordinates to $\interv$, when
using families of orthonormal polynomials or to $[0,2\pi]$ for Fourier series. 

\epubtkImage{}{%
\begin{figure}[ht]
  \centerline{\includegraphics[width=8cm]{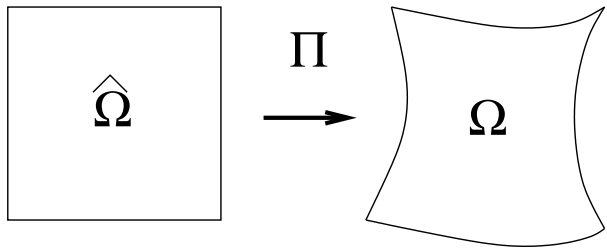}}
  \caption{Regular deformation of the $\interv \times \interv$ square.}
  \label{figure:map_square}
\end{figure}}

An example of how to map the $\interv \times \interv$ domain can be taken from
Canuto~et~al.~\cite{canuto-88}, and is illustrated on
Figure~\ref{figure:map_square}: once known the mappings from the four sides
(boundaries) of $\hat{\Omega}$ to the four sides of $\Omega$, one can
construct a two-dimensional regular mapping $\Pi$, which preserves
orthogonality and simple operators (see Chapter~3.5 of~\cite{canuto-88}).

The case where the boundaries of the considered domain are not known at the
beginning of the computation can also be treated in a spectral way. In the
case where this surface corresponds to the surface of a neutron star, two
approaches have been used. First in Bonazzola~et~al.~\cite{bonazzola-98}, the
star (and therefore the domain) is supposed to be ``star-like'', meaning that
there exists a point from which it is possible to reach any point on the
surface by straight lines that are all contained inside the star. To such a
point is associated the origin of a spherical system of coordinates, so that
it is a spherical domain, which is regularly deformed to coincide with the
shape of the star. This is done within an iterative scheme, at every step once
the position of the surface has been determined. Then, another approach has
been developed in Ansorg~et~al.~\cite{ansorg-03}, using cylindrical
coordinates. It is a square in the plane $(\rho, z)$ which is mapped onto the
domain describing the interior of the star. This mapping involves an unknown
function, which is itself decomposed on a basis of Chebyshev polynomials, so
that its coefficients are part of the global vector of unknowns (as the
density and gravitational field coefficients). 

In the case of binary black hole systems, Scheel~et~al.~\cite{scheel-06} have
developed horizon-tracking coordinates using results from control theory. They
define a control parameter as the relative drift of the black hole position,
and they design a feedback control system with the requirement that the
adjustment they make on the coordinates be sufficiently smooth, so that they
do not spoil the overall Einstein solver. In addition, they use a
dual-coordinate approach, so that they can construct a comoving coordinate map
which tracks both orbital and radial motion of the black holes and allows them
to successfully evolve the binary. The evolutions simulated in that article
are found to be unstable, when using a single rotating-coordinate frame. As a
remark, we note here the work by Bonazzola~et~al.~\cite{bonazzola-07}, where
another option is explored: the so-called stroboscopic technique to match
between an inner rotating domain and an outer inertial one.

\subsubsection{Spatial compactification}
\label{sss:compactification}

As stated above, the mappings can also be used to include spatial infinity
into the computational domain. Such a {\em compactification\/} technique is
not tied to spectral methods and has already been used with finite-differences
methods in numerical relativity by e.g.\ Pretorius~\cite{pretorius-05}.
However, due to the relatively lower number of degrees of freedom necessary to
describe a spatial domain within spectral methods, it is easier within this
framework to use some resources to describe spatial infinity and its
neighborhood. Many choices are possible to do so, either choosing directly a
family of well-behaved functions on an unbounded interval, for example the
Hermite functions (see e.g.\ Section~17.4 in Boyd~\cite{boyd-01}), or
making use of standard polynomial families, but with an adapted mapping. A
first example within numerical relativity was given by
Bonazzola~et~al.~\cite{bonazzola-93}, with the simple inverse mapping
in spherical coordinates
\begin{equation}
  \label{eq:map_unsurr}
  r = \frac{1}{\alpha (x-1)}, \quad x \in \interv.
\end{equation}
This inverse mapping for spherical ``shells'' has also been used by other
authors Kidder and Finn~\cite{kidder-00a},
Pfeiffer~et~al.~\cite{pfeiffer-03a, pfeiffer-03b}, or by
Ansorg~et~al.\ in cylindrical~\cite{ansorg-03} and
spheroidal~\cite{ansorg-04} coordinates. Many more elaborated
techniques are discussed in Chapter~17 of Boyd~\cite{boyd-01}, but to
our knowledge, none has been used in numerical relativity
yet. Finally, it is important to point out that, in general, the
simple compactification of spatial infinity is not well-adapted to
solving hyperbolic PDEs and the above mentioned examples were solving
only for elliptic equations (initial data, see
Section~\ref{s:station}). For instance, the simple wave
equation~(\ref{eq:def_wave}) is not invariant under the
mapping~(\ref{eq:map_unsurr}), as it has been shown e.g.\ by
Sommerfeld (see~\cite{sommerfeld-49}, Section~23.E). Intuitively, it
is easy to see that when compactifying only spatial coordinates for a
wave-like equation, the distance between two neighboring grid points
becomes larger than the wavelength, which makes the wave poorly
resolved after a finite time of propagation on the numerical grid. For
hyperbolic equations, is is therefore usually preferred to impose
physically and mathematically well-motivated boundary conditions at
finite radius (see e.g.\ Friedrich and
Nagy~\cite{friedrich-99}, Rinne~\cite{rinne-06} or Buchman and
Sarbach~\cite{buchman-07}).

\subsubsection{Patching in more than one dimension}
\label{sss:patching}

The multi-domain (or multi-patch) technique has been presented in
section~\ref{ss:multidomain_techniques} for one spatial dimension. In
Bonazzola~et~al.~\cite{bonazzola-99} or Grandcl\'ement~et~al.~\cite{grandclement-01}, the three-dimensional spatial domains consist
of spheres (or star-shaped regions) and spherical shells, across which the
solution can be matched as in one dimensional problems (only through the
radial dependence). In general, when performing a matching in two or three
spatial dimensions, the reconstruction of the global solution across all
domains might need some more care to clearly write down the matching
conditions (see e.g.~\cite{pfeiffer-03b}, where overlapping as well as
non-overlapping domains are used at the same time). For example in two
dimensions, one of the problems that might arise is the counting of matching
conditions for corners of rectangular domains, when such a corner is shared
among more than three domains. In the case of a PDE where matching conditions
must be imposed on the value of the solution, as well as on its normal
derivative (Poisson or wave equation), it is sufficient to impose continuity
of either normal derivative at the corner, the jump in the other normal
derivative being spectrally small (see Chap.~13 of Canuto~et~al.~\cite{canuto-88}).
 
\epubtkImage{}{%
\begin{figure}[ht]
  \centerline{\includegraphics[width=8cm]{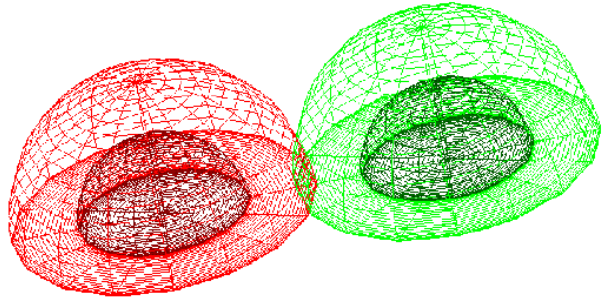}}
  \caption{Two sets of spherical domains describing a binary neutron star
    or black hole system. Each set is surrounded by a compactified domain of
    the type~(\ref{eq:map_unsurr}), which is not displayed}
  \label{figure:grid_system}
\end{figure}}

A now typical problem in numerical relativity is the study of binary
systems (see also Sections~\ref{ss:binary_stars} and
\ref{ss:dyn_binary}) for which two sets of spherical shells have been
used by Gourgoulhon~et~al.~\cite{gourgoulhon-01}, as displayed on
Figure~\ref{figure:grid_system}. Different approaches have been
proposed by Kidder~et~al.~\cite{kidder-00b}, and used by
Pfeiffer~\cite{pfeiffer-03b} and Scheel~et~al.~\cite{scheel-06} where
spherical shells and rectangular boxes are combined together to form a
grid adapted to binary black hole study. Even more sophisticated setups
to model fluid flows in complicated tubes can be found
in~\cite{lovgren-07}.

Multiple domains can thus be used to adapt the numerical grid to the
interesting part (manifold) of the coordinate space; they can be seen as a
technique close to the spectral element method~\cite{patera-84}. Moreover, it
is also a way to increase spatial resolution in some parts of the
computational domain where one expects strong gradients to occur: adding a
small domain with many degrees of freedom is the analog of fixed-mesh
refinement for finite-differences.

\subsection{Spherical coordinates and harmonics}
\label{ss:spherical_coordinates_harmonics}

Spherical coordinates (see Figure~\ref{figure:cart_spher_triad}) are
well-adapted for the study of many problems in numerical relativity. Those
include the numerical modeling of isolated astrophysical single objects, like
a neutron star or a black hole. Indeed, stars' surfaces have spherical-like
shapes and black hole horizons have this topology too, which is best described
in spherical coordinates (eventually through a mapping, see
Section~\ref{sss:mappings}). As these are isolated systems in General
Relativity, the good boundary conditions are imposed at infinity,
requiring a compactification of space, which is here achieved with the
compactification of the radial coordinate $r$ only.

\epubtkImage{}{%
\begin{figure}[ht]
  \centerline{\includegraphics[width=8cm]{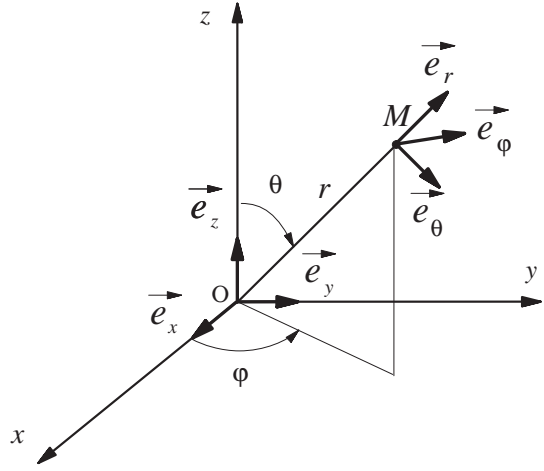}}
  \caption{Definition of spherical coordinates $(r, \th, \ph)$ of a
    point $M$ and associated triad $(\vec{e}_r, \vec{e}_\th,
    \vec{e}_\ph)$, with respect to the Cartesian ones.}
  \label{figure:cart_spher_triad}
\end{figure}} 

When the numerical grid does not extend to infinity, e.g.\ when
solving for a hyperbolic PDE, the boundary defined by $r=\textrm{const}$ is a
smooth surface, on which boundary conditions are much easier to
impose. Finally, {\em spherical harmonics\/}, which are strongly linked with
these coordinates can simplify a lot the solution of Poisson-like or wave-like
equations. On the other hand, there are some technical problems linked with
this set of coordinates, as detailed hereafter, but spectral methods can handle
them in a very efficient way.

\subsubsection{Coordinate singularities}
\label{sss:coordinate_singularities}

The transformation from spherical $(r, \th, \ph)$ to Cartesian
coordinates $(x, y, z)$ is obtained by
\begin{eqnarray}
  x & = & r \sin\th \cos\ph, \label{eq:cart_spher_x}\\
  y & = & r \sin\th \sin\ph, \label{eq:cart_spher_y}\\
  z & = & r \cos\th.\label{eq:cart_spher_z}
\end{eqnarray}
One immediately sees that the origin $r=0 \iff x=y=z=0$ is singular in spherical
coordinates because neither $\th$ nor $\ph$ can be uniquely
defined. The same happens for the $z-$axis, where $\th = 0$ or $\pi$, and $\ph$
cannot be defined. Among the consequences is the singularity of some usual
differential operators, as for instance the Laplace operator
\begin{equation}
  \label{eq:laplace_spher}
  \Delta = \frac{\partial^2}{\partial r^2} + \frac{2}{r}\frac{\partial
  }{\partial r} + \frac{1}{r^2} \left( \frac{\partial^2}{\partial \th^2} +
    \frac{1}{\tan \th} \frac{\partial }{\partial \th} + \frac{1}{\sin^2
      \th} \frac{\partial^2}{\partial \ph^2} \right).
\end{equation}
Here, the divisions by $r$ at the center, or by $\sin\th$ on the $z$-axis look
singular. On the other hand, the Laplace operator, expressed in Cartesian
coordinates is a perfectly regular one and, if it is applied to a {\em
  regular\/} function, it should give a well-defined result. So the same
should be true if one uses spherical coordinates: the
operator~(\ref{eq:laplace_spher}) applied to a regular function should yield a
regular result. This means that a regular function of spherical coordinates
must have a particular behavior at the origin and on the axis, so that the
divisions by $r$ or $\sin \th$ appearing in regular operators are always
well-defined. If one considers an analytic function of the (regular) Cartesian
coordinates $f(x, y, z)$, it can be expanded as a series of powers of $x,y$
and $z$, near the origin
\begin{equation}
  \label{eq:anal_cart}
  f(x,y,z) = \sum_{n, p, q} a_{npq} x^n y^p z^q.
\end{equation}
Replacing the coordinate definitions~(\ref{eq:cart_spher_x})-(\ref{eq:cart_spher_z}) 
into this expression gives
\begin{equation}
  \label{eq:eq:anal_spher}
  f(r, \th, \ph) = \sum_{n, p, q} a_{npq} r^{n+p+q} \cos^q\th
  \sin^{n+p}\th \cos^n\ph \sin^p\ph;
\end{equation}
and rearranging the terms in $\ph$:
\begin{equation}
  \label{eq:anal_spher2}
  f(r, \th, \ph) = \sum_{m, p, q} b_{mpq} r^{|m|+2p+q}
  \sin^{|m|+2p}\th \cos^q \th e^{im\ph}.
\end{equation}
With some transformations of trigonometric functions in $\th$, one can
express the angular part in terms of spherical harmonics $Y_\ell^m(\th,
\ph)$, see  Section~\ref{sss:spherical_harmonics}, with $\ell = |m| + 2p + q$
and obtain the two following regularity conditions, for a given couple
$(\ell, m)$:

\begin{itemize}
  \item near $\th=0$, a regular scalar field is equivalent to $f(\th)
  \sim \sin^{|m|}\th$,
  \item near $r=0$, a regular scalar field is equivalent to $f(r) \sim
  r^\ell$.
\end{itemize}

In addition, the $r$-dependence translates into a Taylor series near the
origin, with the same parity as $\ell$ . More details in the case of polar
(2D) coordinates are given in Chapter~18 of Boyd~\cite{boyd-01}. 

If we go back to the evaluation of the Laplace
operator~(\ref{eq:laplace_spher}), it is now clear that the result is always
regular, at least for $\ell\geq 2$ and $m\geq 2$. We detail the cases of
$\ell=0$ and $\ell=1$, using the fact that spherical harmonics are
eigenfunctions of the angular part of the Laplace operator (see
Equation~(\ref{eq:eigen_lapang})). For $\ell=0$ the scalar field $f$ is reduced to
a Taylor series of only even powers of $r$, therefore the first derivative
contains only odd powers and can be safely divided by $r$. Once decomposed on
spherical harmonics, the angular part of the Laplace
operator~(\ref{eq:laplace_spher}) acting on the $\ell=1$ component reads
$-2/r^2$, which is a problem only for the first term of the Taylor expansion.
On the other hand, this term cancels with the $\frac{2}{r}
\frac{\partial}{\partial r}$, providing a regular result. This is the general
behavior of many differential operators in spherical coordinates: when applied
to a regular field, the {\em full\/} operator gives a regular result, but {\em
  single terms\/} of this operator, may give singular results when computed
separately; these singularities canceling between two different terms.

As this may seem an argument against the use of spherical coordinates, let us
stress that spectral methods are very powerful in evaluating such operators,
keeping everything finite. As an example, we use Chebyshev polynomials in
$\xi$ for the expansion of the field $f(r=\alpha \xi)$, $\alpha$ being a
positive constant. From the recurrence relation on Chebyshev
polynomials~(\ref{eq:recurrence_cheb}), one has
\begin{equation}
  \label{eq:div_x_cheb}
  \forall n>0 , \quad \frac{T_{n+1}(\xi)}{\xi} = 2T_n(\xi) - \frac{T_{n-1}(\xi)}{\xi},
\end{equation}
which recursively gives the coefficients of
\begin{equation}
  \label{eq:div_x_fp}
  g(\xi) = \frac{f(\xi) - f(0)}{\xi}
\end{equation}
form those of $f(\xi)$. The computation of this {\em finite part\/} $g(\xi)$
is always a regular and linear operation on the vector of coefficients. Thus,
the singular terms of a regular operator are never computed, but the result is
the good one, as if the cancellation of such terms had occurred. Moreover,
from the parity conditions it is possible to use only even or odd Chebyshev
polynomials, which simplifies the expressions and saves computer time and
memory. Of course, relations similar to Equation~(\ref{eq:div_x_cheb}) exist
for other families of orthonormal polynomials, as well as relations to divide
by $\sin \th$ a function developed on a Fourier basis. The combination of
spectral methods and spherical coordinates is thus a powerful tool for
accurately describing regular fields and differential operators inside a
sphere~\cite{bonazzola-90}. To our knowledge, this is the first reference
showing that it is possible to solve PDEs with spectral methods inside a
sphere, including the three-dimensional coordinate singularity at the origin.

\subsubsection{Spherical harmonics}
\label{sss:spherical_harmonics}

Spherical harmonics are the pure angular functions
\begin{equation}
  \label{eq:def_ylm}
Y_\ell^m(\th, \ph)  = \sqrt{\frac{2\ell + 1}{4\pi} \frac{(\ell
    -m)!}{(\ell +m)!}} P_\ell^m\l( \cos\th \r) e^{im\ph},
\end{equation}
where $\ell \geq 0$ and $|m|\leq \ell$.
$P_\ell^m(\cos\th)$ are the associated Legendre functions defined by
\begin{equation}
  \label{eq:def_assoc_leg}
  P_\ell^m(x) = \frac{(\ell+m)!}{(\ell-m)!} \frac{1}{2^\ell
    \ell!\sqrt{(1-x^2)^m}} \frac{d^{\ell-m}}{dx^{\ell-m}} \l( 1-x^2 \r)^\ell,
\end{equation}
for $m\geq0$. The relation
\begin{equation}
  \label{eq:def_assoc_leg2}
  P_\ell^{-m}(x) = \frac{(\ell-m)!}{(\ell+m)!} P_\ell^m(x)
\end{equation}
gives the associated Legendre functions for negative $m$;
note that the normalization factors can vary in the literature. This family
of functions have two very important properties. First, they represent an
orthogonal set of regular functions defined on the sphere; thus any regular
scalar field $f(\th,\ph)$ defined on the sphere can be decomposed onto
spherical harmonics
\begin{equation}
  \label{eq:decomp_ylm}
  f(\th,\ph) = \sum_{\ell=0}^{+\infty} \sum_{m=-\ell}^{m=\ell} f_{\ell
    m} Y_\ell^m(\th, \ph).
\end{equation}
Since they are regular, they automatically take care of the coordinate
singularity on the $z$-axis.  Then, they are eigenfunctions of the angular
part of the Laplace operator (noted here $\Delta_{\th\ph}$):
\begin{equation}
  \label{eq:eigen_lapang}
  \forall (\ell,m) \quad \Delta_{\th\ph} Y_\ell^m(\th, \ph):=
  \frac{\partial^2 
    Y_\ell^m}{\partial \th^2} + \frac{1}{\tan \th}\frac{\partial
    Y_\ell^m}{\partial \th} + \frac{1}{\sin^2\th}\frac{\partial^2
    Y_\ell^m}{\partial \ph^2} = -\ell(\ell+1) Y_\ell^m(\th, \ph),
\end{equation}
the associated eigenvalues being $-\ell(\ell+1)$.

The first property makes the description of scalar fields on spheres very
easy: spherical harmonics are used as decomposition basis within spectral
methods, for instance in geophysics or meteorology, and by some groups in
numerical relativity~\cite{bartnik-00, grandclement-01, tichy-06}. However,
they could be more broadly used in numerical relativity, for example for
Cauchy-characteristic evolution or matching~\cite{winicour-05, babiuc-05},
where a single coordinate chart on the sphere might help in matching
quantities. They can also help to describe star-like surfaces being defined by
$r=h(\th, \ph)$, as event or apparent horizons~\cite{nakamura-84, baumgarte-96,
  alcubierre-00}. The search for apparent horizons is also made easier: since
the function $h$ verifies a two-dimensional Poisson-like equation, the linear
part can be solved directly, just by dividing by $-\ell(\ell+1)$ in the
coefficient space.

The second property makes the Poisson equation
\begin{equation}
  \label{eq:poisson}
  \Delta \phi(r,\th,\ph) = \sigma(r,\th,\ph)
\end{equation}
very easy to solve (see  Section~\ref{ss:simple_example}). If the
source $\sigma$ and the unknown $\phi$ are decomposed onto spherical
harmonics, the equation transforms into a set of {\em ordinary\/}
differential equations for the coefficients (see
also~\cite{grandclement-01}):
\begin{equation}
  \label{eq:poisson_lm}
  \forall (\ell, m) \quad \frac{d^2 \phi_{\ell m}}{d r^2} +
  \frac{2}{r}\frac{d \phi_{\ell m}}{d r} - \frac{\ell(\ell+1)\phi_{\ell
      m}}{r^2} = \sigma_{\ell m}.
\end{equation}
Then, any ODE solver can be used for the radial coordinate: spectral methods
of course (see  Section~\ref{ss:sm_for_ODEs}), but other ones have been used too
(see {\it e,g,\/} Bartnik~et~al.~\cite{bartnik-97, bartnik-00}). The
same technique can be used to advance in time the wave equation with an
implicit scheme and Chebyshev-tau method for the radial
coordinate~\cite{bonazzola-90, novak-04}.

The use of spherical harmonics decomposition can be regarded as a basic
spectral method, as the Fourier decomposition. There are therefore publicly
available ``spherical harmonics transforms'' which consist of a Fourier
transform in the $\ph$-direction and a successive Fourier and Legendre
transform in the $\th$-direction. A rather efficient one is the
SpharmonicsKit/S2Kit~\cite{spharmonicskit-04}, but writing one's own functions
is also possible~\cite{lorene}.

\subsubsection{Tensor components}
\label{sss:tensor_components}

All the discussion in
Sections~\ref{sss:coordinate_singularities}--\ref{sss:spherical_harmonics} has
been restricted to scalar fields. For vector, or more generally tensor fields
in three spatial dimensions, a vector basis (triad) must be specified to
express the components. At this point, it is very important to stress out that
the choice of the basis is independent of the choice of coordinates.
Therefore, the most straightforward and simple choice, even if one is using
spherical coordinates, is the Cartesian triad $\l( {\bf e}_x
=\frac{\partial}{\partial x}, {\bf e}_y = \frac{\partial}{\partial y}, {\bf
  e}_z = \frac{\partial}{\partial z} \r)$. With this basis, from a numerical
point of view, all tensor components can be regarded as scalars and therefore,
a regular tensor can be defined as a tensor field whose components with
respect to this Cartesian frame are expandable in powers of $x,y$ and $z$ (as
in Bardeen and Piran~\cite{bardeen-84}). Manipulations and solutions of PDEs for
such tensor fields in spherical coordinates are generalization of the
techniques for scalar fields. In particular, when using the multi-domain
approach with domains having different shapes and coordinates, it is much
easier to match Cartesian components of tensor fields. Examples of use of
Cartesian components of tensor fields in numerical relativity include the
vector Poisson equation~\cite{grandclement-01} or, more generally, the
solution of elliptic systems arising in numerical
relativity~\cite{pfeiffer-03a}. In the case of the evolution of the
unconstrained Einstein system, the use of Cartesian tensor components is the
general option, as it is done by the Caltech/Cornell group~\cite{kidder-01,
  scheel-06}.

The use of an {\em orthonormal spherical basis\/} $\l( {\bf e}_r =
\frac{\partial}{\partial r}, {\bf e}_\th =
\frac{1}{r}\frac{\partial}{\partial \th}, {\bf e}_\ph =
\frac{1}{r\sin\th} \frac{\partial }{\partial \ph} \r)$
(see. Figure~\ref{figure:cart_spher_triad}) requires some
more care, as it is outlined hereafter. The interested reader can also find
some details in the works by Bonazzola~et~al.~\cite{bonazzola-90,
  bonazzola-04}. Nevertheless, there are systems in General Relativity where
spherical components of tensors can be useful: 

\begin{itemize}
\item When doing excision for the simulation of black holes, the boundary
  conditions on the excised sphere for elliptic equations (initial data) may be
  better formulated in terms of spherical components for the shift or the
  3-metric~\cite{cook-02, gourgoulhon-06, jaramillo-07}. In
  particular, the component which is normal to the excised surface is easily identified
  with the radial component.
\item Still in 3+1 approach, the extraction of gravitational radiation in the wave
  zone is made easier if the perturbation to the metric is expressed in
  spherical components, because the transverse part is then straightforward to
  obtain~\cite{thorne-80}. 
\end{itemize}

Problems arise because of the singular nature of the basis itself, in addition
to the spherical coordinate singularities. The consequences are first that
each component is a multi-valued function at the origin $r=0$ or on the
$z$-axis, and then that components of a given tensor are not independent one from
another, meaning that one cannot in general specify each component
independently or set it to zero, keeping the tensor field regular. As an
example, we consider the gradient $V^i = \nabla^i \phi$ of the scalar field
$\phi = x$, where $x$ is the usual first Cartesian coordinate field. This
gradient expressed in Cartesian components is a regular vector field
$V^x = 1, \quad V^y = 0, \quad V^z = 0$.
The spherical components of $\bf{V}$ read
\begin{eqnarray}
  V^r &=& \sin\th \cos\ph, \nonumber\\
  V^\th &=& \cos \th \cos \ph, \nonumber \\
  V^\ph &=& -\sin\ph, \label{eq:grad_spher}
\end{eqnarray}
which are all three multi-defined at the origin, and the last two on the
$z$-axis. In addition, if $V^\th$ is set to zero, one sees that the
resulting vector field is no longer regular: for example the square of its
norm is multi-defined, which is not the good property for a scalar field. As
for the singularities of spherical coordinates, these difficulties can be
properly handled with spectral methods, provided that the decomposition basis
are carefully chosen. 

The other drawback of spherical components is that usual partial differential
operators mix the components. This is due to the non-vanishing connection coefficients
associated with the spherical flat metric~\cite{bonazzola-04}. For example,
the vector Laplace operator ($\nabla_j \nabla^j V^i$) reads
\begin{eqnarray}
    && \frac{\partial^2 V^r}{\partial r^2} + \frac{2}{r}\frac{\partial
      V^r}{\partial r}
    +\frac{1}{r^2} \left( \Delta_{\th\ph} V^r - 2 V^r 
    - 2 \frac{\partial V^\th}{\partial \th} - 2\frac{V^\th}{\tan\th} -
    \frac{2}{\sin\th} \frac{\partial V^\ph}{\partial \ph} \right)  
                    \label{eq:poisson_vr} \\
  && \frac{\partial^2 V^\th}{\partial r^2} + \frac{2}{r} \frac{\partial
    V^\th}{\partial r}
  +\frac{1}{ r^2} \left( \Delta_{\th\ph} V^\th + 2 \frac{\partial
      V^r}{\partial \th}
    - \frac{V^\th}{\sin^2\th} 
    - 2\frac{\cos\th}{\sin^2\th}\frac{\partial V^\ph}{\partial \ph} \right)
     \label{eq:poisson_vt} \\
  && \frac{\partial^2 V^\ph}{\partial r^2} + \frac{2}{r} \frac{\partial
    V^\ph}{\partial r}
  + \frac{1}{r^2} \left( \Delta_{\th\ph} V^\ph 
  + \frac{2}{\sin\th} \frac{\partial V^r}{\partial \ph}
   +2\frac{\cos\th}{\sin^2\th} \frac{\partial V^\th}{\partial \ph} 
   - \frac{V^\ph}{\sin^2\th} \right) ,
   \label{eq:poisson_vp}
\end{eqnarray}
with $\Delta_{\th\ph}$ defined in Equation~(\ref{eq:eigen_lapang}). In particular,
the $r$-component~(\ref{eq:poisson_vr}) of the operator involves the other two
components. This can make the resolution of a vector Poisson equation, which
naturally arises in the initial data problem~\cite{cook-00} of numerical
relativity, technically more complicated and the technique using scalar
spherical harmonics (Section~\ref{sss:spherical_harmonics}) is no longer valid.
One possibility can be to use vector, and more generally
tensor~\cite{mathews-62, zerilli-70, thorne-80, brizuela-06} spherical
harmonics as decomposition basis. Another technique might be to build from the
spherical components regular scalar fields, which can have a similar physical
relevance to the problem. In the vector case, one can think of the following
expressions
\begin{equation}
  \label{eq:def_khi_div_mu}
  \Theta = \nabla_i V^i, \quad \chi = r_iV^i, \quad \mu = r^i \epsilon_{ijk}
  \nabla^j V^k, 
\end{equation}
where ${\bf r} = r {\bf e}_r$ denotes the position vector and
$\epsilon_{ijk}$ the third-rank fully antisymmetric tensor. These scalars are
the divergence, $r$-component and curl of the vector field. The reader can
verify that a Poisson equation for $V^i$ transforms into three equations for
these scalars, expandable onto scalar spherical harmonics. The reason why
these fields may be more interesting than Cartesian components is that they
can have more physical or geometrical meaning.

\subsection{Going further}
\label{ss:going_further}

The development of spectral methods linked with the problems arising in the
field of numerical relativity has always been active and still is now. Among
the various directions of research one can foresee, quite interesting ones
might be the beginning of higher-dimensional studies and the development of
better-adapted mappings and domains, within the spirit of going from pure
spectral methods to spectral elements~\cite{patera-84, benbelgacem-99}.

\subsubsection{More than three spatial dimensions}
\label{sss:manyD}

There have been some interest for the numerical study of black holes in higher
dimensions: as well with compactified extra-dimensions~\cite{sorkin-04}, as in
brane world models~\cite{shiromizu-00, kudoh-05}; recently, some simulations
of the head-on collision of two black holes have already been
undertaken~\cite{yoshino-06}. With the relatively low number of degrees of
freedom per dimension needed, spectral methods should be very efficient in
simulations involving four spatial dimensions, or more. We give here starting
points to implement 4-dimensional (as needed by e.g.\ brane world
models) spatial representation with spectral methods. The simplest approach
is to take Cartesian coordinates $(x, y, z, w)$, but a generalization of
spherical coordinates $(r, \th, \ph, \xi)$ is also possible and necessitates
less computational resources. The additional angle $\xi$ is defined in $[0,
\pi]$, with the following relations with Cartesian coordinates 
\begin{eqnarray}
  x&=& r\sin\th\cos\ph\sin\xi,\nonumber\\
  y&=& r\sin\th\sin\ph\sin\xi,\nonumber\\
  z&=& r\cos\th\sin\xi,\nonumber\\
  w&=& r\cos\xi.\nonumber\label{eq:spher_4D}
\end{eqnarray}
The four-dimensional flat Laplace operator appearing in constraint
equations~\cite{shiromizu-00} reads
\begin{equation}
  \label{eq:lap_4D}
  \Delta_4 \phi = \frac{\partial^2 \phi}{\partial r^2} +
  \frac{3}{r}\frac{\partial \phi}{\partial r} + \frac{1}{r^2} \l(
  \frac{\partial^2 \phi}{\partial \xi^2} + \frac{2}{\tan \xi} \frac{\partial
    \phi}{\partial \xi} + \frac{1}{\sin^2 \xi} \Delta_{\th\ph} \phi \r),
\end{equation}
where $\Delta_{\th\ph}$ is the two-dimensional angular Laplace
operator~(\ref{eq:eigen_lapang}). As in the three-dimensional case, it is
convenient to use the eigenfunctions of the angular part, which are here
\begin{equation}
  \label{eq:eigen_lap3D}
  G_k^\ell(\cos\xi)P_\ell^m(\cos\th)e^{im\ph},
\end{equation}
with $k, \ell, m$ integers such that $|m|\leq \ell \leq k$. $P_\ell^m(x)$ are
the associated Legendre functions defined by Equation~(\ref{eq:def_assoc_leg}).
$G_k^\ell(x)$ are the associated Gegenbauer functions
\begin{equation}
  \label{eq:def_assoc_geg}
  G_k^\ell(\cos\xi) = (\sin^\ell \xi) G_k^{(\ell)}(\cos \xi) \text{ with }
  G_k^{(\ell)}(x) = \frac{d^\ell G_k(x)}{dx^\ell},
\end{equation}
and $G_k(x)$ being the $k$-th Gegenbauer polynomial $C_k^{(\lambda)}$ with
$\lambda = 1$. Since the $G_k$ are also particular case of Jacobi polynomials
with $\alpha = \beta = 1/2$ (see, for example~\cite{korn-61}). Jacobi
polynomials are also solutions of a singular Sturm-Liouville problem, which
ensures fast convergence properties (see  Section~\ref{sss:strum}). The $G_k(x)$
fulfill recurrence relations that make them easy to implement as spectral
decomposition basis, like the Legendre polynomials. These eigenfunctions are
associated with the eigenvalues $-k(k+2)$:
\begin{equation}
\Delta_4 \left( G_k^\ell(\cos\xi)P_\ell^m(\cos\th)e^{im\ph} \right) = -k(k+2)
G_k^\ell(\cos\xi)P_\ell^m(\cos\th)e^{im\ph}.\label{eigen_4D}
\end{equation}
So as in 3+1 dimensions, after decomposing on such a basis, the Poisson
equation turns into a collection of ODEs in the coordinate $r$. This type of
construction might be generalized to even higher dimensions, with the choice
of appropriate type of Jacobi polynomials for every new introduced angular
coordinate.

\newpage

\section{Time-Dependent Problems}
\label{s:time}

From a relativistic point of view, the time coordinate could be treated in the
same way as spatial coordinates and one should be able to achieve spectral
accuracy for the time representation of a space-time function $f(t, x, y, z)$
and its derivatives. Unfortunately, this does not seem to be the case and, we
are neither aware of any efficient algorithm dealing with the time coordinate,
nor of any published successful code solving any of the PDE coming from the
Einstein equations, with the recent exception of the 1+1 dimensional study by
Hennig and Ansorg~\cite{ansorg-06}. Why is time playing such a special role?
It is not obvious to find in the literature on spectral methods a complete and
comprehensive study. A first standard explanation is the difficulty, in
general, to predict the exact time interval on which one wants to study the
time evolution. Then, time discretization errors in both finite-differences
and spectral methods are typically much smaller than spatial ones.
Finally, one must keep in mind that, contrary to finite-differences, spectral
methods are storing all global information about a function on the whole time
interval. Therefore, one reason may be that there are strong memory and CPU
limitations to fully three-dimensional simulations, it is already very CPU and
memory consuming to describe a complete field depending on 3+1 coordinates,
even with fewer degrees of freedom, as it is the case for spectral methods.
But the strongest limitation is the fact that, in the full 3+1 dimensional
case, the matrix representing a differential operator would be of very big
size; it would therefore be very time-consuming to invert it in a general
case, even with iterative methods.

More details on the standard, finite-differences techniques for time
discretization are first given in  Section~\ref{ss:time_discretization}. Due to the
technical complexity of a general stability analysis, we first restrict the
discussion of this section to the eigenvalue
stability~(Section~\ref{ss:time_discretization}), with the following approach:
the eigenvalues of spatial operator matrices must fall within the stability
region of the time-marching scheme. Although this condition is only a
necessary one and, in general, is not sufficient, it provides very useful
guidelines for selecting time-integration schemes. A discussion on the
imposition of boundary conditions in time-dependent problems is given in
 Section~\ref{ss:time_bc}. Section~\ref{ss:stab_conv} then details stability
analysis for spatial discretization schemes, with the examples of heat
and advection equations, before details of a fully-discrete analysis are given
for a simple case (Section~\ref{ss:fully_discrete}).

\subsection{Time discretization}
\label{ss:time_discretization}

There have been very few theoretical developments on spectral time
discretization, with the exception of Ierley~et~al.~\cite{ierley-92},
were the authors have applied spectral methods in time for the study of the
Korteweg de Vries and Burger equations, using Fourier series in space and
Chebyshev polynomials for the time coordinate. They observe a time-stepping
restriction: they have to employ multi-domain and patching techniques (see
 Section~\ref{ss:multidomain_techniques}) for the time interval, with the size of
each sub-domain being roughly given by the Courant--Friedrichs--Lewy (CFL)
condition. Therefore, the most common approach for time representation are
finite-differences techniques, which allow for the use of many
well-established time-marching schemes, and the method of lines (for other
methods, including fractional stepping, see Fornberg~\cite{fornberg-95}).  Let
us write the general form of a first-order in time linear PDE:
\begin{equation}
  \label{eq:time_PDE}
  \forall t\geq 0,\quad \forall x\in[-1,1],\quad \frac{\partial
    u(x,t)}{\partial t} = Lu(x,t),
\end{equation}
where $L$ is a linear operator containing only derivatives with respect to
the spatial coordinate $x$. For every value of the time $t$, the spectral
approximation $u_N(x, t)$ is a function of only one spatial dimension
belonging to some finite-dimensional subspace of the suitable Hilbert space
$\mathcal H$, with the given $L^2_w$ spatial norm, associated for example to the scalar
product and the weight $w$ introduced in  Section~\ref{sss:ortho}. Formally, the
solution of Equation~(\ref{eq:time_PDE}) can be written as:
\begin{equation}
  \label{eq:sol_time_PDE}
  \forall x\in[-1,1],\quad u(x,t) = e^{Lt} u(x,0).
\end{equation}
In practice, to integrate time-dependent problems one can use spectral methods
to calculate spatial derivatives and standard finite-differences schemes to
advance in time.

\subsubsection{Method of lines}
\label{sss:method_of_lines}

 At every instant $t$, one can represent the function $u_N(x,t)$ through a finite set
$U_N(t)$, composed of its time-dependent spectral coefficients, or values at
the collocation points. We note $L_N$ the spectral approximation to the
operator $L$, together with the boundary conditions, if a tau or collocation
method is used. $L_N$ is therefore represented as an $N \times N$ matrix. This
is the so-called method of lines, which allows one to reduce a PDE to some
ODE, after discretization in all dimensions but one. The advantage is that
many ODE integration schemes are known (Runge-Kutta, symplectic integrators,
...) and can be used here. We shall suppose an equally-spaced grid in time,
with the time-step noted $\Delta t$ and $U^J_N = U_N(J \times \Delta t)$. 

In order to step from $U^J_N$ to $U^{J+1}_N$, one has essentially two
possibilities: explicit and implicit schemes. In an {\em explicit scheme\/},
the action of the spatial operator $L_N$ on $\left. U_N^K \right|_{K\leq J}$,
must be computed to explicitly get the new values of the field (either
spatial spectral coefficients or values at collocation points). A simple example is
the {\em forward Euler method\/}:
\begin{equation}
  \label{eq:forward_Euler}
  U_N^{J+1} = U_N^J + \Delta t L_N U_N^J,
\end{equation}
which is first-order and for which, as for any explicit schemes, the time-step
must is limited by the CFL condition. The imposition of boundary conditions is
discussed in  Section~\ref{ss:time_bc}. With an {\em implicit scheme\/} one must
solve for a boundary value problem in term of $U^{J+1}_N$ at each time-step:
it can be performed in the same way as for the solution of the elliptic
equation (\ref{e:test_pb}) presented in  Section~\ref{s:tau}. The simplest example
is the {\em backward Euler method\/}:
\begin{equation}
  \label{eq:backward_Euler}
  U_N^{J+1} = U_N^J + \Delta t L_N U_N^{J+1},
\end{equation}
which can be re-written as an equation for the unknown $U_N^{J+1}$:
$$
\left( I + \Delta t L_N \right) U_N^{J+1} = U_N^J;
$$
with $I$ the identity operator. Both types of schemes have different stability
properties, which can be analyzed as follows.  Assuming that $L_N$ can be
diagonalized in the sense of the definition given in (\ref{sss:spec_space}),
the stability study can be reduced to the study of the collection of scalar
ODE problems
\begin{equation}
  \label{eq:time_scal}
  \frac{\partial U_N}{\partial t} = \lambda_i U_N,
\end{equation}
where $\lambda_i$ is any of the eigenvalues of $L_N$ in the sense of
Equation~(\ref{eq:eigen_tau}).

\subsubsection{Stability}
\label{sss:time_stab}

The basic definition of {\em stability\/} for an ODE integration scheme is that,
if the time-step is lower than some threshold, then $\|U_N^J\| \leq
Ae^{KJ\Delta t}$, with the constants $A$ and $K$ independent of the time-step.
This is perhaps not the most appropriate definition, since in practice one
often deals with bounded functions and an exponential growth in time would not
be acceptable. Therefore, an integration scheme is said to be {\em absolutely
  stable\/} (or asymptotically stable), if $\|U_N^J\|$ remains bounded,
$\forall J\geq 0$. This property depends on a particular value of the product
$\lambda_i \times \Delta t$. For each time integration scheme, the {\em region
  of absolute stability\/} is the set of the complex plane containing all the
$\lambda_i \Delta t$ for which the scheme is absolutely stable.

\epubtkImage{}{%
\begin{figure}[ht]
  \centerline{\includegraphics[width=10cm]{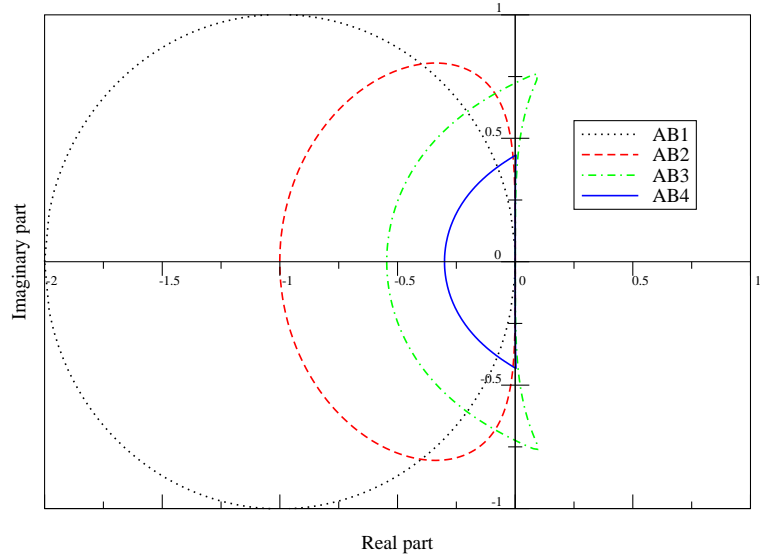}}
  \caption{Regions of absolute stability for the Adams-Bashforth integration
    schemes of order 1 to 4.}
  \label{figure:adams_bashford}
\end{figure}}

Finally, a scheme is said to be $A$-stable if its region of absolute stability
contains the half complex plane of numbers with negative real part. It is
clear that no explicit scheme can be $A$-stable due to the CFL condition. It
has been shown by Dahlquist~\cite{dahlquist-63} that there is no linear
multi-step method of order higher than 2 which is $A$-stable. Thus implicit
methods are also limited in time-step size if more than second-order accurate.
In addition, Dahlquist~\cite{dahlquist-63} shows that the most accurate
second-order $A$-stable scheme is the trapezoidal one (also called
Crank-Nicolson, or second-order Adams-Moulton scheme)
\begin{equation}
  \label{eq:crank_nicholson}
  U_N^{J+1} = U_N^J + \frac{\Delta t}{2} \left( L_N U_N^{J+1} + L_N U_N^{J}
  \right). 
\end{equation}

\epubtkImage{}{%
\begin{figure}[ht]
  \centerline{\includegraphics[width=10cm]{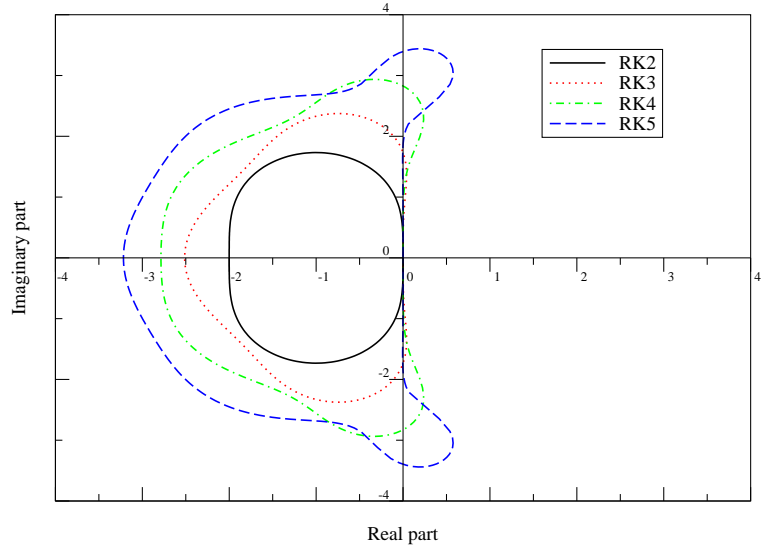}}
  \caption{Regions of absolute stability for the Runge--Kutta integration
    schemes of order 2 to 5. Note that the size of the region is increasing
    with the order.}
  \label{figure:runge_kutta}
\end{figure}}

On Figures~\ref{figure:adams_bashford} and~\ref{figure:runge_kutta}
are displayed the absolute stability regions for the Adams--Bashforth
and Runge--Kutta families of explicit schemes (see for
instance~\cite{canuto-88}). For a given type of spatial linear
operator, the requirement on the time-step usually comes from the
largest (in modulus) eigenvalue of the operator. For example, in the
case of the advection equation on $\interv$, with a Dirichlet boundary
condition
\begin{eqnarray}
  Lu = \frac{\partial u}{\partial x}, \nonumber \\
  \forall t, \quad u(1,t) = 0,
  \label{eq:advec_diri}
\end{eqnarray}
and using a Chebyshev-tau method, one has that the largest eigenvalue of $L_N$
grows in modulus as $N^2$. Therefore, for any of the schemes considered on
Figures~\ref{figure:adams_bashford} and~\ref{figure:runge_kutta}, the
time-step has a restriction of the type
\begin{equation}
  \label{eq:CFL}
  \Delta t \lesssim O(N^{-2}),
\end{equation}
which can be related to the usual CFL condition by the fact that the minimal
distance between two points of a ($N$-point) Chebyshev grid decreases like
$O(N^{-2})$. Due to the above cited {\em Second Dahlquist barrier\/}
\cite{dahlquist-63}, implicit time marching schemes of order higher than two
also have such kind of limitation.

\subsubsection{Spectrum of simple spatial operators}
\label{sss:spec_space}

An important issue in determining the absolute stability of a time-marching
scheme for the solution of a given PDE is the computation of the spectrum
$\left( \lambda_i \right)$ of the discretized spatial operator
$L_N$~(\ref{eq:time_scal}). As a matter of fact, these eigenvalues are those
of the matrix representation of $L_N$, together with the necessary boundary
conditions for the problem to be well-posed (e.g.\ ${\mathcal B}_N u =
0$).  If one notes $b$ the number of such boundary conditions, each eigenvalue
$\lambda_i$ (here, in the case of the tau method) is defined by the existence
of a non-null set of coefficients $\left\{ c_j \right\}_{1 \leq j \leq N}$
such that
\begin{eqnarray}
  (\forall j) \, 1\leq j \leq N-b, \quad (L_N u)_j &=& \lambda_i c_j, \nonumber\\
  {\mathcal B}_N u &=& 0. \label{eq:eigen_tau}
\end{eqnarray}

\epubtkImage{}{%
\begin{figure}[ht]
  \centerline{\includegraphics[width=10cm]{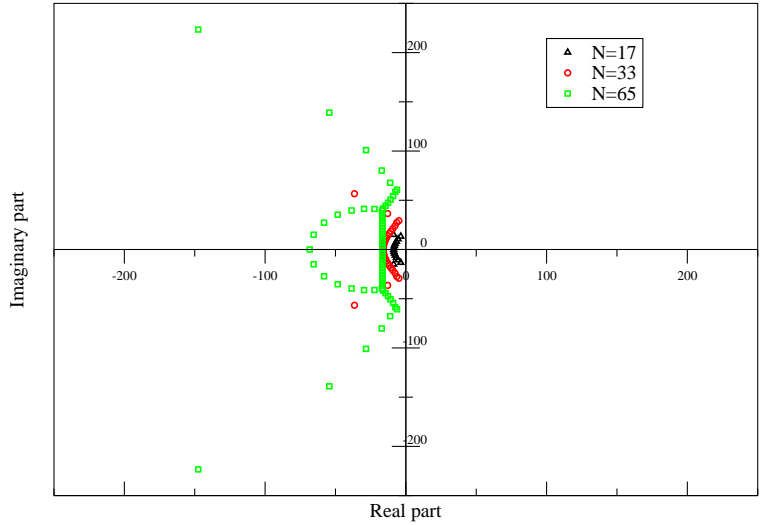}}
  \caption{Eigenvalues of the first derivative-tau
  operator~(\ref{eq:eigen_tau}) for Chebyshev polynomials. The largest
  (in modulus) eigenvalue is not displayed; this one is real, negative
  and goes as $O(N^2)$.}
  \label{figure:eigen_tau}
\end{figure}}

As an example, let us consider the case of the advection equation (first-order
spatial derivative) with a Dirichlet boundary condition, solved with the
Chebyshev-tau method (\ref{eq:advec_diri}). Because of the definition of the
problem~(\ref{eq:eigen_tau}), there are $N-1$ ``eigenvalues'', which can be
computed, after a small transformation, using any standard linear algebra
package. For instance, it is possible, making use of the boundary condition,
to express the last coefficient as a combination of the other ones
\begin{equation}
  \label{eq:bound_eigen}
  c_N = - \sum_{j=1}^{N-1} c_j
\end{equation}
One is thus left with a usual eigenvalue problem for a $(N-1) \times (N-1)$
matrix. Results are displayed on Figure~\ref{figure:eigen_tau} for three
various values of $N$. Real parts are all negative: the eigenvalue which is
not displayed lies on the negative part of the real axis and is much larger in
modulus (it is growing like $O(N^2)$) than the $N-1$ others.

This way of determining the spectrum can be, of course, generalized to any
linear spatial operator, for any spectral basis, as well as to the collocation
and Galerkin methods. Intuitively from CFL-type limitations, one can see that
in the case of the heat equation ($Lu = \partial^2 u / \partial x^2$),
explicit time-integration schemes (or any scheme which is not $A$-stable)
shall have a severe time-step limitation of the type
\begin{equation}
  \label{eq:CFL_sec_order}
  \Delta t \lesssim O(N^{-4}),
\end{equation}
for both Chebyshev or Legendre decomposition basis. Finally, one can decompose
a higher-order in time PDE into a first-order system and then use one of the
above proposed schemes. In the particular case of the wave equation
\begin{equation}
  \label{eq:def_wave}
  \frac{\partial^2 u}{\partial t^2} = \frac{\partial^2 u}{\partial x^2},
\end{equation}
it is possible to write a second-order Crank-Nicolson scheme directly~\cite{novak-04}
\begin{equation}
  \label{eq:wave_CN}
  U^{J+1}_N = 2U^J_N - U^{J-1}_N + \frac{\Delta t^2}{2} \left( \frac{\partial^2
      U^{J+1}_N}{\partial x^2} + \frac{\partial^2 U^{J-1}_N}{\partial x^2} \right).
\end{equation}
Since this scheme is $A$-stable, there is no limitation on the time-step
$\Delta t$, but for explicit or higher-order schemes this limitation would be 
$\Delta t \lesssim O(N^{-2})$, as for advection equation. The solution of such
an implicit scheme is obtained as that of a boundary value problem at each
time-step.

\subsubsection{Semi-implicit schemes}
\label{sss:semi_implicit}

It is sometimes possible to use a combination of implicit and explicit schemes
to loosen a time-step restriction of the type~(\ref{eq:CFL}). Let us consider
as an example the advection equation with non-constant velocity on
$\interv$ 
\begin{equation}
  \label{eq:adv_var}
  \frac{\partial u}{\partial t} = v(x) \frac{\partial u}{\partial x},
\end{equation}
with the relevant boundary conditions, which shall in general depend on the
sign of $v(x)$. If on the one hand the stability condition for explicit time
schemes~(\ref{eq:CFL}) is too strong, and on the other hand an implicit scheme
is too lengthy to implement or to use (because of the non-constant
coefficient $v(x)$), then it is interesting to consider the semi-implicit
two-step method (see also~\cite{gottlieb-77})
\begin{eqnarray}
  \label{eq:semi_impl}
  U_N^{J+1/2} - \frac{\Delta t}{2} L_N^{-}U_N^{J+1/2} &=& U_N^J + \frac{\Delta
    t}{2} \left( L_N - L_N^{-} \right) U_N^J,\nonumber\\
  U_N^{J+1} - \frac{\Delta t}{2} L_N^{+}U_N^{J+1} &=& U_N^{J+1/2} + \frac{\Delta
    t}{2} \left( L_N - L_N^{+} \right) U_N^{J+1/2},
\end{eqnarray}
where $L_N^{+}$ and $L_N^{-}$ are respectively the spectral approximations to
the constant operators $-v(1)\partial / \partial x$ and $-v(-1)\partial /
\partial x$, together with the relevant boundary conditions (if any). This
scheme is absolutely stable if
\begin{equation}
  \label{eq:dt_semi}
  \Delta t \lesssim \frac{1}{N \max |v(x)|}.
\end{equation}
With this type of scheme, the propagation of the wave at the boundary of the
interval is treated implicitly, whereas the scheme is still explicit in the
interior. The implementation of the implicit part, for which one needs to
solve a boundary-value problem, is much easier than for the initial
operator~(\ref{eq:adv_var}) because of the presence of only
constant-coefficient operators. This technique is quite helpful in the case of
more severe time-step restrictions~(\ref{eq:CFL_sec_order}), for example for a
variable coefficient heat equation.

\subsection{Imposition of boundary conditions}
\label{ss:time_bc}

The time-dependent PDE~(\ref{eq:time_PDE}) can be written as a system of ODEs
in time either for the time-dependent spectral coefficients $\left\{ c_i(t)
\right\}_{i=0\dots N}$ of the unknown function $u(x,t)$ (Galerkin or tau
methods), or for the time-dependent values at collocation points $\left\{
  u(x_i, t) \right\}_{i=0 \dots N}$ (collocation method). Implicit time-marching
schemes (like the backward Euler scheme~(\ref{eq:backward_Euler})) are
technically very similar to a succession of boundary-value problems, as for
elliptic equations or Equation~(\ref{e:test_pb}) described in
 Section~\ref{ss:sm_for_ODEs}. The coefficients (or the values at collocation
points) are determined at each new time-step by the inversion of the matrix of
the type $I + \Delta t L$ or its higher-order generalization. To represent a
well-posed problem, this matrix needs in general the incorporation of boundary
conditions, for tau and collocation methods. Galerkin methods are not so
useful if the boundary conditions are time-dependent: this would require the
construction of a new Galerkin basis at each new time-step, which is too
complicated and/or time-consuming. We shall therefore discuss in the following
sections the imposition of boundary conditions for explicit time schemes, with
the tau or collocation methods.

\subsubsection{Strong enforcement}
\label{sss:strong_bc}

The standard technique is to enforce the boundary conditions exactly,
i.e.\ up to machine precision. Let us suppose here that the
time-dependent PDE~(\ref{eq:time_PDE}), which we want to solve, is
well-posed with the boundary condition
\begin{equation}
  \label{eq:time_PDE_BC}
  \forall t\geq 0,\quad u(x=1, t) = b(t),
\end{equation}
where $b(t)$ is a given function. We give here some examples, with the forward
Euler scheme~(\ref{eq:forward_Euler}) for time discretization.

In the {\bf collocation method}, the values of the approximate solution at
(Gauss--Lobatto type) collocation points $\left\{ x_i \right\}_{i=0 \dots N}$
are determined by a system of equations:
\begin{eqnarray}
  \label{eq:BC_colloc}
  \forall i = 0 \dots N-1, \quad U_N^{J+1}(x_i) &=& U_N^J(x_i) + \Delta t \left(
    L_N U_N^J \right)(x=x_i),\\
  U_N^{J+1}(x=x_N=1) &=& b\left( (J+1)\Delta t\right), \nonumber
\end{eqnarray}
the value at the boundary $(x=1)$ is directly set to be the boundary
condition.

In the {\bf tau method}, the vector $U_N^J$ is composed of the $N+1$
coefficients $\left\{ c_i(J\times \Delta t) \right\}_{i=0\dots N}$ at the
$J$-th time-step. If we denote by $\left(L_N U_N^J \right)_i$ the $i$-th
coefficient of $L_N$ applied to $U_N^J$, then the vector of coefficients
$\left\{ c_i \right\}_{i=0\dots N}$ is advanced in time through the
system:
\begin{eqnarray}
  \label{eq:BC_tau}
  \forall i=0\dots N-1,\quad  c_i\left((J+1)\times \Delta t\right) &=& 
    c_i(J\times \Delta t) + \Delta t \left(L_N U_N^J \right)_i\\
    c_N\left((J+1)\times \Delta t\right) &=& b\left((J+1)\Delta t\right) -
    \sum_{k=0}^{N-1} c_k \nonumber, 
\end{eqnarray}
the last equality ensures the boundary condition in the coefficient space.

\subsubsection{Penalty approach}
\label{sss:penalty}

As shown in the previous examples, the standard technique consists in
neglecting the solution to the PDE for one degree of freedom, in
configuration or coefficient space, and using this degree of freedom
in order to impose the boundary condition. However, it is interesting
to try and impose a linear combination of both the PDE and the
boundary condition on this last degree of freedom, as it is shown by
the next simple example. We consider the simple (time-independent)
integration over the interval $x\in [-1,1]$:
\begin{equation}
  \label{eq:penalty_integ}
  \frac{{\rm d}u}{{\rm d}x} = \sin(x-1), \textrm{ and } u(1) = 0,
\end{equation}
where $u(x)$ is the unknown function. Using a standard Chebyshev collocation
method (see Section~\ref{s:colloc}), we look for an approximate
solution $u_N$ as a polynomial of degree $N$ verifying
\begin{eqnarray*}
  \forall i=0\dots N-1,\quad \frac{{\rm d} u_N}{{\rm d}x} (x_i) &=& \sin(x_i-1),\\
  \frac{{\rm d} u_N}{{\rm d}x} (x_N=1) &=& 0, 
\end{eqnarray*}
where the $\left\{ x_i \right\}_{i=0\dots N}$ are the
Chebyshev--Gauss--Lobatto collocation points.

\epubtkImage{}{%
\begin{figure}[ht]
  \centerline{\includegraphics[width=10cm]{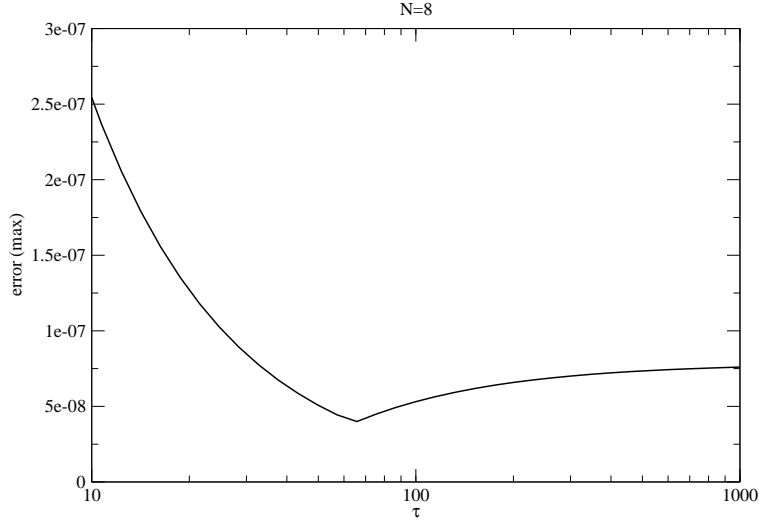}}
  \caption{Behavior of the error in the solution of the differential
    equation~(\ref{eq:penalty_integ}), as a function of the parameter $\tau$
    entering the numerical scheme~(\ref{eq:penalty_system_integ}).}
  \label{figure:penalty}
\end{figure}}

We now adopt another procedure that takes into account the differential
equation at the boundary as well as the boundary condition, with $u_N$
verifying (remember that $x_N=1$):
\begin{eqnarray}
  \label{eq:penalty_system_integ}
  \forall i=0\dots N-1,\quad \frac{{\rm d} u_N}{{\rm d}x} (x_i) &=& \sin(x_i-1),\\
  \frac{{\rm d} u_N}{{\rm d}x} (x_N) - \tau u_N(x_N) &=& \sin(x_N-1), \nonumber
\end{eqnarray}
where $\tau>0$ is a constant; one notices that taking the limit $\tau \to
+\infty$, both system become equivalent. The discrepancy between the numerical
and analytical solutions is displayed on Figure~\ref{figure:penalty}, as a
function of that parameter $\tau$, when using $N=8$. It is clear from that
figure that there exists a finite value of $\tau$ ($\tau_{\rm min} \simeq 70$)
for which the error is minimal and, in particular, lower than the error
obtained by the standard technique. Numerical evidences indicate that
$\tau_{\rm min} \sim N^2$. This is a simple example of {\em weakly imposed\/}
boundary conditions, with a {\em penalty term\/} added to the system. The idea
of imposing boundary conditions up to the order of the numerical scheme has
been first proposed by Funaro and Gottlieb~\cite{funaro-88} and can be
efficiently used for time-dependent problems, as illustrated by the following
example. For a more detailed description, the interested reader can refer to
the review article by Hesthaven~\cite{hesthaven-00}.

Let us consider the linear advection equation
\begin{eqnarray}
  \label{eq:penalty_advect}
  \forall x\in [-1,1], \, \forall t\geq 0,\quad \frac{\partial u}{\partial t}
  &=& \frac{\partial u}{\partial x}\\
  \forall t\geq 0,\quad u(1,t) &=& f(t), 
\end{eqnarray}
where $f(t)$ is a given function. We look for a Legendre collocation method to
obtain a solution, and define the polynomial $Q^-(x)$, which vanishes on the
Legendre-Gauss-Lobatto grid points, except at the boundary $x=1$:
$$
Q^-(x) = \frac{(1+x)P'_N(x)}{2P'_N(1)}.
$$
Thus, the Legendre collocation {\em penalty method\/} uniquely defines a
polynomial $u_N(x, t)$ through its values at Legendre-Gauss-Lobatto
collocation points $\left\{ x_i \right\}_{i=0\dots N}$
\begin{equation}
  \label{eq:penalty_colloc}
 \forall i=0\dots N,\quad \left. \frac{\partial u_N}{\partial
     t}\right|_{x=x_i} = \left. \frac{\partial u_N}{\partial x}\right|_{x=x_i} - \tau
Q^-(x_i) \left( u_N(1,t) - f(t) \right),
\end{equation}
where $\tau$ is a free parameter as in (\ref{eq:penalty_system_integ}). For
all the grid points, except the boundary one, this is the same as the standard
Legendre collocation method ($\forall i=0\dots N-1, \ Q^-(x_i) = 0$). At the
boundary point $x=x_N=1$, one has a linear combination of the advection
equation and the boundary condition. Contrary to the case of the simple
integration~(\ref{eq:penalty_system_integ}), the parameter $\tau$ here cannot
be too small: in the limit $\tau \to 0$, the problem is ill posed and the
numerical solution diverges. On the other hand, we still recover the standard
(strong) imposition of boundary conditions when $\tau \to +\infty$. With the
requirement that the approximation be asymptotically stable, we get for the
discrete energy estimate (see details about the technique in
 Section~\ref{sss:energy} hereafter) the requirement
$$
\frac{1}{2} \frac{\rm d}{{\rm d}t} \|u_N(t)\|^2  = \sum_{i=0}^N u_N(x_i, t)
\left. \frac{\partial u_N}{\partial x} \right|_{x=x_i} w_i - \tau u_N^2(t,
x_N)w_N \leq 0.
$$
Using the property of Gauss-Lobatto quadrature rule (with the
Legendre-Gauss-Lobatto weights $w_i$), and after an integration by parts, the
stability is obtained if
\begin{equation}
  \label{eq:penalty_stability}
  \tau \geq \frac{1}{2w_N} \geq \frac{N(N+1)}{4}.
\end{equation}
It is also possible to treat more complex boundary conditions, as described in
Hesthaven and Gottlieb~\cite{hesthaven-96} in the case of Robin-type boundary
conditions (see  Section~\ref{sss:wrm} for a definition). Specific conditions for
the penalty coefficient $\tau$ are derived but the technique is the same: for
each boundary, a penalty term is added which is proportional to the error on
the boundary condition at the considered time. Thus, non-linear boundary
operators can also be incorporated in an easy way (see e.g.\ the case
of the Burgers equation in~\cite{hesthaven-00}). The generalization to
multi-domain solution is straightforward: each domain is considered as an
isolated one, which requires boundary conditions at every time-step. The
condition is imposed through the penalty term containing the difference
between the junction conditions. This approach has very strong links with the
variational method presented in  Section~\ref{ss:weak} in the case of
time-independent problems. More detailed discussion about {\em weak\/}
imposition of boundary condition is given in Canuto~et~al. (Sec.~3.7
of~\cite{canuto-06} and Sec.~5.3 of~\cite{canuto-07} for multi-domain
methods).

\subsection{Discretization in space: stability and convergence}
\label{ss:stab_conv}

After dealing with temporal discretization, we now turn to another fundamental
question of numerical analysis of initial value problems, which is to find
conditions under which the discrete (spatial) approximation $u_N(x,t)$
converges to the right solution $u(x,t)$ of the PDE~(\ref{eq:time_PDE}) as
$N\to \infty$ and $t \in [0,T]$. The time derivative term is treated formally,
as could also be treated a source term on the right-hand side, that we do not
consider here, for better clarity.

A given spatial scheme to the PDE is said to be {\em convergent\/} if
any numerical approximation $u_N(x,t)$ obtained through this scheme to the
solution $u(x,t)$
\begin{equation}
  \label{eq:def_conv}
  \| P_N u - u_N \|_{L^2_w} \to 0 \text{ as } N \to \infty.
\end{equation}
Two more concepts are helpful in the convergence analysis of numerical
schemes:
\begin{itemize}
\item {\em consistency\/}: an approximation to the PDE~(\ref{eq:time_PDE}) is
  consistent if $\forall v \in {\mathcal H}$  both
\begin{equation}
    \label{eq:def_consistent}
    \left.
    \begin{array}{l}
    \| P_N (Lv - L_N v)  \|_{L^2_w} \to 0 \\
    \| P_N v - v_N \|_{L^2_w} \to 0 
  \end{array}
  \right\}
  \text{ as } N \to \infty;
  \end{equation}
\item {\em stability\/}: with the formal notations of
  Equation~(\ref{eq:sol_time_PDE}), an approximation to the
  PDE~(\ref{eq:time_PDE}) is stable if
\begin{equation}
    \label{eq:def_stable}
    \forall N, \quad \|e^{L_Nt}\| = \sup_v \frac{\| e^{L_Nt}v \|_{L^2_w}}{\|v \|_{L^2_w}}
    \leq C(t),
  \end{equation}
where $C(t)$ is independent of $N$ and bounded for $t \in [0, T]$.
\end{itemize}

\subsubsection{Lax--Richtmyer theorem}
\label{sss:Lax_Richtmyer}

The direct proof of convergence of a given scheme is usually very difficult to
obtain. Therefore, a natural approach is to use the {\em Lax--Richtmyer
  equivalence theorem\/}: ``a consistent approximation to a well-posed linear
problem is stable if and only if it is convergent''. Thus, the study of
convergence of discrete approximations can be reduced to the study of their
stability, assuming they are consistent. Hereafter, we sketch out
the proof of this equivalence theorem.

The time-evolution PDE~(\ref{eq:time_PDE}) is approximated by
\begin{equation}
  \label{eq:approx_time_PDE}
  \frac{\partial u_N}{\partial t} = L_N u_N.
\end{equation}
To show that stability implies convergence, we subtract it to the exact
one~(\ref{eq:time_PDE})
$$
\frac{\partial \left(u-u_N\right)}{\partial t} = L_N \left( u - u_N \right) +
Lu - L_N u, 
$$
and obtain after integration (the dependence on the space coordinate $x$ is skipped)
\begin{equation}
  \label{eq:Lax_Richt1}
  u(t) - u_N(t) = e^{L_Nt} \left[ u(0) - u_N(0) \right] + \int_0^t
  e^{L_N(t-s)} \left[ Lu(s) - L_Nu(s) \right] {\rm d}s.
\end{equation}
Using the stability property~(\ref{eq:def_stable}), the norm (${L^2_w}$) of
this equation implies
\begin{equation}
  \label{eq:Lax_Richt2}
  \| u(t) - u_N(t) \| \leq C(t)  \| u(0) - u_N(0) \| + \int_0^t C(t-s) \|
  Lu(s) - L_Nu(s) \| {\rm d}s.
\end{equation}
Since the spatial approximation scheme is consistent and $C(t)$ is a bounded
function independent of $N$, for a given $t\in [0,T]$ the left-hand side goes
to zero as $N \to \infty$, which proves the convergence.

Conversely, to show that convergence implies stability, we use the triangle
inequality to get
$$
0 \leq \left| \left\|e^{L_Nt}u \right\| - \left\|e^{Lt}u \right\| \right| \leq
\left\| e^{L_Nt}u - e^{Lt}u \right\| \to 0.
$$
From the well-posedness $\|e^{Lt}u\|$ is bounded and therefore
$\|e^{L_Nt}u\|$ is bounded too, independently of $N$.

The simplest stability criterion is the {\em von Neumann stability
  condition\/}: if we define the adjoint $L^*$ of the operator $L$, using the
inner product, with weight $w$ of the Hilbert space
$$
\forall (u,v) \in {\mathcal H}, \quad \left(u, Lv\right)_w = \left(L^*u, v \right)_w,
$$
then the matrix representation $L_N^*$ of $L^*$ is also the adjoint of the
matrix representation of $L_N$. The operator $L_N$ is said to be normal if it
commutes with its adjoint $L^*_N$. The von Neumann stability condition is
that for normal operators, if there exists a constant $K$ independent of $N$,
such that
\begin{equation}
  \label{eq:neumann_stab}
  \forall i,  1 \leq i \leq N, \quad  {\rm Re}(\lambda_i) < K,
\end{equation}
with $\left(\lambda_i\right)$ being the eigenvalues of the matrix $L_N$, then
the scheme is stable. This condition provides an operational technique for
checking the stability of normal approximations. Unfortunately, spectral
approximations using orthogonal polynomials have in general strongly
non-normal matrices $L_N$ and therefore, the von Neumann condition cannot be
applied. Some exceptions include Fourier-based spectral approximations for
periodic problems.

\subsubsection{Energy estimates for stability}
\label{sss:energy}

The most straightforward technique for establishing the stability of spectral
schemes is the {\em energy method\/}: it is based on choosing the approximate
solution itself as a test function in the evaluation of
residual~(\ref{e:residual}). However, this technique only provides a
sufficient condition and in particular, crude energy estimates indicating that
a spectral scheme might be unstable may be very misleading for non-normal
evolution operators (see the example in Section~8 of Gottlieb and
Orszag~\cite{gottlieb-77}). 

Some sufficient conditions on the spatial operator $L$ and its
approximation $L_N$ are used in the literature to obtain energy
estimates and stability criteria. Some of them are listed hereafter:

\begin{itemize}
\item If the operator $L$ is {\em semi-bounded\/}:
\begin{equation}
    \label{eq:def_semi_bounded}
    \exists \gamma, \quad L + L^* \leq \gamma I,
  \end{equation}
  where $I$ is the identity operator;
\item In the parabolic case, if $L$ satisfies the {\em coercivity condition\/}
  (see also Chap.~6.5 of Canuto~et~al.~\cite{canuto-06}\footnote{Note
    the difference in sign convention between~\cite{canuto-06} and here}):
\begin{equation}
    \label{eq:def_coercivity}
    \exists A>0, \forall (u,v),\quad  \left|(Lu, v) \right| \leq A \| u\| \| v\|, 
  \end{equation}
  and the {\em continuity condition\/}:
\begin{equation}
    \label{eq:def_continuity}
    \exists \alpha >0, \forall u, \quad (Lu, u) \leq - \alpha\| u \|^2;
  \end{equation}
\item In the hyperbolic case, if there exists a constant $C>0$ such that
\begin{equation}
    \label{eq:cond_hyp}
    \forall u, \quad \left\| Lu \right\| \leq C \| u \|,
  \end{equation}
  and if the operator verifies the {\em negativity condition\/}:
\begin{equation}
    \label{eq:def_negativity}
    \forall u, \quad \left( Lu, u \right) \leq 0.
  \end{equation}
\end{itemize}

As an illustration, we consider hereafter a Galerkin method applied to the
solution of Eq.(\ref{eq:time_PDE}), where the operator $L$ is semi-bounded,
following the definition ~(\ref{eq:def_semi_bounded}). The discrete solution
$u_N$ is such that the residual (\ref{e:residual}) estimated on the
  approximate solution $u_N$ itself verifies
\begin{equation}
    \label{eq:galerkin_semi_bounded}
    \left( \frac{\partial u_N}{\partial t} - L u_N, u_N \right)_w = 0.
  \end{equation}
  Separating the time-derivative and the spatial operator:
$$
\frac{1}{2}\frac{\rm d}{\rm dt} \|u_N(t) \|_w^2 = \frac{1}{2} \left( (L +
  L^*)u_N(t), u_N(t)\right)_w,
$$
which shows that the ``energy'' 
\begin{equation}
\label{eq:energy_semi_bounded}
\|u_N(t)\|^2 \leq e^{\gamma t} \| u_N(0) \|^2
\end{equation}
grows at most exponentially with time. Since $u_N(t) = e^{L_Nt} u_N(0)$ for
any $u_N(0)$, we obtain 
\begin{equation}
  \label{eq:stable_semi_bounded}
  \left\| e^{L_Nt} \right\| \leq e^{\frac{1}{2}\gamma t}
\end{equation}
which gives stability and therefore convergence, provided that the
approximation is consistent (thanks to Lax--Richtmyer theorem).

\subsubsection{Examples: heat equation and advection equation}
\label{sss:time_examples}

\paragraph{Heat equation\\}

\noindent We first study the linear heat equation
\begin{equation}
  \label{eq:linear_heat}
  \frac{\partial u}{\partial t} - \frac{\partial^2 u}{\partial x^2} = 0,
  \textrm{ with } -1< x < 1, t>0,
\end{equation}
homogeneous Dirichlet boundary conditions
\begin{equation}
  \label{eq:heat_Dirichlet_BC}
  \forall t\geq 0, \quad u(-1, t) = u(1,t) = 0,
\end{equation}
and initial condition
\begin{equation}
  \label{eq:heta_initial}
  \forall -1 \leq x \leq 1, \quad u(x, 0) = u^0(x).
\end{equation}
In the semi-discrete approach, the Chebyshev collocation method for this
problem (see Section~\ref{s:colloc}) can de devised as follows: the spectral
solution $u_N(t>0)$ is a polynomial of degree $N$ on the interval $[-1, 1]$,
vanishing at the endpoints. On the other Chebyshev-Gauss-Lobatto collocation
points $\left\{ x_k \right\}_{k=1\dots N-1}$ (see  Section~\ref{sss:cheby}),
$u_N(t)$ is defined through the collocation equations
\begin{equation}
  \label{eq:heat_colloc}
  \forall k=1\dots N-1, \quad \frac{\partial u}{\partial t}(x_k,t)  -
  \frac{\partial^2 u}{\partial x^2}(x_k, t)  = 0,
\end{equation}
which are time ODEs (discussed in  Section~\ref{ss:time_discretization}) with the
initial conditions
\begin{equation}
  \label{eq:heat_colloc_ini}
  \forall k=0 \dots N, \quad u_N(x_k, 0) = u^0(x_k).
\end{equation}

The stability of such a scheme is now discussed, with the computation of an
energy bound to the solution. Multiplying the $k$-th
equation of the system (\ref{eq:heat_colloc}) by $u_N(x_k, t)w_k$, where
$\left\{w_k\right\}_{k=0\dots N}$
are the discrete weights for the Chebyshev-Gauss-Lobatto quadrature
(Section~\ref{sss:cheby}), and summing over $k$; one gets:
\begin{equation}
  \label{eq:heat_demo1}
  \frac{1}{2} \frac{\rm d}{\rm dt} \sum_{k=0}^N \left( u_N(x_k, t) \right)^2
  w_k - \sum_{k=0}^N \frac{\partial^2 u_N}{\partial x^2}(x_k, t) u_N(x_k, t)
  w_k = 0.
\end{equation}
Boundary points ($k=0, N$) have been included in the sum since $u_N$ is zero
there from the boundary conditions. The product $u_N \times \partial^2 u_N /
\partial x^2$ is a polynomial of degree $2N-2$, so the quadrature formula is
exact
\begin{equation}
\sum_{k=0}^N \frac{\partial^2 u_N}{\partial x^2}(x_k, t) u_N(x_k, t)
  w_k = \int_{-1}^1 \frac{\partial^2 u_N}{\partial x^2}(x_k, t) u_N(x_k, t)
  w(x) {\rm d}x;
\end{equation}
and integrating by parts twice, one gets the relation
\begin{equation}
\int_{-1}^1 \frac{\partial^2 u_N}{\partial x^2}(x_k, t) u_N(x_k, t)
  w(x) {\rm d}x = \int_{-1}^1 \left(\frac{\partial u_N}{\partial x} \right)^2
  w(x) {\rm d}x - \frac{1}{2} \int_{-1}^1u_N^2 \frac{\partial^2 w}{\partial x^2}
  {\rm d}x.
\end{equation}
By the properties of the Chebyshev weight 
\begin{equation}
\frac{\partial^2 w}{\partial x^2} - \frac{2}{w} \left( \frac{\partial
    w}{\partial x} \right)^2 = 0 \textrm{ and } \frac{\partial^2 w}{\partial x^2}
= \left( 1+2x^2 \right)w^5,
\end{equation}
it is possible to show that 
\begin{equation}
\int_{-1}^1 u_N^2 \frac{\partial^2 w}{\partial x^2} {\rm d}x \leq 3
\int_{-1}^1 u_N^2 w^5 {\rm d}x \leq 6 \int_{-1}^1 \frac{\partial^2
  u_N}{\partial x^2}(x_k, t) u_N(x_k, t) w(x) {\rm d}x  ;
\end{equation}
and thus that
\begin{equation}
\int_{-1}^1 \frac{\partial^2 u_N}{\partial x^2}(x_k, t) u_N(x_k, t) w(x) {\rm
  d}x \geq \frac{1}{4} \int_{-1}^1 \left(\frac{\partial u_N}{\partial x}
\right)^2 w(x) {\rm d}x \geq 0.
\end{equation}
Therefore, integrating over the time interval $[0, t]$ the
relation~(\ref{eq:heat_demo1}), one obtains the result that
\begin{equation}
  \label{eq:heat_major_ut}
  \sum_{k=0}^N \left( u_N(x_k, t) \right)^2 w_k \leq \sum_{k=0}^N \left(
    u^0(x_k) \right)^2 w_k \leq 2 \max_{x\in [0,1]} |u^0(x)|^2 .
\end{equation}
The left-hand side represents the discrete norm of $u_N(t)^2$, but
since this is a polynomial of degree $2N$, one cannot apply the
Gauss--Lobatto rule. Nevertheless, it has been shown (see e.g.\
Section~5.3 of Canuto~et~al.~\cite{canuto-06}) that discrete and
$L^2_w$-norms are uniformly equivalent, therefore:
\begin{equation}
  \label{eq:heat_energy_bound}
  \int_{-1}^1 \left( u_N(x,t)\right)^2 w(x) \leq 2 \max_{x\in [0,1]} |u^0(x)|^2,
\end{equation}
which proves the stability of the Chebyshev-collocation method for the heat
equation. Convergence can again be deduced from the Lax--Richtmyer theorem, but
a detailed analysis (cf.\ Section~6.5.1 of
Canuto~et~al.~\cite{canuto-06}) shows that the numerical solution
obtained by the method described here converges to the true solution
and one can obtain the convergence rate. If the solution $u(x,t)$ is
$m$-times differentiable with respect to the spatial coordinate $x$
(see Section~\ref{sss:convergence}) the energy norm of the error
decays like $N^{1-m}$. In particular, if the solution is ${\mathcal
  C}^\infty$, the error decays faster than any power of $N$.

\paragraph{Advection equation\\}

\noindent We now study the Legendre-tau approximation to the simple
advection equation
\begin{equation}
  \label{eq:advect_linear}
  \frac{\partial u}{\partial t} + \frac{\partial u}{\partial x} = 0,
  \textrm{ with } -1< x < 1, t>0,
\end{equation}
homogeneous Dirichlet boundary condition
\begin{equation}
  \label{eq:advect_Dirichlet_BC}
  \forall t\geq 0, \quad u(-1, t) = 0,
\end{equation}
and initial condition
\begin{equation}
  \label{eq:advect_initial}
  \forall -1 \leq x \leq 1, \quad u(x, 0) = u^0(x).
\end{equation}
If we seek the solution as the truncated Legendre series:
$$
  u_N(x,t) = \sum_{i=0}^N a_i(t) P_i(x)
$$
by the tau method, then $u_N$ satisfies the equation:
\begin{equation}
  \label{eq:advect_tau}
  \frac{\partial u_N}{\partial t} + \frac{\partial u_N}{\partial x} = \tau_N(t) P_N(x).
\end{equation}
Equating coefficients of $P_N$ on both sides of~(\ref{eq:advect_tau}), we get
$$
\tau_N = \frac{{\rm d}a_N}{{\rm d}t}.
$$
Applying the $L^2_w$ scalar product with $u_N$ to both sides of
Equation~(\ref{eq:advect_tau}), we obtain
$$
\frac{1}{2} \frac{\partial}{\partial t} \left( \|u_N \|^2 - a_N^2 \right) =
-\int_{-1}^1 u \frac{\partial u_N}{\partial x} {\rm d}x =-\frac{1}{2} u_N^2(1)
\leq 0;
$$
which implies the following inequality:
\begin{equation}
  \label{eq:advect_energy}
  \frac{\rm d}{{\rm d} t} \sum_{i=0}^{N-1} a_i^2 \leq 0.
\end{equation}
Finally, $a_N(t)$ is bounded because it is determined in terms of $\left\{ a_i
\right\}_{i=0\dots N-1}$ from the boundary
condition~(\ref{eq:advect_Dirichlet_BC}), and thus stability is proved. In the
same way as before, for the heat equation, it is possible to derive a bound
for the error $\|u(x,t) - u_N(x,t)\|$, if the solution $u(x,t)$ is $m$-times
differentiable with respect to the spatial coordinate $x$: the energy norm of
the error decays like $N^{1-m}$ (see also Section~6.5.2 of
Canuto~et~al.~\cite{canuto-06}). In particular, if the solution is
${\mathcal C}^\infty$, the error decays faster than any power of $N$.

\subsection{Fully-discrete analysis}
\label{ss:fully_discrete}

Stability issues have been discussed separately for time
(Section~\ref{sss:time_stab}) and space (Section~\ref{ss:stab_conv})
discretizations. The global picture ({\em fully discrete analysis\/}), taking
into account both discretizations is, in general, very difficult to study.
However, it is possible in some particular cases to address the problem and in
the following lines, we shall perform a fully discrete analysis of the advection
equation~(\ref{eq:advect_linear}), when using a Legendre collocation method in
space and a forward Euler scheme in time. With the notations of
 Section~\ref{ss:time_discretization}, the scheme writes
\begin{equation}
  \label{eq:advect_full}
  \forall x \in [-1, 1],\quad  U_N^{J+1}(x) = U_N^J(x) - \Delta t
  \frac{\partial U_N^J}{\partial x} + \Delta t
  \left. \frac{\partial U_N^J}{\partial x}\right|_{x=-1}
  \frac{P_{N+1}(x)}{P_{N+1}(-1)},
\end{equation}
where the last term imposes the boundary condition $\forall J, \quad
U_N^{J}(x=-1) = 0$. We consider this relation at the Legendre--Gauss collocation
points $\left( \left\{x_i\right\}_{i=0\dots N}\right)$, which are zeros of
$P_{N+1}(x)$; the square of this expression taken at these collocation points gives
$$
  \forall i \in [0, N],\quad  \left(U_N^{J+1}(x_i)\right)^2 = \left( U_N^J(x_i)
  \right)^2 + \Delta t^2 \left( \left. \frac{\partial U_N^J}{\partial x}\right|_{x=x_i}
  \right)^2 - 2\Delta t\, U_N^J(x_i) \left. \frac{\partial U_N^J}{\partial
      x}\right|_{x=x_i}.
$$
we multiply by $(1-x_i)w_i$, where the $\left\{ w_i \right\}_{i=0\dots N}$ are
the Legendre--Gauss weights, and sum over $i$ to obtain
\begin{eqnarray*}
  \sum_{i=0}^N(1-x_i)\left(U_N^{J+1}(x_i)\right)^2 w_i &=& \sum_{i=0}^N
  (1-x_i) \left( U_N^J(x_i) \right)^2 w_i - 2\Delta t \sum_{i=0}^N (1-x_i)
  U_N^J(x_i)w_i \left. \frac{\partial U_N^J}{\partial x}\right|_{x=x_i}  \\
  &+& \Delta t^2\, \sum_{i=0}^N \left( \left. \frac{\partial U_N^J}{\partial
        x}\right|_{x=x_i} \right)^2  (1-x_i) w_i. 
\end{eqnarray*}
For stability we need that a certain discrete energy of $U_N^J$ be bounded in
time: 
\begin{equation}
  \label{eq:full_stability}
   \sum_{i=0}^N(1-x_i)\left(U_N^{J+1}(x_i)\right)^2 w_i \leq \sum_{i=0}^N
  (1-x_i) \left( U_N^J(x_i) \right)^2 w_i,
\end{equation}
which means that
$$
\Delta t^2\, \sum_{i=0}^N \left( \left. \frac{\partial U_N^J}{\partial
        x}\right|_{x=x_i} \right)^2  (1-x_i) w_i - 2\Delta t \sum_{i=0}^N (1-x_i)
  U_N^J(x_i)w_i \left. \frac{\partial U_N^J}{\partial x}\right|_{x=x_i} \leq 0.
$$
With the exactness of the Legendre--Gauss quadrature rule for polynomials of degree
lower than $2N+1$, we have that 
$$
\sum_{i=0}^N \left( \left. \frac{\partial U_N^J}{\partial
        x}\right|_{x=x_i} \right)^2  (1-x_i) w_i = \int_{-1}^1 \left(
    \frac{\partial U_N^J}{\partial x} \right)^2 (1-x) {\rm d}x,
$$
and, with an additional integration by parts 
$$
\sum_{i=0}^N (1-x_i) U_N^J(x_i)w_i \left. \frac{\partial U_N^J}{\partial
    x}\right|_{x=x_i} = \int_{-1}^1 (1-x) U_N^J \frac{\partial U_N^J}{\partial
  x} {\rm d}x = \frac{1}{2} \int_{-1}^1 \left(U_N^J(x)\right)^2 {\rm d}x.
$$
The stability condition obtained from energy analysis translates into an upper
bound for the time-step, which can be seen as an accurate estimate of the CFL
restriction on the time-step:
\begin{equation}
  \label{eq:full_time_step}
  \Delta t \leq \frac{\int_{-1}^1 \left(U_N^J(x)\right)^2 {\rm
      d}x}{\int_{-1}^1 \left(
      \frac{\partial U_N^J}{\partial x} \right)^2 (1-x)\, {\rm d}x} \simeq {\mathcal
    O}\left( \frac{1}{N^2} \right).
\end{equation}

\subsubsection{Strong stability preserving methods}

The above fully-discrete analysis must in principle be performed for every
time-marching scheme. Therefore, it is very convenient to have a way of
extending results from first-order Euler method to higher-order
methods. {\em Strong stability preserving\/} Runge--Kutta and multi-step
methods preserve such kind of stability properties, including the case of
non-linear stability analysis. A general review on the subject has been done by
Shu~\cite{shu-02}, and we list some results hereafter.

If we consider the general time ODE:
\begin{equation}
  \label{eq:time_ODE}
  \frac{{\rm d}U_N}{{\rm d}t} = L_N U_N,
\end{equation}
arising from the spatial discretization of the PDE~(\ref{eq:time_PDE}), we
suppose that, after discretization in time using the first-order forward Euler
scheme, the strong stability requirement $\| U_N^{J+1} \| \leq \| U_N^J \|$
gives a CFL restriction of the type~(\ref{eq:full_time_step})
\begin{equation}
  \label{eq:SSP_CFL}
  \Delta t \leq \Delta t_{FE}.
\end{equation}
We can then write an $s$-stage Runge--Kutta method in the form
\begin{eqnarray*}
  U_N^{(0)} &=& U_N^J\\
  U_N^{(i)} &=& \sum_{k=0}^{i-1} \left( \alpha_{i,k} + \Delta t\, \beta_{i,k}
    L_N U_N^{(k)} \right), \quad i=1, \dots s\\
  U_N^{J+1} &=& U_N^{(s)},
\end{eqnarray*}
and see that, as long as $\alpha_{i,k} \geq 0$ and $\beta_{i,k} \geq 0$, all
the intermediate stages are simply convex combinations of forward Euler
operators. If this method is strongly stable for $L_N$, under the
condition~(\ref{eq:SSP_CFL}), then the intermediate stages can be bounded and
the Runge--Kutta scheme is stable under the CFL condition
\begin{equation}
  \label{eq:SSP_CFL_RK}
  \Delta t \leq c \Delta t_{FE}, \quad c = \min_{i,k}
  \frac{\alpha_{i,k}}{\beta_{i,k}}. 
\end{equation}

In the same manner, one can devise strong stability preserving explicit
multi-step methods of the form
$$
U_N^{J+1} = \sum_{i=1}^s \left( \alpha_i U_N^{J+1-i} + \Delta t\, \beta_i L_N
  U_N^{J+1-i} \right),
$$
which can also be cast into convex combinations of forward Euler steps and
therefore, these multi-step methods are also stable, provided that
\begin{equation}
  \label{eq:SSP_CFL_multi_step}
  \Delta t \leq c \Delta t_{FE}, \quad c = \min_i \frac{\alpha_i}{\beta_i}.
\end{equation}
Examples of useful coefficients for Runge-Kutta and multi-step strong
stability preserving methods can be found in~\cite{shu-02,
  hesthaven-07}. Optimal such methods are those for which the CFL coefficient
$c$ is as large as possible.

\subsection{Going further: High-order time schemes}
\label{ss:going_further_time} 

When using spectral methods in time-dependent problems, it is sometimes
frustrating to have so accurate numerical techniques for the evaluation of
spatial derivatives, and the integration of elliptic PDEs, whereas the time
derivatives, and hyperbolic PDEs, do not benefit from the spectral
convergence.  Some tentative studies are being undertaken in order to
represent also the time interval by spectral methods~\cite{ansorg-06}. In this
spherically symmetric study of the wave equation in Minkowski space-time,
Hennig and Ansorg have applied spectral methods to both spatial and time
coordinates. Moreover, they have used a conformal compactification of Minkowski
space-time, making the wave equation singular at null infinity. They have
obtained nicely accurate and spectrally convergent solution, even to a
non-linear wave equation. If these techniques can be applied in general
three-dimensional simulations, it would really be a great improvement.

Nevertheless, there are other, also more sophisticated and accurate
time-integration techniques that are currently investigated for several stiff
PDEs~\cite{kassam-05}, among which Korteweg--de~Vries and nonlinear
Schr\"odinger equations~\cite{klein-06}. Many such PDEs share the properties
of being stiff (very different time-scales/ characteristic frequencies) and
combining low-order non-linear terms with higher-order linear terms. Einstein
evolution equations can also be written in such a way~\cite{bonazzola-04}. Let
us consider a PDE
\begin{equation}
  \label{eq:stiff_PDE}
  \frac{\partial u}{\partial t} = Lu + {\mathcal N}u,
\end{equation}
with the notations of  Section~\ref{sss:method_of_lines} and $\mathcal N$ being a nonlinear
spatial operator. Following the same notations and within spectral
approximation, one recovers
\begin{equation}
  \label{eq:stiff_PDE_discret}
  \frac{\partial U_N}{\partial t} = L_N U_N + {\mathcal N}_N U_N.
\end{equation}
We detail hereafter five methods to solve this type of ODEs (see also~\cite{kassam-05}):
\begin{itemize}
\item{\bf Implicit-explicit} techniques use some explicit multi-step scheme to
  advance the nonlinear part ${\mathcal N}_N$, and an implicit one for the linear one.
\item{\bf Split-step} are effective when the equation splits into two equations
    which can be directly integrated (see~\cite{klein-06} for examples with
    the nonlinear Schr\"odinger and Korteweg-de Vries equations).
  \item{\bf Integrating factor} is a change of variable that allows for the exact
    solution of the linear part
\begin{equation}
      \label{eq:integ_factor}
      V_N = e^{-L_Nt} U_N,
    \end{equation}
and to use an explicit multi-step method for the integration of the new
nonlinear part
\begin{equation}
  \label{eq:integ_factor2}
  \frac{\partial V_N}{\partial t} = e^{-L_N t} {\mathcal N}_N e^{L_Nt} V_N. 
\end{equation}
\item {\bf Sliders} can be seen as an extension of the implicit-explicit
  method described above. In addition to splitting to problem into a linear and
  nonlinear part, the linear part itself is split into two or three regions
  (in Fourier space), depending on the wavenumber. Then, different numerical
  schemes are used for different groups of wavenumbers: implicit schemes for
  high wavenumbers and explicit high-order methods for the low
  wavenumbers. This method is restricted to Fourier spectral techniques in
  space. 
\item {\bf Exponential time-differencing} have been known for some time in
  computational electrodynamics. These methods are similar to the integrating factor
  technique, but one considers the {\em exact\/} equation over one time-step
\begin{equation}
    \label{eq:exponential_time_differencing}
    U_N^{J+1} = e^{L_N\Delta t} U_N^J + e^{L_N\Delta t} \int_0^{\Delta t}
    e^{-L_N \tau} {\mathcal N}_N(U_N(N\Delta t + \tau), N\Delta t+\tau) d\tau.
  \end{equation}
Various orders for these schemes come from the approximation order of the
integral. For example Kassam and Trefethen~\cite{kassam-05} consider a
fourth-order Runge--Kutta type approximation to this integral, where the
difficulty comes from the accurate computation of functions which suffer from
cancellation errors.
\end{itemize}

\newpage

\section{Stationary Computations and Initial Data}
\label{s:station}

\subsection{Introduction}

In this section, we restrict ourselves to problems where time does not appear
explicitly. This is especially the case for systems which are stationary, like
neutron stars in rotation or binary systems in circular orbits. The
computation of initial data also falls into this class, given that it consists
in finding a particular solution of Einstein equations {\em at a given time}
only. Indeed, when using the standard 3+1 decomposition of spacetime, the
initial data that are passed to the evolution equations cannot be totally
arbitrary and must satisfy a set of equations called Einstein's constraint
equations. For more details on the initial data problem we refer to the review
by G.B.~Cook~\cite{cook-00}. So, in treating the problems considered here, one
can forget about the issues specific to time presented in Sec.
\ref{s:time}.

It must be said that spectral methods are not the only technique that has been
successfully used to generate stationary spacetimes. The papers
\cite{baumgarte-98, uryu-00, cook-94, matzner-99} give some examples of this
fact, especially in the case of binary systems, for neutron stars or black
holes. More references can be found in~\cite{cook-00}.

\subsection{Single compact stars}

The computation of the structure of stationary compact stars dates back to
1939 with the famous solution of Tolman--Oppenheimer--Volkoff. During the last
years, the need for accurate models has been more pressing especially with the
coming online of the gravitational wave detectors which could help to probe
the interior of such compact stars. Isolated stars in rotation are essentially
axisymmetric but some physical effects can induce a symmetry breaking that
could lead to the emission of gravitational waves. In the following, we shall
review some computations that aim at including some of those effects, like
the spontaneous symmetry breaking, the inclusion of magnetic field, the effect
of exotic dense matter, mainly with strange quarks or the influence of an
interior composed of two different superfluids.

\subsubsection{Formalisms}
\label{sss:formalisms}

The first computation of models of relativistic rotating stars in general
relativity, by means of spectral methods, is presented in~\cite{bonazzola-93}.
The equations are solved in spherical coordinates (see
Section~\ref{ss:spherical_coordinates_harmonics}). Doing so, the
fields only depend on the azimuthal angle $\theta$ and on the radius
$r$. The fields are expanded onto spherical harmonics with respect to
the angle and onto Chebyshev polynomials with respect to $r$. The use of
spherical harmonics gives a natural way of dealing with the coordinate
singularity on the $z$-axis. In~\cite{bonazzola-93} the whole space is
divided into two spherical domains, the outer one extending up to
infinity by making use of the compactification in $1/r$ seen in
Section~\ref{sss:compactification}. With such setting, Einstein
equations reduce to a set of four elliptic equations with sources
extending up to infinity that are solved using a version of the
algorithm based on matching with the homogeneous solutions (presented
in Section~\ref{ss:hom}), for each spherical harmonics. The system is
complete once a description of the matter is given. The simplest
choice is to consider a polytropic fluid, with or without magnetic
field. The system is solved by iteration.

In the paper~\cite{bonazzola-93}, a particular emphasis is put on the various
methods to measure the accuracy of the code. For non-rotating stars, the error
is found to decrease exponentially, as the number of coefficients increases
(see Figures~5 and 6 of~\cite{bonazzola-93}). However, for fast-rotating
configurations, the error only decays as a power-law (see Figure~7
of~\cite{bonazzola-93}). This comes from the fact that quantities like
the energy density are no longer ${\mathcal C}^\infty$ across the
star's surface. Nevertheless, the results are in good agreement (to
the level of a 0.1\%) with those obtained by other numerical methods,
as can be seen in~\cite{nozawa-98}.

Spectral convergence can be recovered by using surface-adapted coordinates as
first done in~\cite{bonazzola-96}. A regular mapping of the numerical
coordinates to the physical ones is introduced, so that the surface of the star
lies at the boundary between two domains (see Section~\ref{sss:mappings}). For
polytropes with $\gamma<2$, this is sufficient to recover spectral convergence
(see Figures~5 and 6 of~\cite{bonazzola-98}). However, for $\gamma>2$, some
quantities are still diverging at the surface but the convergence can be made
closer and closer to the spectral one by analytically regularizing the density
(see Section~IV of~\cite{bonazzola-98}). Doing so, the error decreases as a
power-law but the decrease can be made arbitrary fast at the cost of
increasing the number of operations and so the computational time.

Up to 2006, the neutron stars were computed using quasi-isotropic coordinates.
However, in order to use those configurations as initial data for evolutionary
codes, it may be useful to allow for other choices. Among the possible gauges,
the Dirac one is one of the most
promising~\cite{bonazzola-04}. In~\cite{lin-06} models of rotating
neutron stars, in the Dirac gauge are computed, for both polytropic
and realistic equations of state. Contrary to the quasi-isotropic
coordinates, the use of this new gauge implies to solve one
tensor-like Poisson equation. Configurations obtained with the two
different formalisms are shown to be in very good agreement.

\subsubsection{Rotating neutron star models}

Even before adapted mappings were available, interesting results could be
obtained. In two papers~\cite{salgado-94, salgado-94b}, models of rotating
neutron stars with various equations of state have been computed. Among the
most surprising findings, let us mention the existence of supra-massive stars.
Those stars do not connect to the non-rotating limit. Indeed, their high mass
can only be supported by the presence of a centrifugal force. One of the
remarkable features of such stars is the fact that they actually spin-up when
they lose angular momentum, contrary to what is observed for normal stars.
This effect can also be seen for neutron stars containing hyperons and thus a
softer equation of state~\cite{zdunik-04}. Let us mention that, in this case,
the stability analysis of the configurations required the great precision that
spectral methods with adapted coordinates could provide.

It is known that isolated pulsars spin down due to magnetic braking. As the
rotational frequency decreases, it is possible that the star will encounter a
transition from one state of matter to another. Stationary rotating models
have been used to determine the properties of such
transitions~\cite{zdunik-06, zdunik-07, zdunik-08}. A puzzling result
is that the amount of energy released in a first order phase
transition does note depend on the orbital velocity of the star and is
the same as for non-rotating ones. This is shown to be the case for
both weak~\cite{zdunik-07} and strong~\cite{zdunik-08} first order
transitions.

\subsubsection{Spontaneous symmetry breaking}

It is known that stars can undergo a spontaneous symmetry breaking when
rotating fast enough. When such a phenomenon occurs, triaxial configurations
are formed that are potential emitters of gravitational waves. The departure
from axisymmetry is studied in two papers by the Meudon
group~\cite{bonazzola-96b, bonazzola-98b}. The idea of the method is to start
from an axisymmetric neutron star configuration and to follow the growth or
decay of triaxial instabilities. Well-established results in the Newtonian
regime are recovered and this work presents the first results in general
relativity, for various equations of states. For a few of them, the frequency
at which symmetry-breaking occurs lies in the frequency band of the LIGO and
Virgo detectors.

In 2002, this work has been extended~\cite{gondek-02} by making use of
surface-fitting coordinates. This enabled the authors to obtain results in the
incompressible case by properly dealing with discontinuities lying at the
surface of the star. Classical results in the incompressible case are thus
recovered and it is found that the inclusion of relativity has only a moderate
effect. Indeed the critical ratio between the kinetic energy and the absolute
gravitational one $T/\l|W\r|$ at which the triaxial instability occurs is only
30\% larger for relativistic stars, with respect to their classical
counterparts.

If relativistic effects only slightly stabilize the stars, the same is not
true for differential rotation. Indeed, in~\cite{saijo-06}, the authors study
various rotation profiles and equations of state using the same technique as
in~\cite{bonazzola-96b, bonazzola-98b} to determine the onset of instability.
It appears that the critical value of $T/\l|W\r|$ can be almost twice as high
as for uniformly rotating stars.

\subsubsection{Configurations with magnetic field}

Even if magnetic fields are strong in neutron stars, the structure of the
objects is not affected until it reaches huge values, of the order of at least
$10^{10} {\rm T}$. In~\cite{bocquet-95}, models of rapidly rotating stars with
poloidal fields are constructed, for various equations of state. The magnetic
effects are taken into account consistently by solving the appropriate Maxwell
equations, also by means of spectral methods. The maximum mass of highly
magnetized neutrons stars is found to be higher from 13 to 29 \% than for the
non-magnetized stars. The magnetic field induces an additional pressure which
can help to support more massive stars, thus explaining this increase.

The presence of a magnetic field can also lead to a deformation of the neutron
star. Such deformation could lead to the formation of a triaxial
configuration, which would then emit gravitational
wave. In~\cite{bonazzola-96} the emitted signal is computed. Typically
the system radiates at two frequencies: $\Omega$ and $2\Omega$ where
$\Omega$ is the angular velocity of the star.

In a more recent work by the Meudon group~\cite{novak-03}, magnetized
configurations have been computed using coordinates matched to the surface of
the star, thus making the computation much more accurate. Gyromagnetic ratios
of rapidly rotating neutron stars of various equations of state are obtained.
The limit of a ratio $g=2$, value for a charged black hole, is never reached.

\subsubsection{Strange stars}

It is possible that the fundamental state of nuclear matter is not the
ordinary matter but rather a plasma of deconfined quarks $u$, $d$ and $s$,
called {\em strange matter}. If this is the case, neutron stars would rather
be strange stars. The main difference between those two types of compact stars
is that strange ones are somewhat smaller and thus more compact. In
particular, they would support higher rotation rates. There is a strong
density jump at the surface of a strange star and surface-fitting coordinates
are required in order to deal with it.

Fast rotating models of strange stars are computed
in~\cite{gourgoulhon-99,gondek-00}. Due to higher compactness, it is
found that strange stars can rotate significantly faster than their
neutron star counterparts. The attained $T/\l|W\r|$ can be twice as
large. As in the neutron star case, supermassive configurations that
spin-up with angular momentum loss are found. The influence of strange
matter on the emission of gravitational waves is studied
in~\cite{gondek-03} where viscosity effects and triaxial instabilities
are carefully taken into account.

It is believed that millisecond pulsars have been spun-up by accreting matter
from a companion. However, the details of this mechanism depend on the nature
of the compact object. In~\cite{zdunik-02}, the differences between accretion
onto a neutron star and onto a strange star are investigated, using 2D
stationary models computed by spectral methods.

\subsubsection{Quasi-periodic oscillations}

Quasiperiodic oscillations (QPOs) are observed in the kHz regime and are
believed to be the signature of matter falling onto a compact object. In the
standard picture, the frequency of the QPOs, is that of the last stable
orbit around the compact object. Let us mention that the presence of a last
stable orbit around an extended body is not an effect of relativity but can
also be seen in the Newtonian regime, as shown in~\cite{zdunik-01}.

The precise structure of the accreting object has a great influence on the
QPO. In a series of papers~\cite{zdunik-00, gondek-01, amsterdamski-02,
  bejger-07}, comparisons are made between observations and various compact
stars models that could account for QPOs.

Using a multi-domain approach, strange stars with a crust can
also be computed~\cite{zdunik-01b}, one domain describing the interior of the
star and another one the crust. It is shown that the presence of the crust
could change the value of the QPO by up to 20\%.

\subsubsection{More complex configurations}
\label{sss:complex}

In this section, objects in more exotic configurations are presented. This is
an illustration of both the complexity of the neutron stars physics and the
ability of spectral methods to deal with complicated systems.

The observation of glitches in isolated pulsars is consistent with the
presence of a superfluid interior. The simplest model considers two fluids,
one composed of neutrons and the other one of protons and electrons, both
components being superfluids. However, those two components could have
different velocities, in particular different rotation rates. Such
configurations are computed in~\cite{prix-05}. A multi-domain setting is
crucial to be able to follow the two different fluids because the components
do not have the same shape. Among the various results obtained, let us mention
the confirmation of the existence of prolate-oblate configurations.

Neutron stars are usually assumed to be at zero-temperature. However this
approximation is no longer true for newborn neutron stars, just after the
supernova. Models of newborn neutron stars in uniform rotations are 
constructed in~\cite{goussard-97} using an extension of the code
developed in~\cite{bonazzola-93}. Various hypothesis about the
interior (different lepton numbers, isothermal versus isentropic) are
considered. Sequences of fixed baryon mass and angular momentum are
constructed. Such sequences approximate the evolution of the
proto-neutron star into a cold neutron star. The results have been
extended to differentially rotation proto-neutron stars
in~\cite{goussard-98}.

The effect of finite temperature is also investigated in~\cite{villain-04}. 
The authors found that newborn neutron
stars are unlikely to undergo the bar mode instability but that the secular
ones could take place and result in a significant emission of gravitational
waves. Another interesting result of~\cite{villain-04} is the existence of
toroidal-like configurations, which appear for a broad range of parameters and
before the mass-shedding limit. In such cases, the surface of the star is
highly deformed and surface-fitting coordinates are required.

Axisymmetric rotating neutron stars have also been computed by a code
developed by M.~Ansorg and collaborators~\cite{ansorg-02, ansorg-03} . This
code is based on Lewis-Papapetrou coordinates $\l(\rho, \xi\r)$ , which are
closely related to the usual cylindrical coordinates. Typically, space is
decomposed into two domains: one for the interior of the star and another to
the exterior which extends up to spatial infinity. Compactification of space
and surface-fitting mappings are used. Both variables are expanded on
Chebyshev polynomials. Instead of solving the equations harmonics by harmonics
and iterate, as is done by the Meudon group, the equations are written with a
collocation method (see Section~\ref{s:colloc}) and solved as one single system.
The price to pay is that the size of the system is somewhat larger (i.e.\ in
$m^2$, $m$ being the total number of coefficients for each coordinates). The
system is solved by means of the Newton-Raphson's method. At each step, the
linear system is solved using iterative techniques with preconditioning. With
this setting, impressive accuracy is reached.

The coordinates used in~\cite{ansorg-02, ansorg-03} are more general than the
ones used by the Meudon group, especially with respect to their
surface-fitting capabilities. They can account for more complicated
configurations and, in particular, highly distorted matter distribution can be
obtained. This is shown in~\cite{ansorg-03b, ansorg-03c}, where relativistic axisymmetric
toroidal configurations of matter, known as the Dyson rings, are computed.
Such rings are obtained up to the mass-shedding limit. Transition to the limit
of an extreme Kerr black hole is also discussed.

\subsection{Single black holes}
\label{ss:single_hole}

Compared to the compact star case, single black holes have not been very much
studied. This is probably because the structure of a stationary black hole is
somewhat simpler than the one of a compact star. However, as it will be seen,
there are still properties that must be investigated.

Spacetimes containing a single black hole constitute a good benchmark for
numerical methods, a lot of results being known
analytically. In~\cite{kidder-00a}, the authors have implemented a
spectral solver and applied it to various test problems. The solver
itself is two-dimensional and thus applicable only to axisymmetric
systems. There is a single domain that consists of the whole space
outside a sphere of given radius (i.e.\ the black hole). Space is
compactified by using the standard variable $1/r$. The two physical
variables $\l(r, \theta\r)$ are mapped onto squares in ${\mathbb R}^2$
and then expanded on Chebyshev polynomials. The equations are written
using a 2-dimensional collocation method (see Section~\ref{s:colloc})
and the resulting system is solved by an iterative algorithm (here
Richardson's method with preconditioning). This solver is applied to
solve the Einstein's constraint equations for three different systems:
i) a single black hole ii) a single black hole with angular momentum
iii) a black hole plus Brill waves. In all three cases, spectral
convergence is achieved and accuracy of the order of $10^{-10}$ is
reached with 30 points in each dimension.

A black hole is somewhat simpler than a neutron star, in the sense that there is
no need for a description of matter (no equation of state for instance).
However, in some formalisms, the presence of a black hole is enforced by
imposing non-trivial solution on some surfaces (typically spheres). The basic
idea is to demand that the surface is a trapped surface. Such surfaces are
known to lie inside event horizons and so are consistent with the presence of
a black hole. Discussions about such boundary conditions can be found
in~\cite{cook-02}. However, in non-stationary cases, the set of
equations to be used is not easy to derive. The authors
of~\cite{jaramillo-07} implemented various sets of boundary conditions
to investigate their properties. Two different and independent
spectral codes are used. Both codes are very close to those used in
the case of neutron stars, one of them being based on {\sc Lorene}
library~\cite{lorene} (see Section~\ref{sss:formalisms}) and the other
one has been developed by M.~Ansorg and shares a lot a features
with~\cite{ansorg-02, ansorg-03}. Such numerical tools have proved
useful in clarifying the properties of some sets of boundary
conditions that could be imposed on black hole horizons.

The reverse problem is also important in the context of numerical relativity.
In some cases one needs to know if a given configuration contains a trapped
surface and if it can be located, at each time-step. Several algorithms have
been proposed in the past to find the locus where the expansion of the
outgoing light rays vanishes (thus defining the trapped surface). Even if the
term is not used explicitly, the first application of spectral expansions to
this problem is detailed in~\cite{baumgarte-96}. The various fields are
expanded on a basis of symmetric trace-free tensors. The algorithm is applied
to spacetimes containing one or two black holes. However results seem to
indicate that high order expansions are required to locate the horizons with a
sufficient precision.

More recently, another code~\cite{lin-07} using spectral method has been used
to locate apparent horizons. It is based on the {\sc Lorene} library with its
standard setting i.e. a multi-domain decomposition of space and spherical
coordinates (see Section~\ref{sss:formalisms} for more details). The horizon
finder has been successfully tested on known configurations, like Kerr-Schild
black holes. The use of spectral methods makes it both fast and accurate. Even
if the code is using only one set of spherical coordinates (hence its
presentation in this section), it can be applied to situations with more than
one black hole, like the well-known Brill--Lindquist data~\cite{brill-63}.

\subsection{Rings around black holes}

The problem of uniformly rotating rings surrounding a black hole can be viewed
as an intermediate step between one body, axisymmetric configurations and the
two body problem. Indeed, even if one has to deal with two components, the
problem is still axisymmetric. In~\cite{ansorg-05b}, configurations of a black
hole surrounded by a uniformly rotating ring of matter are computed in
general relativity. The matter is assumed to be a perfect fluid. To solve the
equations, space is divided into five computational domains. One of them
describes the ring itself, another one the region around the black hole and
another extends up to infinity. One of the difficulties is that the surface of
the ring is not know a priori and so the domains must be dynamically adapted
to its surface. Cylindrical-type coordinates are used and, in each domain, are
mapped onto squares of numerical coordinates. The actual mappings depend on
the domain and can be found in Section~IV of~\cite{ansorg-05b}.

Numerical coordinates are expanded onto Chebyshev polynomials. The system to
be solved is obtained by writing Einstein equations in the collocation space
including regularity conditions on the axis and appropriate boundary
conditions on both the horizon of the black hole and at spatial infinity. As
in~\cite{ansorg-02, ansorg-03}, the system is solved iteratively, using
Newton--Raphson's method.

Both Newtonian and relativistic configurations are computed. The ratio between
the mass of the black hole and the mass of the ring is varied from 0 (no black
hole) up to 144. The inner mass shedding of the ring can be obtained. One of
the most interesting results is the existence of configurations for which the
ratio $J_c/M_c^2$ of the black hole angular momentum and the square of its
mass exceeds one, contrary to what can be achieved for an isolated black hole.

\subsection{Binary compact stars}
\label{ss:binary_stars}

\subsubsection{Formalism}
\label{sss:bns_form}

Systems consisting of two compact objects are known to emit gravitational
waves. Due to this emission, no closed orbits can exist and the objects follow
a spiral-like trajectory. It implies that such systems have no symmetries that
can be taken into account and full time-evolutions should be made. However,
when the objects are relatively far apart, the emission of gravitational waves
is small and the inspiral slow. In this regime, one can hope to approximate
the real trajectory with a sequence of closed orbits. Moreover, the emission
of gravitational waves is known to efficiently circularize eccentric orbits so
that only circular orbits are usually considered.

So, a lot of efforts have been devoted to the computation of circular orbits in
general relativity. This can be done by demanding that the system admit an
helical Killing vector $\partial_t + \Omega \partial_\varphi$, $\Omega$ being
the orbital angular velocity of the system. Roughly speaking, this means that
advancing in time is equivalent to turning the system around its axis.
Working in the corotating frame, one is left with a time-independent problem.

Additional approximations must be made in order to avoid any diverging
quantities. Indeed, when using the helical symmetry, the system has an
infinite lifetime and can fill the whole space with gravitational waves, thus
causing quantities like the total mass to be infinite. To deal with that,
various techniques can be used, the simplest one consisting in preventing
the appearance of any gravitational waves. This is usually done by demanding
that the {\em spatial metric} be conformally flat. This is not a choice of
coordinates but a true approximation that has a priori no reason to be
verified. Indeed, even for a single rotating black hole, one can not find
coordinates in which the spatial 3-metric is conformally flat, so that we do
not expect it to be the case for binary systems. However comparisons with
post-Newtonian results or non-conformally flat results tend to indicate that
this approximation is relatively good.

Under these assumptions, Einstein equations reduce to a set of five elliptic
equations for the lapse, the conformal factor and the shift vector. Those
equations encompass both the Hamiltonian and momentum constraint equations and
the trace of the evolution equations. To close the system, one must provide a
description of the matter. It is commonly admitted that the fluid is
irrotational, the viscosity in neutron stars being too small to synchronize
the motion of the fluid with the orbital motion. It follows that the motion of
the fluid is described by an additional elliptic equation for the potential of
the flow. The matter terms entering the equations via the stress-energy tensor
can then be computed, once an equation of state is given. An evolutionary
sequence can be obtained by varying the separation between both stars.

\subsubsection{Numerical procedure}
\label{sss:bns_numeric}

Up to now, only the Meudon group has solved those equations by means of
spectral methods in the case of two neutron stars. Two sets of domains are
used, one centered on each star. Each set consists of spherical-like domains
that match the surface of the star and extend up to infinity. Functions are
expanded onto spherical harmonics with respect to the angles $\l(\theta,
\varphi\r)$ and Chebyshev polynomials with respect to the radial coordinates.
Each Poisson equation $\Delta N = S_N$ is split into two parts $\Delta N_1 =
S_{N_1}$ and $\Delta N_2 = S_{N_2}$, such that $S_N = S_{N_1} + S_{N_2}$ and
$N = N_1 + N_2$. The splitting is of course not unique and only requires that
$S_{N_i}$ is mainly centered around the star $i$ so that it is well described
by spherical coordinates around it. The equation labeled $i$ is then solved
using the domains centered on the appropriate star. The splittings used for
the various equations can be found explicitly in Section~IV-C
of~\cite{gourgoulhon-01}.

The elliptic equations are solved using the standard approach by the Meudon
group found in~\cite{grandclement-01}. For each spherical harmonic, the
equation is solved using a Tau-method and the matching between the various
domains is made using the homogeneous solutions method (see
Section~\ref{ss:hom}). The whole system of equations is solved by
iteration and most of the computational time is spent when quantities
are passed from on set of domains to the other one by means of a
spectral summation (this requires $N^6$ operations, $N$ being the
number of collocation points in one dimension). A complete and precise
description of the overall procedure can be found
in~\cite{gourgoulhon-01}.

\subsubsection{Binary neutron stars}

The first sequence of irrotational neutron star binaries computed by spectral
means is shown in~\cite{bonazzola-99}. Both stars are assumed to be polytropes
with an index $\gamma=2$. The results are in good agreement with those
obtained, simultaneously, by other groups with other numerical techniques (see
for instance~\cite{baumgarte-98, uryu-00}). One of the important points that
has been clarified by~\cite{bonazzola-99} concerns the evolution of the
central density of the stars. Indeed, at the end of the nineties, there was a
claim that both stars could individually collapse to black holes before
coalescence, due to the increase of central density as the two objects spiral
towards each other. Should that have been true, this would have had a great
impact on the emitted gravitational wave signal. However it turned out that
this was coming from a mistake in the computation of one of the matter term.
The correct behavior, confirmed by various groups and in particular
by~\cite{bonazzola-99}, is a decrease in the central density as the
stars get closer and closer (see Figure~I of~\cite{bonazzola-99}).

If the first sequence computed by spectral methods is shown
in~\cite{bonazzola-99}, the complete description and validation of the
method are given in~\cite{gourgoulhon-01}. Convergence of the results
with respect to the number of collocation points is exhibited. Global
quantities like the total energy or angular momentum are plotted as a
function of the separation and show remarkable agreement with results
coming from analytical methods (see Figures~8 to 15
of~\cite{gourgoulhon-01}). Relativistic configurations are also shown
to converge to the Newtonian counterparts when the compactness of the
stars is small (Figures~16 to 20 of~\cite{gourgoulhon-01}).

Newtonian configurations of compact stars with various equations of state are
computed for both equal masses~\cite{taniguchi-01} and various mass
ratios~\cite{taniguchi-02}. One of the main results of the
computations concerns the nature of the end point of the sequence. For
equal masses, the sequence ends at contact for synchronized binaries
and at mass shedding for irrotational configurations. This is to be
contrasted with the non-equal mass case where the sequence always ends
at the mass shedding limit of the smallest object.

Properties of the sequences in the relativistic regime are discussed
in~\cite{taniguchi-02b, taniguchi-03}. In~\cite{taniguchi-02b}
sequences with $\gamma=2$ are computed, for various compactness and
mass ratios, for both synchronized and irrotational binaries. The
nature of the end point of the sequences is discussed and similar
behavior to the Newtonian regime is observed. The existence of a
configuration of maximum binding energy is also discussed. Such
existence could have observational implications because it could be an
indication of the onset of a dynamical instability. Sequences of
polytropes with various indexes ranging from 1.8 to 2.2 are discussed
in~\cite{taniguchi-03}. In particular, the authors are lead to conjecture that,
if a configuration of maximum binding energy is observed in Newtonian regime,
it is also observed in conformal relativity for the same set of parameters.

In~\cite{faber-02} the authors derive, from the sequences computed
in~\cite{taniguchi-02b}, a method to constrain the compactness of the
stars from the observations. Indeed, from
results~\cite{taniguchi-02b}, one can easily determine the energy
emitted in gravitational waves per interval of frequency (i.e.\ the
power-spectrum of the signal). For large separation, that is for small
frequencies, the curves follow the Newtonian one. However, there is a
break frequency at the higher end (see Figure~2
of~\cite{faber-02}). The location of this frequency depends mainly on
the compactness of the stars. More precisely, the more compact the
stars are, the higher the break frequency is. Should such frequency be
observed by the gravitational wave detectors, this could help to put
constraints on the compactness of the neutron stars and thus on the
equation of state of such objects.

\subsubsection{Extensions}

The framework of~\cite{gourgoulhon-01} is applied to more realistic neutron
stars in~\cite{bejger-05}. In this work, the equations of state are more
realistic than simple polytropes. Indeed, three different equations are
considered for the interior, all based on different microscopic models. The
crust is also modeled. For all the models, the end point of the evolution
seems to be given by the mass shedding limit. However, the frequency at which
the shedding occurs depends rather strongly on the EOS. The results are in
good agreement with post-Newtonian ones, until hydrodynamic effects begin to
be dominant. This occurs at frequencies in the range 500--1000~Hz, depending
on the EOS.

Sequences of binary strange stars have also been computed~\cite{limousin-05}.
Contrary to the neutron star case, the matter density does not vanish at the
surface of the stars and one really needs to use surface-fitting domains to
avoid any Gibbs phenomenon that would spoil the convergence of the overall
procedure. Sequences are computed for both synchronized and irrotational
binaries and a configuration of maximum binding energy is attained in both
cases. This is not surprising: the strange stars are more compact than the
neutron stars and are less likely to be tidally destroyed before reaching the
extremum of energy, making it easier to attain dynamical instability.  More
detailed results on both neutron star and strange star binaries are discussed
in the follow-up papers~\cite{gondek-07,gondek-08}.

All the works presented above are done in the conformal flatness
approximation. As already stated in Section~\ref{sss:bns_form} this is only an
approximation and one expects that the true conformal 3-metric will depart
from flatness. However, in order to maintain asymptotic flatness of spacetime,
one needs to get rid of the gravitational wave content. One such waveless
approximation is presented in~\cite{shibata-04} and implemented
in~\cite{uryu-06}. Two independent codes are used, one of them being
an extension of the work described in~\cite{gourgoulhon-01}. The
number of equations to be solved is then greater than in conformal
flatness (one has to solve for the conformal metric), but the
algorithms are essentially the same. It turned out that the deviation
from conformal flatness is rather small. The new configurations are
slightly further from post-Newtonian results than the conformally flat
ones, which is rather counter-intuitive and might be linked to a
difference in the definition of the waveless approximations.

\subsection{Binary black hole systems}
\label{ss:binary_holes}

\subsubsection{Digging the holes}

If the computation of binary black holes in circular orbits has a lot of
common features with the neutron star case, there are also some differences
that need to be addressed. In at least one aspect, black holes are much
simpler objects because they are solution of Einstein equations without
matter. So the whole issue of investigating various equations of state is
irrelevant and there is no need to solve any equation for the matter. However,
there is a price to pay and one must find a way to impose the presence of
holes in the spacetime. Two main ideas have been proposed.

In the {\em puncture method}, the full spacetime contains three asymptotically
flat regions. One is located at $r=\infty$ and the two others at two
points $M1$ an $M2$ which are called the punctures. The presence of the flat regions
near the punctures is enforced by demanding that some quantities, like the conformal factor,
diverge at the those points (typically like $1/r$). The discontinuities are 
taken out analytically and
the equations are solved numerically for the regular parts, in the whole
space. This idea dates back to the work of Brill and
Lindquist~\cite{brill-63}, at least in the case of black holes
initially at rest.The puncture approach has been successfully applied to the
computation of quasi-circular orbits by means of spectral methods in \cite{ansorg-04}.

The {\em apparent horizon} method relies on initial works by
Misner~\cite{misner-63} and Lindquist~\cite{lindquist-63}. In this
case, the space has only two asymptotically flat regions. One can show
that this is equivalent to solving Einstein's equations outside two
spheres on which boundary conditions must be imposed. The boundary
conditions are based on the concept of trapped surface and apparent
horizons. The physical state of the black holes are precisely encoded
in the boundary conditions.

\subsubsection{First configurations}

The first configurations of binary black holes computed by means of spectral
methods can be found in~\cite{grandclement-02}. The formalism and various
hypothesis are given in the companion paper~\cite{gourgoulhon-02}. The
assumptions are very similar to those used for binary neutron stars
(see Section~\ref{sss:bns_form}). Helical symmetry is enforced and
conformal flatness assumed. The holes are described by the apparent
horizon technique. However, the boundary conditions used have been
shown to be only approximately valid, up to a rather good
accuracy. This effect is discussed in the original
paper~\cite{grandclement-02} and further explored by Cook
in~\cite{cook-02}. The numerical techniques are very similar to the
ones employed for binary neutron star configurations (see
Section~\ref{sss:bns_numeric}). Two sets of spherical domains are
used, one for each black hole. Boundary conditions are imposed on the
surface between the nucleus and the first shell. Both sets extend up
to infinity using a compactification in $1/r$.

For the first time, a good agreement was found between numerical results and
post-Newtonian ones. A detailed comparison can be found in~\cite{damour-02}. In
particular, the location of the minimum of energy is shown to coincide at the
level of a few percent. This improvement with respect to previous numerical
works is mainly due to a difference in the physical hypothesis (i.e. the use of
helical symmetry). One important product of~\cite{grandclement-02} is the use
of a new criterion to determine the appropriate value of the orbital angular
velocity $\Omega$. Indeed, for neutron stars, this is done by demanding that
the fluid of both stars be in equilibrium~\cite{gourgoulhon-01}. This, of
course, is not applicable for black holes. Instead,
in~\cite{gourgoulhon-02, grandclement-02}, it is proposed to find $\Omega$ by
demanding that the ADM mass and the Komar-like mass coincide. Without going
into to much details, this amounts to demanding that, far from the binary and
at first order in $1/r$, the metric behave like the Schwarzschild one. It is
shown in~\cite{gourgoulhon-02} that it can be linked to a relativistic virial
theorem. Since then, it has been shown that this criterion could also be used for
neutron stars~\cite{taniguchi-03} and that it was equivalent to the use of a
variational principle called {\em the effective potential
  method}~\cite{cook-94, baumgarte-00, pfeiffer-00, caudill-06}, where
the binding energy is minimized with respect to $\Omega$.

\subsubsection{Further investigations}\label{further}

More recently, two other spectral codes have been developed in the
context of binary black holes and successfully applied to address some
of the issues raised by the work of~\cite{gourgoulhon-02,
  grandclement-02}.

One of those codes is due to the Caltech/Cornell group by H.~Pfeiffer and
collaborators and is described extensively in~\cite{pfeiffer-03a,
  pfeiffer-03b}. The code is multi-domain and two main types of
domains are used i) square domains where each Cartesian-like
coordinate is expanded onto Chebyshev polynomials and ii) spherical
domains where spherical harmonics are used for the angles $\l(\theta,
\varphi\r)$ and Chebyshev polynomials for the radial coordinate. Space
can be compactified by a standard use of the variable $1/r$. The two
types of domains can be set up in various manners to accommodate the
desired geometry. When using both square and spherical domains, there
are regions of space that belong to more than one domain. This is to
be contrasted with work by the Meudon group where domains are only
touching but not overlapping. The algorithm of~\cite{pfeiffer-03a}
solves differential equations by using a multi-dimensional collocation
method. The size of the resulting system is roughly equal to the
number of collocation points. It is then solved iteratively via a
Newton--Raphson algorithm with line search. At each step of the Newton--Raphson method,
the linear system is solved by means of an iterative scheme (typically GMRES).
This inner iterative solver requires careful preconditioning to work properly.
Various tests are passed by the code in~\cite{pfeiffer-03a}, where elliptic equations
and systems are solved in either spherical or bispherical
topologies. In the cases presented the error decays spectrally.

In~\cite{pfeiffer-02} the code is used to investigate different ways of
solving the constraint equations. Three different decompositions are used: the
conformal TT one, the physical TT one and the thin-sandwich decomposition.
When solving for the constraint equations only, one also needs to precise some
{\em freely specifiable} variables, which describe the physical state of the
system. In~\cite{pfeiffer-02}, those specifiable variables are fixed using a
superposition of two Kerr--Schild black holes. The net result
of~\cite{pfeiffer-02} is that global quantities, like the total
energy, are very sensitive to the choice of decomposition. The
variation of total energy can be as large as 5\%, which is the order
of the energy released by gravitational waves. It is also shown that
the choice of extrinsic curvature tensor is more crucial than the one
of conformal metric, in accordance with an underlying result
of~\cite{grandclement-02}. Let us precise that
the equations derived form the helical Killing vector approach 
in~\cite{gourgoulhon-02, grandclement-02} are equivalent to the ones
obtained by making use of the thin-sandwich decomposition of the constraints.
The freely specifiable variables are obtained by both the imposition of the helical Killing symmetry
and by solving an additional equation for the lapse function (resulting in the
so-called {\em extended thin-sandwich} formalism).

In~\cite{cook-04} the boundary conditions based on the apparent horizon
formalism~\cite{cook-02} are implemented and tested numerically in the case of
one or two black holes. In the latter case, the main difference
with~\cite{grandclement-02}, lies in the use of more elaborate and
better boundary conditions on the horizons of the holes. By allowing
for non-vanishing lapse on the horizons, the authors of~\cite{cook-04}
solve the constraint equations exactly. This is to be contrasted
with~\cite{grandclement-02}, where the momentum constraint equation
was only solved up to some small correction. Anyway, both results show
a rather good agreement.  This is not surprising because the
correction used by the Meudon group was known to be small (see
Figures~10 and 11 of~\cite{grandclement-02}). More results are
presented in~\cite{caudill-06}, for both corotating and irrotational
black holes. An important result of~\cite{caudill-06} is the
comparison of the two criteria for determining the orbital angular
velocity $\Omega$. They indeed show that the {\em the effective
  potential method} first introduced in~\cite{cook-94} and the method
based on the virial theorem proposed in~\cite{gourgoulhon-02} are in
very good agreement.

By slightly extending the boundary conditions used in~\cite{caudill-06}, the
authors of~\cite{pfeiffer-07} proposed to reduce the eccentricity of the binary
black hole configurations. This is done by giving the holes a small radial
velocity by modifying the boundary condition on the shift vector. The code and
other equations are the same as in~\cite{caudill-06}. Time evolution of the
obtained initial data shows indeed that this technique can reduce the
eccentricity of the binary. However, the effect on the emitted gravitational
wave is small and probably unimportant.

Another application of the Caltech/Cornell solver can be found in~\cite{lovelace-08},
where the emphasis is put on nearly maximum spinning black holes. Initial
data are constructed for both single and binary black holes. Three families of
initial data are investigated. Using a formalism based on the Kerr--Schild
spacetime, the authors are able to reach spins as large as $a=0.9998$. Such
nearly maximum spinning black holes may be relevant from the astrophysical
point of view. Evolutions of those data are also discussed.

The other spectral code used to compute configuration of binary black holes is
due to M.~Ansorg~\cite{ansorg-05}. It shares a lot of features with the code
developed by the same author in the context of rotating stars~\cite{ansorg-02,
  ansorg-03} already discussed in Section~\ref{sss:complex}. Space is decomposed
in two domains. One of them lies just outside the horizons of the holes and
bispherical-like coordinates are used. The other domain extends up to infinity
and an appropriate mapping is used in order to avoid the singularity of the
bispherical coordinates at spatial infinity (see Section~IV of~\cite{ansorg-05}).
The angle of the bispherical coordinates (i.e.\ the angle around the x-axis
joining the two holes) is expanded onto Fourier series and the two other
coordinates onto Chebyshev polynomials. Like in~\cite{ansorg-05b,
  pfeiffer-03a}, the partial differential equations are solved using a
collocation method and the resulting system is solved by Newton-Raphson's
method. Once again the linear system coming from the Jacobian is solved by an
iterative scheme with preconditioning. The code is used to compute essentially
the same configuration as those shown in~\cite{caudill-06}. An interesting
point of~\cite{ansorg-05} is the detailed investigation of convergence of the
results when increasing the resolution. As can bee seen in Figure~4 of
\cite{ansorg-05}, the error starts by decreasing exponentially but, for high
number of points, it seems that the error only follows a power-law. This is an
indication that some non-$\mathcal{C}^\infty$ fields must be present. It is
conjectured in~\cite{ansorg-05} that this comes from logarithm terms that can
not be dealt with properly with a compactification in $1/r$. The same kind of
effect is investigated in some details in~\cite{grandclement-01}, where some
criteria for the appearance of such terms are discussed.

A code very similar to the one used in~\cite{ansorg-05} has also been used to
compute spacetimes with black holes using the puncture
approach~\cite{ansorg-04}. Given that the black holes are no longer
described by their horizons, one do not need to impose inner boundary
conditions. The absence of this requirement enabled the author
of~\cite{ansorg-04} to use a single domain to describe the whole
space, from the puncture up to infinity. The other features of the
spectral solver are the same as in~\cite{ansorg-05}. This method has
been successfully applied to the computation of binary black hole
configurations in the puncture framework. The authors have, in
particular, investigated high mass ratios between the bodies and
compared their results with the ones given in the test-mass limit
around a Schwarzschild black hole. The discrepancy is found to be of
the order of 50\% for the total energy. It is believed that this comes
from the fact that the mass of each puncture cannot be directly
related to the local black hole mass (see discussion in Section~VII
of~\cite{ansorg-04}).

Let us finally mention that the algorithms developed by M.~Ansorg
in~\cite{ansorg-02, ansorg-03, ansorg-04,ansorg-05} have all been
unified in~\cite{ansorg-07} to accommodate any type of
binaries. Various domain decompositions are exhibited that can be used
to represent neutron stars, excised black holes or puncture black
holes, with compactification of space. The algorithms are shown to be
applicable to limiting cases like large mass ratios.

\subsection{Black hole-neutron star binaries}

Until recently the binaries consisting of a neutron star and a black hole
received fewer attention than the other types of systems. It was believed, and
this was partly true, that this case could easily be handled once the cases of
binary neutron stars and binary black holes were understood. However, such
binaries are of evident observational interest and could be the most promising
source of gravitational waves for the ground-based
detectors~\cite{belczynski-02}.

The first application of spectral methods to the black hole-neutron star
binaries can be found in~\cite{taniguchi-05}. The main approximation is to
consider that the black hole is not influenced by the neutron star.
Technically, this means that the Einstein equations are split into two parts
(i.e.\ like for binary neutron stars~\ref{sss:bns_numeric}). However the part
of the fields associated to the black hole are fixed to their analytical
value. As the fields are not solved for the black hole part, the results
should depend on the actual splitting, the equations being non-linear. The
part of the fields associated with the neutron star are solved using the
standard setting for the Meudon group. Of course, this whole procedure is only
valid if the black hole is much more massive than the neutron star and this is
why~\cite{taniguchi-05} is limited to mass ratios of 10. In this particular
case, it is shown that the results depend to the level of a few percent on the
choice of splitting, which is a measure of the reached accuracy. It is also
shown that the state of rotation of the star (i.e.\ synchronized or
irrotational) has little influence on the results.

In~\cite{taniguchi-06} the equations of the extended thin-sandwich formulation
are solved consistently. Like for the binary neutron star case, two sets of
spherical coordinates are used, one centered around each object. The freely
specifiable variables are derived from the Kerr--Schild approach.
Configurations are obtained with a moderate mass ratio of 5. However the
agreement with post-Newtonian results is not very good and the data seem to be
rather noisy (especially the deformation of the star).

Quasi-equilibrium configurations based on a helical Killing vector and
conformal flatness have been obtained independently by~\cite{grandclement-06}
and~\cite{taniguchi-07}. Both codes are based on the {\sc Lorene}
library~\cite{lorene} and use two sets of spherical coordinates. They
differ mainly in the choice of boundary conditions for the black
hole. However, it is shown in the erratum of~\cite{grandclement-06}
that the results match pretty well and are in very good agreement with
post-Newtonian results. Mass ratios ranging from 1 to 10 are obtained
in~\cite{taniguchi-07} and the emitted energy spectra are
estimated. The work of~\cite{taniguchi-07} has been extended
in~\cite{taniguchi-08} where the parameter space of the binary is
extensively explored. In particular, the authors determine whether the
end-point of the sequences is due to an instability or to the
mass-shedding limit. It turns out that the star is more likely to
reach the mass-shedding limit if it is less compact and if the mass
ratio between the black hole and the star is important, as expected.

More recently, the Caltech/Cornell group has applied the spectral solver
of~\cite{pfeiffer-03a, pfeiffer-03b} in order to compute black
hole-neutron stars configurations~\cite{foucart-08}. Some extensions have been
made to enable the code to deal with matter by making use of surface fitting
coordinates. Thanks to the domain decomposition used (analogous to the
one of~\cite{pfeiffer-03a, pfeiffer-03b}), the authors
of~\cite{foucart-08} can reach an estimated accuracy of $5 \cdot
10^{-5}$, which is better than the precision of previous works (by
roughly an order of magnitude). Configurations with one spinning black
hole and configurations with reduced eccentricity are also presented,
in the line of~\cite{pfeiffer-07}.

\subsection{Spacetimes with waves}

The work~\cite{pfeiffer-05} presents a method to produce initial data
configuration containing waves. Given a fixed background metric, it shows how
to superimpose a given gravitational wave content. The equations are solved
numerically using a multi-domain spectral code based
on~\cite{pfeiffer-03a, pfeiffer-03b}. Space is covered by various spherical-like
shells and is described up to infinity. When no black hole is present, the
origin is covered by a square domain because regularity conditions at the
origin, in spherical coordinates, are not handled
by~\cite{pfeiffer-03a, pfeiffer-03b}. Such setting is used to generate spacetimes
containing i) pure quadrupolar waves ii) flat space with ingoing pulse and
iii) a single black hole superimposed with an ingoing quadrupolar wave.

\subsection{Hyperboloidal initial data}

If the 3+1 decomposition is the most widely used for numerical relativity,
some other possibilities have been proposed, with possibly better features. In
particular, one can vary the foliation of spacetime to get {\em hyperboloidal
  data}. With such a setting, at infinity spacetime is foliated by light cones
instead of spatial hypersurfaces, which makes extraction of gravitational
waves in principle easier.

In~\cite{frauendiener-99} Frauendiener is interested in generating hyperboloidal
initial data sets from data in physical space. The technique proceeds in two
steps. First a non-linear partial differential equation (the Yamabe equation)
must be solved to determine the appropriate conformal factor $\omega$. Then,
the data are constructed by dividing some quantities by this $\omega$. This
second step involves an additional difficulty: $\omega$ vanishes at infinity
but the ratios are finite and smooth. It is demonstrated
in~\cite{frauendiener-99} that spectral methods can deal with those
two steps. Some symmetry is assumed so that the problem reduces to a
2-dimensional one. The first variable is periodic and expanded onto
Fourier series whereas Chebyshev polynomials are used for the other
one. The Yamabe equation is solved using an iterative scheme based on
Richardson's iteration procedure. The construction of the fields,
hence the division by a field vanishing at infinity, is then handled
by making use of the non-local nature of the spectral expansion (i.e.\
by working in the coefficient space; see Section~4
of~\cite{frauendiener-99} for more details).

\newpage

\section{Dynamical Evolution of Relativistic Systems}
\label{s:dynamical_evolutions}

The modeling of time-dependent physical systems is traditionally the ultimate
goal in numerical simulation. Within the field of numerical relativity, the
need for studies of dynamical systems is even more pronounced because of the
seek for gravitational wave patterns. Unfortunately, as presented in
Section~\ref{ss:time_discretization}, there is no efficient spectral time
discretization yet and one normally uses finite-order time-differentiation
schemes. Therefore, in order to get high temporal accuracy, one needs to use
high-order explicit time marching schemes (e.g.\ fourth or sixth-order
Runge--Kutta~\cite{boyle-07}). This requires quite some computational power and
might explain why, except for gravitational collapse~\cite{gourgoulhon-91,
  novak-98}, very few studies using spectral methods have dealt with dynamical
situations until the Caltech/Cornell group began to use spectral methods in
numerical relativity, in the beginning of years 2000~\cite{kidder-00b,
  kidder-01}. This group now have a very-well developed pseudo-spectral
collocation code ``Spectral Einstein Code'' (SpEC), for the solution of full
three-dimensional dynamical Einstein equations.

In this section, we review the status of the numerical simulations using
spectral methods in some fields of General Relativity and Relativistic
Astrophysics. Although we may give at the beginning of every section a very
short introduction to the context of the relevant numerical simulations, it is
not our point to detail them since dedicated reviews exist for most of the
themes presented here and the interested reader should consult them for
physical details and comparisons with other numerical and analytical
techniques. Among the systems which have been studied, one can find
gravitational collapse~\cite{fryer-03} (supernova core collapse or collapse of
a neutron star to a black hole), oscillations of relativistic
stars~\cite{stergioulas-03, kokkotas-99} and evolution of ``vacuum''
spacetimes. These include the cases of pure gravitational waves or scalar
fields, evolving in the vicinity of a black hole or as (self-gravitating)
perturbations of Minkowski flat spacetime.  Finally, we shall discuss the
situation of compact binaries~\cite{postnov-06, blanchet-06} spectral
numerical simulations.

\subsection{Single Stars}
\label{ss:single_stars}

The numerical study of the evolution of stars in General Relativity involves
two parts: first one has to solve for the evolution of matter (relativistic
hydrodynamics, see~\cite{font-03}), and second one must compute the new
configuration of the gravitational field. Whereas, spectral-methods based
codes are now able to study quite well the second part (see
 Section~\ref{ss:einstein_system_in_vacuum}), the first part has not benefited
from so many efforts from the groups using spectral methods in the past
decade. One is facing the paradox: spectral methods have been primarily
developed for the simulation of hydrodynamic systems (see
 Section~\ref{ss:sm_in_physics}) but they are not often used for relativistic
hydrodynamics. This might be understood as a consequence of the general
problem of spectral methods to deal with discontinuous fields and supersonic
flows: the Gibbs phenomenon (see Section~\ref{sss:convergence}).  Relativistic
flows in astrophysics are often supersonic and therefore contain shocks.
Although some techniques have been devised to deal with them in
one-dimensional studies (see e.g.~\cite{bonazzola-91}), there have
been no multi-dimensional convincing work. Other problems coming from
multi-dimensional relativistic hydrodynamics which can spoil the rapid
convergence properties are the density sharp profiles near neutron star
surfaces. These can imply a diverging or discontinuous radial derivative of
the density, thus slowing down the convergence of the spectral series.

\subsubsection{Supernova core collapse}
\label{sss:core_collapse}

The physical scenario studied here is the formation of a neutron star from the
gravitational collapse of degenerate stellar core. This core can be thought as
to be the iron core of a massive star at the end of its evolution (standard
mechanism of type II supernova).  The degeneracy pressure of the electrons can
no longer support the gravity and the collapse occurs. When the central
density reaches nuclear values, the strong interaction stiffens the equation
of state, stopping the collapse in the central region and a strong shock is
generated. This mechanism has been long thought to be a powerful source of
gravitational radiation, but recent simulations show that the efficiency is
much lower than previously estimated~\cite{dimmelmeier-07, shibata-04}. The
first numerical study of this problem was the spherically symmetric approach
by May and White~\cite{may-66}, using artificial viscosity to damp the
spurious numerical oscillations caused by the presence of shock waves in the
flow solution. Nowadays, state-of-the-art codes use more sophisticated
High-Resolution Shock-Capturing (HRSC) schemes or High-Resolution Central
(HRC) Schemes (for details about these techniques, see the review by
Font~\cite{font-03}). The first axisymmetric fully (general) relativistic
simulations of the core collapse scenario have been done by
Shibata~\cite{shibata-03b}, Shibata and Sekiguchi~\cite{shibata-04}, which
then used HRSC schemes and a parametric equation of state. More recently,
magnetohydrodynamic effects have been taken into account in the axisymmetric
core collapse by Shibata~et~al.~\cite{shibata-06b}, using HRC schemes.
Three-dimensional core collapses simulations, including more realistic
equation of state and deleptonization scheme have been performed within the
{\sc cactus-carpet-whisky}~\cite{cactus, baiotti-05} framework by
Ott~et~al.~\cite{ott-07, ott-07b}. These simulations have been
compared with those of the {\sc CoCoNuT} code (see hereafter). A more
detailed historical presentation can
be found in the {\it Living Review\/} by Fryer and New~\cite{fryer-03}.

The appearance of a strong hydrodynamic shock is, in principle, a serious
problem to numerical models using spectral methods. Nevertheless, a first
preliminary study in spherical symmetry and in Newtonian theory of gravity has
been undertaken in 1986 by Bonazzola and Marck~\cite{bonazzola-86}, with the
use of ``natural'' viscosity. The authors showed a mass conservation to a
level better than $10^{-4}$ using one domain with only 33 Chebyshev
polynomials. In 1993, the same authors performed the first three-dimensional
simulation (still in Newtonian theory) of the pre-bounce
phase~\cite{bonazzola-93b}, giving a computation of the gravitational wave
amplitude, which was shown to be lower than standard estimates. Moreover, they
showed that for a given mass, the gravitational wave amplitude depends only
on the deformation of the core.  These three-dimensional simulations were made
possible thanks to the use of spectral methods, particularly for the solution
of the Poisson equation for the gravitational potential.

Shock waves give thus difficulties to spectral codes and have been either
smoothed with spectral vanishing viscosity~\cite{guo-01}, or ignored with the
code stopping before their appearance. Another idea developed first between
the Meudon and Valencia groups was then to use some more appropriate
techniques for the simulation of shock waves: namely the High-Resolution
Shock-Capturing techniques, also known as Godunov methods (see {\it Living
  Reviews\/} by Mart\'{\i} and M\"uller~\cite{marti-03}, and by
Font~\cite{font-03}). On the other hand, one wants to keep the fewer degrees
of freedom required by spectral methods for an accurate-enough description of
fields, in particular for the solution of elliptic equations or for the
representation of more regular fields, like the gravitational one. Indeed,
even in the case where a hydrodynamic shock is present, since it only appears
as a source for the metric in Einstein's equations, the resulting
gravitational field is at least ${\cal C}^1$ and the spectral series do
converge, although slower than in the smooth case. Moreover, in a multi-domain
approach, if the shock is contained within only one such domain, it is then
necessary to increase resolution in only this particular domain and it is
still possible to keep the overall number of coefficients lower than the
number of points for the HRSC scheme. The combination of both types of methods
(HRSC and spectral) was first achieved in 2000 by Novak and
Ib\'a\~nez~\cite{novak-00}. They studied a spherically symmetric core collapse
in tensor-scalar theory of gravity, which is a more general theory than
General Relativity and allows {\it a priori\/} for monopolar gravitational
waves. The system of PDEs to be solved resembles the General Relativity one,
with the addition of a scalar non-linear wave equation for the monopolar
dynamical degree of freedom. It was solved by spectral methods, whereas the
relativistic hydrodynamics equations were solved by Godunov techniques. Two
grids were used, associated to each numerical technique, and interpolations
between both were done at every time-step.  Although strong shocks were
present in this simulation, they were sharply resolved with HRSC techniques
and the gravitational field represented through spectral methods did not
exhibit any Gibbs-like oscillations. Monopolar gravitational waveforms could
thus be given. In collaboration with the Garching-hydro group, this numerical
technique has been extended in 2005 to three-dimensions, but in the so-called
conformally flat approximation of General Relativity (see
Sections~\ref{ss:binary_stars} and~\ref{ss:binary_holes}) by
Dimmelmeier~et~al.~\cite{dimmelmeier-05}. This approach using spectral methods
for the gravitational field computation is now sometimes referred as
``Marriage des Maillages'' (French for ``grid wedding'') and is currently
employed by the core-collapse code {\sc CoCoNuT} of
Dimmelmeier~et~al.~\cite{dimmelmeier-02, dimmelmeier-05} to study general
relativistic simulations to a proto-neutron star, with a microphysical
equation of state as well as an approximate description of
deleptonization~\cite{dimmelmeier-07}.

\subsubsection{Collapse to a black hole}
\label{sss:ns_collapse}

The stellar collapse to a black hole has been a widely studied subject,
starting with spherically symmetric computations: in the case of dust (matter
with no pressure), an analytical solution by Oppenheimer and
Snyder~\cite{oppenheimer-39} has been found in 1939. Pioneering numerical
works by Nakamura and Sato~\cite{nakamura-81, nakamura-82} studied the
axisymmetric general relativistic collapse to a black hole; Stark and
Piran~\cite{stark-85} gave the gravitational wave emission from such collapse,
in the formalism by Bardeen and Piran~\cite{bardeen-84}. Fully general
relativistic collapse simulations in axisymmetry have also been performed by
Shibata~\cite{shibata-00}, and the first three-dimensional calculations of the
gravitational-wave emission in the collapse of rotating stars to black holes
has been done by Baiotti~et~al.~\cite{baiotti-05}. Recently,
Stephens~et~al.~\cite{stephens-07} have developed an evolution code
for the coupled Einstein Maxwell-MHD equations, with the application
to the collapse to a black hole of a magnetized, differentially
rotating neutron stars.

To our knowledge, all studies of the collapse to a black hole which used
spectral methods are currently restricted to spherical symmetry. However, in
this case and contrary to the core-collapse scenario, there is {\it a
  priori\/} no shock wave appearing in the evolution of the system and
spectral methods are highly accurate also at modeling the hydrodynamics.
Thus, assuming spherical symmetry, the equations giving the gravitational
field are very simple, first because of the Birkhoff's theorem, which gives
the gravitational field outside the star, and then from the absence of any
dynamical degree of freedom in the gravitational field. For example, when
choosing the radial (Schwarzschild) gauge and polar slicing, Einstein
equations, expressed within 3+1 formalism, turn into two first-order
constraints which are simply solved by integrating with respect to the radial
coordinate (see~\cite{gourgoulhon-91}).

In the work by Gourgoulhon~\cite{gourgoulhon-91}, a Chebyshev tau-method is used. The
evolution equations for the relativistic fluid variables are integrated with
a semi-implicit time scheme and a quasi-Lagrangian grid: the boundary of the
grid is comoving with the surface of the star, but the grid points remains the
usual Gauss-Lobatto collocation points (Section~\ref{sss:gauss_quad}). Due to the
singularity-avoiding gauge choice, the collapsing star ends in the
``frozen-star'' state, with the collapse of the lapse. This induces strong
gradients on the metric potentials, but the code is able to follow the
collapse down to very small values of the lapse, at less than $10^{-6}$. The
code is very accurate at determining whether a star at equilibrium is
unstable, by triggering the physical instability from numerical noise at very
low level. This property has later been used by
Gourgoulhon~et~al.~\cite{gourgoulhon-95} to study the stability of
equilibrium configurations of neutron stars near the maximal mass,
taking into account the effect of weak interaction processes. The
addition of some inward velocity field to initial equilibrium
configurations enabled Gourgoulhon~\cite{gourgoulhon-92} to partially
answer the question about the minimal mass of black holes: can the
effective mass-energy potential barrier associated with stable
equilibrium states be penetrated by stars with substantial inward
radial kinetic energy? In~\cite{gourgoulhon-92}, Gourgoulhon found the
possibility to form a black hole with a starting neutron star which
was 10\% less massive than the usual maximal mass.

The spectral numerical code developed by Gourgoulhon~\cite{gourgoulhon-91} has
been extended to also simulate the propagation of neutrinos, coming from
thermal effect and non-equilibrium weak interaction processes. With this tool,
Gourgoulhon and Haensel~\cite{gourgoulhon-93} have simulated the neutrino
bursts coming from the collapse of neutron stars, with different equations of
state. Another modification of this spectral code has been done by
Novak~\cite{novak-98}, extending the theory of gravity to tensor-scalar
theories. This allowed for the simulation of monopolar gravitational waves
coming from the spherically symmetric collapse of a neutron star to a black
hole~\cite{novak-98}. From a technical point of view, the solution of a
non-linear wave equation on curved spacetime has been added to the numerical
model. It uses an implicit second-order Crank--Nicolson scheme for the linear
terms and an explicit scheme for the non-linear part. In addition, as for the
hydrodynamics, the wave equation is solved on a grid, partly comoving with
the fluid. The evolution of the scalar field showed that the collapsing
neutron star ``expelled'' all of its scalar charge before the appearance of
the black hole.

\subsubsection{Relativistic stellar pulsations}
\label{sss:relativistic_oscillations}

Oscillations of relativistic stars are often studied as a time-independent,
linear eigenvalue problem~\cite{kokkotas-99}. Nevertheless, numerical
approaches via time evolutions have proved to bring interesting results, as
obtained by Font~et~al.~\cite{font-02} for the first quasi-radial mode
frequencies of rapidly rotating stars in full general relativity. Nonlinear
evolution of the gravitational radiation driven instability in the $r$-modes
of neutron stars has been studied by many authors (for a presentation of the
physical problem, see Section~13 of~\cite{anderson-07}). In particular, the first
study of nonlinear $r$-modes in rapidly rotating relativistic stars, via
three-dimensional general-relativistic hydrodynamic evolutions has been done
by Stergioulas and Font~\cite{stergioulas-01}. Different approaches doing
numerical hydrodynamic simulations in Newtonian gravity have been performed
by Lindblom~et~al.~\cite{lindblom-02}, with an additional braking
term, as by Villain and Bonazzola~\cite{villain-02} (see hereafter).

Because of their very high accuracy, spectral methods are able to track
dynamical instabilities in the evolution of equilibrium neutron star
configurations, as shown in the previous section with the works of
Gourgoulhon~et~al.~\cite{gourgoulhon-91, gourgoulhon-95}. In these works, when
the initial data represented a stable neutron star, some oscillations
appeared, which corresponded to the first fundamental mode of the star. As
another illustration of the accuracy, let us mention the work by
Novak~\cite{novak-98b}, who followed the dynamical evolution of {\em
  unstable\/} neutron stars in tensor-scalar theory of gravity. The
instability is linked with the possibility for these stars to undergo some
``spontaneous scalarization'', meaning that they could gain a very high scalar
charge, whereas the scalar field would be very weak (or even null) outside the
star. Thus, for a given number of baryons there would be three equilibria for
a star: two stable ones with high scalar charges (opposite in sign) and an
unstable one with a weak scalar charge. Starting from this last one, the
evolution code described in~\cite{novak-98} was able to follow the transition
to a stable equilibrium, with several hundreds of damped oscillations for the
star. This damping is due to the emission of monopolar gravitational waves,
which carry away the star's kinetic energy.  The final state corresponds to
the equilibrium configuration, independently computed by a simple code solving
the generalized Tolman--Oppenheimer--Volkoff system with a scalar field, up to
1\% error, after more than 50,000 time-steps. These studies could be
undertaken with spectral methods because in these scenarios the flow remains
subsonic and one does not expect any shock to be formed.

It is therefore quite natural to try and simulate stellar pulsations using
spectral methods. Unfortunately, there have been only a few such studies,
which are detailed hereafter. The work by
Lockitch~et~al.~\cite{lockitch-03} has studied the inertial modes of
slowly rotating stars in full general relativity. They wrote down
perturbation equations in the form of a system of ordinary
differential equations, thanks to a decomposition onto vector and
tensor spherical harmonics. This system is then a nonlinear eigenvalue
problem for the dimensionless mode frequency in the rotating
frame. Equilibrium and perturbation variables are then expanded onto a
basis of Chebyshev polynomials, taking into account the coordinate
singularity at the origin and parity requirements. The authors were
therefore able to determine the values of the mode frequency making
the whole system singular and looked for eigenfunctions, through their
spectral decomposition. They found that inertial modes were slightly
stabilized by relativistic effects.

A different and maybe more natural approach, namely the time integration of
the evolution equations, has been undertaken by
Villain~et~al.~\cite{villain-02, villain-05} with a spectral
hydro code, in spherical coordinates. The code solves the linearized
Euler or Navier--Stokes equations, with the anelastic
approximation. This approximation, which is widely used in other
fields of astrophysics and atmospheric physics, consists in neglecting
acoustic waves by assuming that time derivatives of the pressure and the
density perturbations are negligible.  It allows for a characteristic time
which is not set by acoustic propagation time, but is much longer and the
time-step can be chosen so as to follow the inertial modes themselves. In
their 2002 paper~\cite{villain-02}, the authors study inertial modes
(i.e.\ modes whose restoring force is the Coriolis force, among which
the $r-$modes~\cite{anderson-07}) in slowly rotating polytropes with $\gamma=2$,
in the linear regime.  First, this is done in the framework of Newtonian gravity,
where the anelastic approximation implies that the Eulerian perturbations of
the gravitational potential do not play any role in the velocity
perturbations. Second, they study the relativistic case, but with the
so-called Cowling approximation, meaning again that the metric perturbations
are discarded. In both regimes and trying different boundary conditions for
the velocity field at the surface of the star, they note the appearance of a
polar part of the mode and the ``concentration of the motion'' near the
surface, showing up in less than 15 periods of the linear $r-$mode. A more
recent work~\cite{villain-05} deals with the study of gravity modes, in
addition to inertial modes, in neutron stars. The interesting point of this
work is the use of quite a realistic equation of state for nuclear matter,
which is valid even when the beta equilibrium is broken. The authors were thus
able to show that the coupling between polar and axial modes is increasing
with the rotation of the star, and that the coupling of inertial modes with
gravity modes in non-barotropic stars can produce fast energy exchanges
between polar and axial parts of the fluid motion. From a numerical point of
view, one of the key ingredients is the solution of the vector heat equation,
coming from the Euler or Navier--Stokes equations. This is done by a
poloidal-toroidal~\cite{boronski-07} decomposition of the velocity field onto
two scalar potentials, which is very natural within spectral methods.
Moreover, to ensure the correct analytical behavior at the origin, all scalar
quantities are projected at each time-step to a modified Legendre function
basis.

More recently, a complete non-linear study of rotating star pulsations has
been set by Dimmelmeier~et~al.~\cite{dimmelmeier-06}. They used the
general-relativistic code CoCoNuT (see above,
Section~\ref{sss:core_collapse}) in axial symmetry, with a HRSC
hydrodynamic solver, and spectral methods for the simplified Einstein
equations (conformally flat three-metric). They noted that the
conformal flatness condition did not have much effect on the dynamics,
when comparing with the Cowling approximation. Nevertheless, they
found that differential rotation was shifting the modes to lower
frequencies and they confirmed the existence of the mass-shedding
induced damping of pulsations.

\subsection{Vacuum and black hole evolutions}
\label{ss:einstein_system_in_vacuum}

If one wants to simulate the most interesting astrophysical sources of
gravitational radiation, one must have a code able to follow, in a stable
manner, gravitational waves themselves on a background spacetime. It has been
observed by all numerical relativity groups that the stability of a numerical
code, which solves Einstein field equations, does not only depend on the
numerical algorithm, but also on the particular formulation of the equations.
Successes in the simulations of binary systems of compact objects in General
Relativity (see Section~\ref{ss:dyn_binary}) are also due to numerous studies and
advances in the formulations of Einstein equations. The methods known at
present that work for the numerical evolution of binaries are the {\em
  generalized harmonic\/} coordinates~\cite{friedrich-85, garfinkle-02,
  pretorius-05} and the so-called {\em BSSN\/} (for
Baumgarte--Shapiro--Shibata--Nakamura~\cite{baumgarte-99, shibata-95}). In
particular, these two formulations of the field equations have the important
property that {\em constraint violating modes\/} (see discussion hereafter)
stay at a reasonable level during the evolution. Generalized harmonic gauge
needs constraint damping terms, with a particular method suited for harmonic
evolution which was proposed by Gundlach~et~al.~\cite{gundlach-05},
that enabled Pretorius to evolve black hole spacetimes~\cite{pretorius-05,
  pretorius-05b}.

It is therefore a crucial step to devise such a stable formulation, and more
particularly with spectral methods, because they add very little numerical
dissipation and thus, even the smallest instability is not dissipated away and
can grow up to unacceptable level. The situation becomes even more complicated
with the setup of an artificial numerical boundary at a finite distance from
the source, needing appropriate boundary conditions to control the physical
wave content, and possibly to limit the growth of unstable modes. All these
points have been extensively studied since 2000 by the Caltech/Cornell groups
and their pseudospectral collocation code SpEC~\cite{kidder-00a, kidder-01,
  scheel-02, scheel-04, lindblom-04, holst-04, kidder-05, lindblom-06,
  boyle-07}; they have been followed in 2004 by the Meudon
group~\cite{bonazzola-04} and in 2006 by Tichy~\cite{tichy-06}.

Next, it is necessary to be able to evolve black holes. Successful simulation
of binary black holes have been performed using the so-called black-hole
puncture technique~\cite{campanelli-06, baker-06}. Unfortunately, the
dynamical part of Einstein fields are not regular at the puncture points and
it seems difficult to regularize them so as to avoid any Gibbs-like phenomenon
using spectral methods. Therefore punctures are not generally used for
spectral implementations; instead the excision technique is employed, removing
part of the coordinate space inside the apparent horizon. There is no need for
boundary condition on this new artificial boundary, provided that one uses
free-evolution scheme~\cite{scheel-02}, solving only hyperbolic equations. In
the considered scheme, and also for hydrodynamic equations, one does not need
to impose any boundary condition, nor do any special treatment on the excision
boundary, contrary to finite difference techniques, where one must construct
special one-sided differencing stencils. On the other hand, with a constrained
scheme, elliptic-type equations are to be solved~\cite{bonazzola-04} and, as
for initial data (see Sections~\ref{ss:single_hole} and~\ref{ss:binary_holes})
boundary conditions must be provided e.g.\ on the apparent horizon,
from the dynamical horizon formalism~\cite{gourgoulhon-06b}.

Finally, good outer boundary conditions must
be devised, which are at the same time mathematically well-posed, consistent
with the constraints and prevent as much as possible reflections of outgoing
waves. In that respect, quite complete boundary conditions have been obtained
by Buchman and Sarbach~\cite{buchman-07}.

\subsubsection{Formulation and boundary conditions}
\label{sss:formulation_bc}

Several formulations have been proposed in the literature for the numerical
solution of Einstein equations, using spectral methods. The standard one is
the 3+1 (a.k.a.\ Arnowitt--Deser--Misner -- ADM) formalism of general
relativity~\cite{adm, york-79} (for a comprehensive introduction, see the
lecture notes by Gourgoulhon~\cite{gourgoulhon-07}), which has been
reformulated into the BSSN~\cite{baumgarte-99, shibata-95} for better
stability. But first, let us mention an alternative {\em characteristic\/}
approach based on expanding null hypersurfaces foliated by metric 2-spheres
developed by Bartnik~\cite{bartnik-97}. This formalism allows for a simple
analysis of the characteristic structure of the equations and uses the
standard ``edth'' ($\eth$) operator on $S^2$ to express angular derivatives.
Therefore, Bartnik and Norton~\cite{bartnik-00} used spin-weighted spherical
harmonics (see  Section~\ref{sss:spherical_harmonics}) to numerically describe
metric fields.

Coming back to the 3+1 formalism, Einstein's equations split into two subsets
of equations. First, the {\em dynamical\/} equations specifying the way the
gravitational field evolves from one time-slice to the next; then, the {\em
  constraint\/} equations which must be fulfilled on each time-slice.  Still,
it is well-known that for the Einstein system, as well as for the Maxwell's
equations of electromagnetism, if the constraints are verified on the initial
time-slice, then the dynamical equations guarantee that they shall be verified
in the future of that time-slice. Unfortunately, when numerically doing such
{\em free\/} evolution, i.e.\ solving only for the dynamical equations,
small violations of the constraints due to round-off errors appear to grow
exponentially (for an illustration with spectral methods, see
e.g.~\cite{scheel-02, tichy-06}). The opposite strategy is to discard
some of the evolution equations, keeping the equations for the two
physical dynamical degrees of freedom of the gravitational field, and
to solve for the four constraint equations: this is a {\em
  constrained\/} evolution~\cite{bonazzola-04}.

The advantages of the free evolution schemes are that they usually allow for a
writing of the Einstein's equations in the form of a strongly- or
symmetric-hyperbolic system, for which there are many mathematical theorems of
existence or well-posedness. In addition, it is possible to analyze such
systems in terms of characteristics, which can give very clear and
easy-to-implement boundary conditions~\cite{kidder-05}. Using
finite-differences numerical methods, it is also very CPU-time consuming to
solve for constraint equations, which are of elliptic type, but this is not
the case with spectral methods. On the other hand, constrained evolution
schemes have by definition the advantage of not being subject to
constraint-violation modes. Besides, the equations describing stationary
space-times are usually elliptic and are naturally recovered when taking the
steady-state limit of such schemes. Finally, elliptic PDEs usually do not
exhibit instabilities and are known to be well-posed. To be more precise,
constrained evolution using spectral methods has been implemented by the Meudon
group~\cite{bonazzola-04}, within the framework of BSSN formulation.
Free-evolution schemes have been used by Tichy~\cite{tichy-06} (with the BSSN
formulation) and by the Caltech/Cornell group, which has developed their
Kidder--Scheel--Teukolsky (KST) scheme~\cite{kidder-01} and have later used the
Generalized-Harmonic (GH) scheme~\cite{lindblom-06}. The KST scheme is in fact
a 12-parameters family of hyperbolic formulations of Einstein's equations,
which can be fine-tuned in order to stabilize the evolution of e.g.\
black hole spacetimes~\cite{scheel-02}. 

Even when doing so, constraint-violating modes grow exponentially and
basically three ways of controlling their growth have been studied by the
Caltech/Cornell group. First, the addition of multiples of the constraints to
the evolution system in a way to minimize this growth. The parameters linked
with these additional terms are then adjusted to control the evolution of the
constraint norm. This generalized version of the dynamical constraint control
method used by Tiglio~et~al.~\cite{tiglio-04}, has been presented by
Lindblom~et~al.~\cite{lindblom-04}, and tested on a particular representation
of the Maxwell equations. Second, the same authors devised constraint
preserving boundary conditions from those of
Calabrese~et~al.~\cite{calabrese-03}, where the idea was to get maximally
dissipative boundary conditions on the {\em constraint evolution\/}
equations~\cite{lindblom-04, kidder-05}. This second option appeared to be
more efficient, but still did not completely eliminate the instabilities.
Finally, bulk constraint violations cannot be controlled by
constraint-preserving boundary conditions alone, so
Holst~et~al.~\cite{holst-04} derived techniques to project at each time-step
the solution of the dynamical equations onto the constraint sub-manifold of
solutions. This method necessitates the solution of a covariant inhomogeneous
Helmholtz equation to determine the optimal projection. Nevertheless, the most
efficient technique seems to be the use of the GH formulation, which also
incorporates multiples of the constraints thus exponentially suppressing bulk
constraint violation, together with constraint-preserving boundary
conditions~\cite{lindblom-06}.

Boundary conditions are not only important for the control of the
constraint-violation modes in free evolutions. Because they cannot be imposed
at spatial infinity (see Section~\ref{sss:compactification}), they must be
completely transparent to gravitational waves and prevent any physical wave
from entering the computational domain. A first study of interest for
numerical relativity has been done by Novak and Bonazzola~\cite{novak-04},
where gravitational waves are considered in the wave zone, as perturbations of
flat spacetime. The specificity of gravitational waves is that they start at
the quadrupole level ($\ell=2$) in terms of spherical harmonics expansion.
Standard radiative boundary conditions (known as Sommerfeld boundary
conditions~\cite{sommerfeld-49}) being accurate only for the $\ell=0$
component, a generalization of these boundary conditions has been done to
include quadrupolar terms~\cite{novak-04}. They strongly rely on the spectral
decomposition of the radiative field in terms of spherical harmonics and on
spherical coordinates. More specific boundary conditions for the Einstein
system, in order to control the influx of the radiative part of the Weyl
tensor, have been devised by Kidder~et~al.~\cite{kidder-05} for the KST
formulation, generalizing earlier work by Stewart~\cite{stewart-98} and
Calabrese~et~al.~\cite{calabrese-03}. They were then adapted to the GH
formulation by Lindblom~et~al.~\cite{lindblom-06}. Rinne~\cite{rinne-06} has
studied the well-posedness of the initial-boundary value problem of the GH
formulation of Einstein equations. He has considered first-order boundary
conditions which essentially control the incoming gravitational radiation
through the incoming fields of the Weyl tensor. He has also tested the
stability of the whole system with a collocation pseudo-spectral code
simulation a Minkowski or Schwarzschild perturbed spacetime. Generalizing
previous works, a hierarchy of absorbing boundary conditions has been
introduced by Buchman and Sarbach~\cite{buchman-07}, which have then been
implemented in the Caltech/Cornell SpEC code by Ruiz~et~al.~\cite{ruiz-07},
together with new sets of absorbing and constraint-preserving conditions in
the generalized harmonic gauge. Ruiz~et~al. have shown that their second-order
conditions can control the incoming gravitational radiation, up to a certain
point. In addition, they have analyzed the well-posedness of the
initial-boundary value problems arising from these boundary conditions.
Rinne~et~al.~\cite{rinne-07} have compared various methods for treating outer
boundary conditions. They have used the SpEC code to estimate the reflections
caused by the artificial boundary, as well as the constraint violation it can
generate.

\subsubsection{Gauges and wave evolution}
\label{sss:gauges_waves}

The final ingredient before performing a numerical simulation of the dynamical
Einstein system is the gauge choice. For example, the analytical study of the
linearized gravitational wave in vacuum has been done with the harmonic gauge,
for which the coordinates $\left\{ x^\mu \right\}$ verify the scalar covariant
wave equation
\begin{equation}
  \label{eq:harmonic_gauge}
  H_\mu = g_{\mu\nu} \nabla_\sigma \nabla^\sigma x^\nu = 0.
\end{equation}
This is the definition of the form $H_\mu$, where $g_{\mu\nu}$ is the metric
and $\nabla_\sigma$ the associated covariant derivative. Recent works by the
Caltech/Cornell group used the GH formulation in which the gauge choice is achieved
through the specification of $H_\mu$ as an arbitrary function of $\left\{
  x^\mu \right\}$ and $g_{\mu\nu}$, which can be set for instance to its
initial value~\cite{scheel-06}. Still, it is with the KST formulation, and
with lapse and shift set from the analytic values, that
Boyle~et~al.~\cite{boyle-07} have submitted the Caltech/Cornell SpEC
code to the so-called ``Mexico City tests''~\cite{alcubierre-04}. These are a
series of basic numerical relativity code tests to verify their accuracy
and stability, including small amplitude linear plane wave, gauge wave and
Gowdy spacetime evolutions. These tests have been passed by the
Caltech-Cornell code, using Fourier basis for all three Cartesian coordinates,
and a fourth-order Runge-Kutta time-stepping scheme. In the particular case of
the linear plane wave, they exhibited the proper error behavior, as the square of
the wave amplitude, because all non-linear terms are neglected in this
test. The authors have also shown that the use of filtering of the spherical
harmonics coefficients was very effective in reducing nonlinear aliasing
instabilities. Gauge drivers for the GH formulation of Einstein equations have
been devised by Lindblom~et~al.~\cite{lindblom-08}. They provide an
elegant way of imposing gauge conditions that preserve hyperbolicity for many
standard gauge conditions. These drivers have been tested with the SpEC code.

Within the constrained formulation of Einstein's equations, the Meudon group has
introduced a generalization of the Dirac gauge to any type of spatial
coordinates~\cite{bonazzola-04}. Considering the conformal 3+1 decomposition of
Einstein's equations, the Dirac gauge requires that the conformal 3-metric
$\tilde{\gamma}^{ij}$ (such that $\det \tilde{\gamma}^{ij}=1$) be
divergence-free with respect to the flat 3-metric (defined as the asymptotic
structure of the 3-metric and with the associated covariant derivative $\bar{D}$)
\begin{equation}
  \label{eq:dirac_gauge}
  \bar{D}_i \tilde{\gamma}^{ij} = 0.
\end{equation}
Time coordinate is set by the standard maximal slicing condition. These
conditions turn to be {\em dynamical\/} gauge conditions: the lapse and the
shift are determined through the solution of elliptic PDEs at each time-step.
With this setting, Bonazzola~et~al. have studied the propagation of a
three-dimensional gravitational wave, i.e.\ the solution of the fully
nonlinear Einstein equations in vacuum. Their multidomain spectral code based
on the {\sc Lorene} library~\cite{lorene} was able to follow the wave using
spherical coordinates, including the (coordinate) singular origin, and to let
it out of the numerical grid with transparent boundary
conditions~\cite{novak-04}. Evolution was performed with a second-order
semi-implicit Crank--Nicolson time scheme, and the elliptic system of
constraint equations was solved iteratively. Since only two evolution
equations were solved (out of six), the other were used as error indicators
and proved the awaited second-order time convergence. A preliminary analysis
of the mathematical structure of the evolution part of this formalism done by
Cordero~et~al.~\cite{cordero-08} has shown that the choice of Dirac's
gauge for the spatial coordinates guarantees the strongly hyperbolic character
of that system as a system of conservation laws.

\subsubsection{Black hole spacetimes}
\label{sss:bh_spacetimes}

As stated at the beginning of Section~\ref{ss:einstein_system_in_vacuum}, the
detailed strategy to perform numerical simulations of black hole spacetimes
depends on the chosen formulation. With the characteristic approach, Bartnik
and Norton~\cite{bartnik-00} modeled gravitational waves propagating on a
black hole spacetime, in spherical coordinates but with a null coordinate
$z=t-r$. They interestingly combined a spectral decomposition on spin-weighted
spherical harmonics for the angular coordinates and an eighth-order scheme
using spline convolution to calculate derivatives in the $r$ or $z$ direction.
Integration in these directions was done with a fourth- or eighth-order
Runge--Kutta method. For the spectral part, they had to use Orszag's 2/3
rule~\cite{canuto-88} for anti-aliasing. This code achieved $10^{-5}$ as
global accuracy and was able to evolve the black hole spacetime up to $z=55M$.
More recently, Tichy has evolved a Schwarzschild black hole in Kerr--Schild
coordinates in the BSSN formulation, up to $t \simeq 100M$~\cite{tichy-06}. He
used spherical coordinates in a shell-like domain, excising the interior
of the black hole.  The expansion functions are Chebyshev polynomials for the
radial direction, and Fourier series for the angular ones.

Most successful simulations in this domain have been performed by the
Caltech/Cornell group, who seem to be able to stably evolve forever not only a
Schwarzschild, but also a Kerr black hole perturbed by a gravitational wave
pulse~\cite{lindblom-06}, using their GH formulation with constraint damping
and constraint-preserving boundary conditions. However, several attempts have been
reported by this group before that, starting with the spherically symmetric
evolution of a Schwarzschild black hole by
Kidder~et~al.~\cite{kidder-00b}. Problems had arisen when trying
three-dimensional simulations of such physical systems with the new
parameterized KST formalism~\cite{kidder-01}. Using spherical
coordinates in a shell-like domain, the authors decomposed the fields
(or Cartesian components for tensor fields) on a Chebyshev radial base
and scalar spherical harmonics. The integration in time was done using
a fourth-order Runge--Kutta scheme and the gauge variables were
assumed to keep their analytical initial values. The evolution was
limited by the growth of constraint-violating modes at $t\sim
1000M$. With a fine-tuning of the parameters of the KST scheme,
Scheel~et~al.~\cite{scheel-02} have been able to extend the lifetime
for the numerical simulations to about $8000M$. On the other hand,
when studying the behavior of a dynamical scalar field on a fixed Kerr
background, Scheel~et~al.~\cite{scheel-04} managed to get nice results
on the late time decay of this scalar field. They had to eliminate high-frequency numerical
instabilities, with a filter on the spherical harmonics basis, following again
Orszag's 2/3 rule~\cite{canuto-88} and truncating the last third of coefficients.
It is interesting to note that no filtering was necessary on the radial
(Chebyshev) basis functions. Somehow more complicated filtering rule has been
applied by Kidder~et~al.~\cite{kidder-05}, when trying to limit the
growth of constraint-violation in three-dimensional numerical evolutions of
black hole spacetimes, with appropriate boundary conditions. They have set to
zero the spherical harmonics terms with $\ell \geq \ell_{\rm max} -3$ in the
{\em tensor\/} spherical harmonics expansion of the dynamical fields. The
stable evolutions reported by Lindblom~et~al.~\cite{lindblom-06} thus
might be due to the following three ingredients:

\begin{itemize}
  \item GH formalism, exponentially suppressing all small
  short-wavelength constraint violations,
  \item constraint-preserving boundary conditions,
  \item filtering of spherical harmonics spectral coefficients.
\end{itemize}

\subsection{Binary systems}
\label{ss:dyn_binary}

As seen in Section~\ref{ss:einstein_system_in_vacuum}, not many groups using
spectral methods are able to solve all the three-dimensional Einstein
equations in a stable way. When dealing with black holes, the situation is
even worse. Only very recently, the Caltech/Cornell group succeeded in
following 16 orbits, merger and ring-down of an equal-mass non-spinning binary
black hole system~\cite{scheel-08}. Moreover, we report on three recent
partial works on the field using spectral methods, dealing with each type of
binary system (neutron stars and/or black holes) and leave space for future
studies on this rapidly evolving field. We note, of course, that successful
numerical evolutions of such systems have been performed with other numerical
methods, which we very briefly summarize here. First successful
fully-relativistic simulation of binary neutron stars has been obtained by
Shibata~et~al.~\cite{shibata-03, shibata-05} and now, more groups are also
able to study such systems: the Louisiana State University (LSU)
group~\cite{anderson-08} and the Albert Einstein Institute (AEI, Golm)
group~\cite{baiotti-08}. We also mention here the simulations with more
detailed microphysics by Oechslin and Janka~\cite{oechslin-07}, although with
the conformally flat approximation, and those of Liu~et~al.~\cite{liu-08}
evolving magnetized neutron star binaries. Shibata and
Ury\=u~\cite{shibata-06, shibata-07} have successfully evolved black
hole-neutron star binaries, using the puncture technique for the modeling of
the black hole. As far as black hole binary systems are concerned, after many years
of hard work and codes evolving the binary system for a restricted time, a
first stable simulation up to the merger phase has been performed by
Pretorius~\cite{pretorius-05b} who used general harmonic coordinates together
with constraint-damping terms and a compactification of spatial infinity. He
also used the excision technique for a region surrounding the singularity
inside the horizon. This first success was followed a few moths later by the
Texas/Brownsville group~\cite{campanelli-06} and the NASA/Goddard
group~\cite{baker-06}, using very different techniques, namely BSSN with
moving punctures and ``1+log'' slicing together with ``$\Gamma$-driver'' shift
condition. These techniques have rapidly become standards for many other
groups, which are now able to stably evolve binary black holes, as the AEI/LSU
collaboration~\cite{pollney-07}, group in Jena~\cite{gonzalez-07}, at
Pennsylvania State University~\cite{herrmann-07} and Florida Atlantic
University~\cite{tichy-07}. The results have reached high-confidence level and
it was possible to compare gravitational waveforms obtained numerical
evolution to post-Newtonian template families~\cite{pan-08}.

\subsubsection{Binary neutron stars}
\label{sss:bin_NS}

Numerical simulations of the final stage of inspiral and merger of binary
neutron stars has been performed by Faber~et~al.~\cite{faber-04}, who
have used spectral methods in spherical coordinates (based on {\sc Lorene}
library~\cite{lorene}) to solve the Einstein equations in the conformally flat
approximation (see Sections~\ref{s:station} and~\ref{sss:core_collapse}).  The
hydrodynamic evolution has been computed using a Lagrangian smoothed particle
hydrodynamics (SPH) code. As for the initial conditions, described in
 Section~\ref{ss:binary_stars}, the equations for the gravitational field reduce,
in the case of the conformally flat approximation, to a set of five non-linear
coupled elliptic (Poisson-type) PDEs. The considered fields (lapse, shift and
conformal factor) are ``split'' into two parts, each component being
associated to one of the stars in the binary. Although this splitting is not
unique, the result obtained is independent from it because the equations with
the complete fields are solved iteratively, for each time-step. Boundary
conditions are imposed to each solution of the field equations at radial
infinity, thanks to a multidomain decomposition and a $u=1/r$ compactification
in the last domain. The authors used $\sim 10^5$ SPH particles for each run,
with an estimated accuracy level of 1--2\%. Most of the CPU time was spent in
calculating the values of quantities known from their spectral representation,
at SPH particle positions. Another difficulty has been the determination of
the domain boundary containing each neutron star, avoiding any Gibbs
phenomenon.  Because the conformally flat approximation discards gravitational
waves, the dissipative effects of gravitational radiation back reaction were
added by hand. The authors used the slow-motion approximation~\cite{wilson-96}
to induce a shrinking of the binary systems, and the gravitational waves were
calculated with the lowest-order quadrupole formulas. The code has passed many
tests and, in particular, they have evolved several quasi-equilibrium binary
configurations without adding the radiation reaction force with resulting
orbits that were nearly circular (change in separation lower than 4\%. The
code was thus able to follow irrotational binary neutron stars, including
radiation reaction effects, up to the merger and the formation of a
differentially rotating remnant, which is stable against immediate
gravitational collapse for reasonably stiff equations of state. All the
results agreed pretty well with previous relativistic calculations.

\subsubsection{Black hole-neutron star binaries}
\label{sss:bin_BHNS}

A similar combination of numerical techniques has been used by
Faber~et~al.~\cite{faber-06} to compute the dynamical evolution of
merging black hole-neutron star binaries. In addition to the
conformally flat approximation and similarly to
Taniguchi~et~al.~\cite{taniguchi-06}, the authors considered only the
case of extremely large mass ratio between the black hole and the
neutron star, holding thus the black hole position fixed and
restricting the spectral computational grid to a neighborhood of the
neutron star. The black hole surrounding metric was thus supposed to
keep the form of a Schwarzschild black hole in isotropic
coordinates. The neutron star was restricted to low compactness (only
a few percents) in order to have systems that disrupt well outside the
last stable orbit. The system was considered to be in corotation and,
as for binary neutron stars, the gravitational radiation reaction was
added by hand. As stated above, the numerical methods used SPH
discretization to treat dynamical evolution of matter, and the
spectral library {\sc Lorene}~\cite{lorene} to solve the Einstein
field Poisson-like equations in the conformally flat approximation.
But here, the spectral domains associated with the neutron star did
not extend to radial infinity (no compactified domain) and approximate
boundary conditions were imposed, using multipole expansion of the
fields. The main reason is that the black hole central singularity
could not be well described on the neutron star grid.

The authors have studied the evolution of neutron star-black hole binaries
with different polytropic indices for the neutron star matter equation of
state, the initial data being obtained as solutions of the conformal
thin-sandwich decomposition of Einstein equations. They found that, at
least for some systems, the mass transfer from the neutron star to the black
hole plays an important role in the dynamics of the system. For most of these
systems, the onset of tidal disruption occurred outside the last stable orbit,
contrary to what had been previously proposed in analytical models. Moreover,
they have not found any evidence for significant shocks within the body of the
neutron star. This star possibly expanded during the mass loss, eventually
loosing mass outward and inward, provided that it was not too far within the
last stable orbit. Although the major part of released matter remained bound
to the black hole, a significant fraction could be ejected with sufficient
velocity to become unbound from the binary system.

\subsubsection{Binary black holes}
\label{sss:bin_BH}

Encouraging results concerning binary black holes simulations with spectral
methods have been first obtained by Scheel~et~al.~\cite{scheel-06}. They
have used two coordinate frames to describe the motion of black holes in the
spectral grid. Indeed, when using excision technique (punctures are not
regular enough to be well represented by spectral methods), excision
boundaries are fixed on the numerical grid. This can cause severe problems
when, due to the movement of the black hole, the excision surface can become
timelike and the whole evolution problem is ill-posed in the absence of
boundary conditions. So one solution seems to be the use of comoving
coordinates, but the authors report that the GH formulation they use appear to
be unstable with this setting. They therefore consider a first system of
inertial coordinates (with respect to spatial infinity) to define the tensor
components in the triad associated with these coordinates; and a second system
of comoving (in some sense) coordinates. In the case of their binary black
hole tests~\cite{scheel-06}, they define the comoving coordinates dynamically,
with a feedback control system that adjusts the moving coordinate frame to
control the location of each apparent horizon center. 

The spectral code uses 44 domains of different types (spherical and
cylindrical shells, rectangular blocks) to describe the binary system. Most of
the numerical strategy to integrate Einstein equations is taken from their
tests on the GH formulation by Lindblom~et~al.~\cite{lindblom-06} and
have already been detailed in  Section~\ref{sss:formulation_bc}. the important
technical ingredient detailed by Scheel~et~al.~\cite{scheel-06} is the
particular filtering of tensor fields in terms spherical harmonics. The
dual-coordinate-frame representation can mix the tensor spherical harmonic
components of tensors. So, in their filtering of the highest-order tensor
spherical harmonic coefficients, the authors had to take into account this
mixing by transforming spatial tensors to a rotating frame tensor spherical
harmonic basis before filtering and then transforming back to inertial frame
basis. This method allowed them to evolve binary black hole spacetimes for
more than four orbits, until $t\gtrsim 600 M_{\rm ADM}$. 

However, a central problem has been the capability of the code to follow the
merger phase, and even though the code was able to compute the inspiral quite
accurately, it used to fail just before the holes merged. The problem was
that, when both black holes were coming close to each other, their horizons
became extremely distorted and strong gradients would develop in the dynamical
fields. This has been explained as a gauge effect, coming from the incapacity
of the gauge condition to react and change the geometry when the two black
holes begin to interact strongly, and can be seen as a coordinate singularity
developing close to the merger. Nevertheless, a cure has been found as
explained in Scheel~et~al.~\cite{scheel-08}. The original gauge is kept until
some given time and then smoothly changed to a new one, based on the gauge
treatment by Pretorius~\cite{pretorius-05, pretorius-05b} (for the lapse): the
gauge source function is evolved through a damped, driven wave equation, which
drives the lapse toward unity and the shift vector toward zero near the
horizons. Thus the horizons expand in coordinate space and the dynamical
fields are smoothed out near the horizons, preventing gauge singularities from
developing. With this transition of the gauge condition, the evolution of the
black holes can be tracked until the formation of a common horizon
encompassing both black holes. Then, the evolution of this single-distorted
dynamical black hole is achieved by first interpolating all variables onto a
new computational domain containing only one excised region, then by choosing
a new comoving coordinate system, and finally by modifying again the gauge
condition to drive the shift vector to a time-independent state.  

This new gauge conditions have allowed Scheel~et~al.~\cite{scheel-08} to
follow the inspiral during 16 orbits, then the merger and the ring-down phase
of an equal-mass non-spinning binary black hole system. They were able to
compute the mass and the spin of the final black hole with very high accuracy
($10^{-5}$ and $10^{-4}$ relative accuracy for the mass and spin
respectively), and to extract the physical waveform up accurately to 0.01
radian in phase. This is the first spectral numerical simulation of the full
evolution of a binary black hole system.

\newpage

\section{Conclusions}
\label{s:conclusions}

We would like to conclude our overview of spectral methods in
numerical relativity by pointing a few items that we feel are
particularly interesting.

\subsection{Strengths and weaknesses}
\label{ss:strengths_and_weaknesses}

The main advantage of spectral methods, especially with respect to finite
difference ones, is the very rapid convergence of the numerical approximation
to the real function. This implies that very good accuracy can usually be
reached with only a moderate number of points. This obviously makes the codes
both faster and less demanding in memory. Various examples of convergence can
be found in Section~\ref{section:theoretical_foundations}. However, this rapid
convergence is only achieved for ${\mathcal C}^\infty$ functions. Indeed, when
the functions are less continuous, spurious oscillations appear and the
convergence only follows a power-law. In the case of discontinuous functions,
this is known as the Gibbs phenomenon (see the extreme case of
Figure~\ref{figure:gibbs}). Gibbs-like phenomena are very likely to
prevent codes from converging or to make time evolutions unstable. So
spectral methods must rely heavily on domain decomposition of space
and the domains must be chosen so that the various discontinuities lie
at the boundaries. Because of this, spectral methods are usually more
difficult to implement than standard finite differences (see for
instance the intricate mappings of~\cite{ansorg-07}). The situation is
even more complicated when the surfaces of discontinuities are not
known in advance or have complicated shapes.

Spectral methods are also very efficient at dealing with problems that are
related to coordinate singularities. Indeed, if the basis functions fulfill
the regularity requirements, then all the functions will automatically satisfy
them. In particular, it makes the use of spherical coordinates much easier
than with other methods, as explained in
Section~\ref{ss:spherical_coordinates_harmonics}.

Another nice feature is the fact that a function can be represented either by
its coefficients or its values at the collocation points. Depending on the
operation one has to perform, it is easier to work with the one representation or
the other. When working in the coefficients space, one takes full advantage of
the non-locality of the spectral representation. A lot of operations that
would be difficult otherwise can then be easily performed, like computing the
ratio of two quantities vanishing at the same point (see for
instance~\cite{frauendiener-99}).

\subsection{Combination with other methods}
\label{ss:combination_with_other_methods}

Spectral methods have also demonstrated that they can be a valuable tool when
combined with other methods. For instance, when shocks are present, spectral
methods alone have trouble dealing with discontinuities at the shock
interface. However, this can be efficiently dealt with using Godunov methods.
Such a combination has already been successfully applied to the simulation of
the oscillations of compact stars in~\cite{dimmelmeier-06} and of core
collapse~\cite{ott-07}.

Spectral methods have also been used in conjunction with a description of the
fluid based on SPH (smoothed particle hydrodynamics) in the case of binary
neutron stars~\cite{faber-04} and for the merger of one neutron star and one
black hole~\cite{faber-06}. In both cases, the fluid is described by an
ensemble of particles on which forces are applied. Such technique can account
for complicated fluid configurations, like the spiral arms that are formed
during the merger. Such arms would be tricky to follow by means of spectral
methods alone. On the other hand, the equations describing gravity are nicely
solved by spectral solvers.

\subsection{Future developments}
\label{ss:future_developments}

Finally, we would like to point out a few of the directions of work
that could lead to interesting results. Of course, we are not aware of
what the other groups have planed for the future.

Appropriate choice of coordinates is evidently important. However, for
binary systems, rather few results have been using the natural choice
of the bispherical coordinates. So far, variations of such coordinates
have only been used by M.~Ansorg and collaborators and only in the
context of initial data~\cite{ansorg-04,ansorg-05,ansorg-07}. We
believe that applying those coordinates, or similar coordinates, to
evolutionary codes could lead to interesting results, in terms of both
speed and accuracy.

The application of spectral methods to theories more complicated the general
relativity is also something that can be imagined. One of the possible fields
of application is the study of branes, where there is an additional dimension to
spacetime. The fact that spectral methods are accurate with relatively few
degrees of freedom, makes them a good candidate to simulate systems with
extra-dimensions. The addition of gauge fields is also something that could be
studied with spectral methods, to investigate the possibility of ``hairy''
black holes, for instance. Of course, those are just a few leads on what the
future applications of spectral methods to the fields of relativity might be.

\newpage


\bibliography{refs}

\end{document}